\documentclass[nofootinbib,aps, twocolumn,floatfix,superscriptaddress,prb]{revtex4}
\bibpunct{[}{]}{;}{n}{}{}

\usepackage[greek, english]{babel}
\usepackage{bbm} 
\usepackage{amsmath, amsfonts, amssymb, mathtools}    % need for subequations}   % need for figures
\usepackage{color}      % use if color is used in text
\usepackage{subfigure}  % use for side-by-side figures
\usepackage{epstopdf}
\usepackage{hyperref}   % use for hypertext \elinks, including those to external documents and URLs
\usepackage{tabularx}
\usepackage{dcolumn}

\usepackage{physics}
\usepackage{slashed} 
\usepackage{tensor}
\newcommand{\moy}[1]{\left \langle #1 \right  \rangle}
\newcommand{\moye}[1]{\langle #1 \rangle}
\def\man{\mathcal{M}}

\def\O{\mathcal{O}}

\def\psib{\bar{\psi}}
\def\Psib{\bar{\Psi}}
\def\dag{\dagger}
\def\pa{\partial}

\def\half{\frac{1}{2}}
\def\ddi#1{\left ( \dd \mathbf{#1} \right )}
\DeclareMathOperator\hc{h.c.}
\DeclareMathOperator\sgn{sgn}

\DeclareMathOperator\SU{SU}
\DeclareMathOperator\SO{SO}
\DeclareMathOperator\U{U}

\def\Z{\ensuremath{\mathbb{Z}}}

\def\R{\ensuremath{\mathbb{R}}}
\def\C{\ensuremath{\mathbb{C}}}

\def\Id{\ensuremath{\mathbbm{1}}}
\usepackage{cancel}
\newcommand{\uds}[1]{\underset{#1}}

\def\lb{\left(}
\def\rb{\right)}
\def\lc{\left[}
\def\rc{\right]}

\def\l.{\left.}
\def\r.{\right.}

\def\beq{\begin{equation}}
\def\eeq{\end{equation}}
\def\bsp{\begin{split}}
\def\esp{\end{split}}
\def\bea{\begin{eqnarray}}
\def\eea{\end{eqnarray}}
\def\beano{\begin{eqnarray*}}
\def\eeano{\end{eqnarray*}}

\newcommand{\eqn}[1]{\begin{align}#1\end{align}}

\newcommand{\eqna}[2]{\begin{alignat}{#1}#2\end{alignat}}

\newcommand{\pmatr}[1]{\begin{pmatrix}#1\end{pmatrix}}

\newcommand{\bmatr}[1]{\begin{bmatrix}#1\end{bmatrix}}
\newcommand{\bmatrs}[1]{\begin{bmatrix*}[l]#1\end{bmatrix*}}
\def\nn{\nonumber}

\definecolor{gris}{rgb}{0.4,0.4,0.4}

\definecolor{cadmiumgreen}{rgb}{0.0, 0.42, 0.24}

\newcommand{\dia}[1]{\raisebox{-.5\height}{\includegraphics[width=0.1\linewidth]{#1}}}
\newcommand{\diagen}[2]{\raisebox{-.5\height}{\includegraphics[width=#1\linewidth]{#2}}}
\def\u{\uparrow}
\def\d{\downarrow}

\def\s{s}
\def\v{v}
\def\uff{\bm Q}
\def\dff{- \bm Q}

\def\S{\Sigma}

\def\a{\bm{a}}

\DeclareMathOperator\CP{CP}
\DeclareMathOperator\mQED{QED}
\DeclareMathOperator\WF{WF}

\DeclareMathOperator\GN{GN}
\DeclareMathOperator\GNY{GNY}
\DeclareMathOperator\cHGNY{cHGNY}
\DeclareMathOperator\cHGN{cHGN}

\def\C{\mathcal{C}}

\def\T{\mathcal{T}}

\def\K{f}

\def\lag{\mathcal{L}}

\usepackage{feyn}
\def\nb{N_b}
\def\nf{N_f}

\def\sl{\slashed}
\providecommand{\qg}[4]{\delta^{#1#2}\delta^{#3#4} - \delta^{#1#3}\delta^{#2#4} + \delta^{#1#4}\delta^{#2#3}}
\providecommand{\qgd}[4]{\delta^{#1#2}\delta^{#3#4} + \delta^{#1#3}\delta^{#2#4} + \delta^{#1#4}\delta^{#2#3}}
\usepackage{wasysym} %hexagon symbol

\def\veps{\varepsilon}

\newcommand*\chem[1]{\ensuremath{\mathrm{#1}}}

\usepackage{bm}

\def\zero{``zero''}

\usepackage{lipsum}

\def\mg{M_q}
\def\m{M} 
\def\g{q}
\def\O{\mathcal{O}}
\def\S{\mathcal{S}}

\begin{document}
\title{Transition from a Dirac spin liquid to an antiferromagnet: Monopoles in a QED$_3$-Gross-Neveu theory}
\author{\'Eric Dupuis}
\affiliation{D\'epartement de physique, Universit\'e de Montr\'eal, Montr\'eal (Qu\'ebec), H3C 3J7, Canada}
\author{M.B. Paranjape}
\affiliation{D\'epartement de physique, Universit\'e de Montr\'eal, Montr\'eal (Qu\'ebec), H3C 3J7, Canada}
\affiliation{Centre de Recherches Math\'ematiques, Universit\'e de Montr\'eal; P.O. Box 6128, Centre-ville Station; Montr\'eal (Qu\'ebec), H3C 3J7, Canada}
\author{William Witczak-Krempa}
\affiliation{D\'epartement de physique, Universit\'e de Montr\'eal, Montr\'eal (Qu\'ebec), H3C 3J7, Canada}
\affiliation{Centre de Recherches Math\'ematiques, Universit\'e de Montr\'eal; P.O. Box 6128, Centre-ville Station; Montr\'eal (Qu\'ebec), H3C 3J7, Canada}

\date{\today}

\begin{abstract}
We study the quantum phase transition from a Dirac spin liquid to an antiferromagnet driven by condensing
monopoles with spin quantum numbers.  
We describe the transition in field theory by tuning a fermion interaction to condense a spin-Hall mass,  which in turn allows the appropriate monopole operators to proliferate and confine the fermions.  We compute various critical exponents at the quantum critical point (QCP), including the scaling dimensions of monopole operators by using the state-operator correspondence of conformal field theory. We find that the degeneracy of monopoles in QED$_3$ is lifted and yields a non-trivial monopole hierarchy at the QCP. In particular, the lowest monopole dimension is found to be smaller than that of QED$_3$ using a large $N_f$ expansion where 
$2N_f$ is the number of fermion flavors.  For the minimal magnetic charge, this dimension is $0.39N_f$ at leading order.
%We also give an analytic approximation of the scaling dimensions for large magnetic charge.  
We also study the QCP between Dirac and chiral spin liquids, which allows us to test a conjectured duality to a bosonic
CP$^1$ theory.
Finally, we discuss the implications of our results for quantum magnets on the Kagome lattice.
\end{abstract}
\maketitle
\tableofcontents

\section{Introduction}
Quantum spin liquids (QSLs) are strongly correlated phases of matter characterized by long-range entanglement, fractionalized excitations and, in some cases, topological order \cite{wen_quantum_2007, balents_spin_2010, zhou_quantum_2017}. QSLs can arise in frustrated antiferromagnets where important quantum fluctuations lead to a highly entangled and non-magnetic ground state. In recent years, many candidate materials that may realize a QSL have been identified~\cite{wwk_correlated_2014, norman_colloquium_2016, savary_quantum_2016,  zhou_quantum_2017, takagi_kitaev_2019}. 

The fractionalized excitations of a QSL are said to be deconfined as they don't appear in ordered phases. One important aspect to better understand the fractionalized aspect of these phases of matter is to characterize their transition to confined phases, that is to characterize confinement-deconfinement phase transitions.   In this respect, the $\U(1)$ Dirac spin liquid (DSL) or algebraic spin liquid, which potentially describes certain two-dimensional QSLs at low energy, is an interesting example.  
This theory corresponds to quantum electrodynamics in $2+1$ dimensions $(\mQED_3)$ with typically $2 \nf = 4$ massless fermions, called spinons, and an emergent $\U(1)$ gauge field. The $\U(1)$ gauge field is compact given the underlying lattice, and for this reason the spectrum of the DSL contains topological disorder operators known as monopole operators.  These are the operators that may drive confinement.  In a pure compact U(1) gauge theory, monopoles proliferate and confine the gauge field \cite{polyakov_quark_1977}. The presence of massless fermions may screen the monopoles and prevent the confinement given a sufficiently large number of fermion flavors \cite{borokhov_topological_2003}. The stability of a QSL  is thus determined by the relevance of monopole operators. Even if the spin liquid is intrinsically stable, monopole operators may still drive confinement if the fermions are gapped out at a phase transition. This is the situation considered in this paper.

The DSL phase has been used to describe quantum magnets in many contexts. On the triangular lattice, variational Monte Carlo (VMC) studies \cite{kaneko_gapless_2014, iqbal_triangular_2016} have shown that the ground state of a $J_1-J_2$ Heisenberg antiferromagnet in the range $ 0.07 < J_2/J_1 < 0.15$ is given by the DSL. VMC studies \cite{ran_projected-wave-function_2007,  iqbal_projected_2011,  iqbal_gapless_2013, iqbal_spin_2015} and other numerical methods \cite{he_signatures_2017, liao_gapless_2017} also favor a DSL as the ground state of an Heisenberg antiferromagnet on the Kagome lattice. These results are not yet firmly established as contradicting studies find gapped spin liquids in both these contexts.  The transition to a confined phase through monopole condensation  was proposed for the DSL on the square lattice by Ghaemi and Senthil in Ref.~\cite{ghaemi_neel_2006}. A certain class of monopole operators with spin quantum  numbers may also give the correct order parameter for the $\bm q =0$ antiferromagnet on the Kagome lattice \cite{hermele_properties_2008}. 

Topological disorder operators such as the monopole operators play an important role in other contexts. For example, they are involved in the physics of deconfined quantum critical points (dQCPs)  \cite{senthil_quantum_2004,  wang_deconfined_2018}. The prototypical case study is the quantum critical point of the N\'eel-VBS phase transition  on the square lattice which is described by the bosonic CP$^{1}$ theory where the condensation monopole operators which have lattice quantum numbers allows this non-Landau phase transition. The properties of the monopole operators in the CP$^{1}$ theory have been studied numerically in Refs.~\cite{ribhu_fate_2013, sreejith_scaling_2015, pujari_transitions_2015}. It is also important to note  that dQCPs correspond to strongly correlated systems whose description may be reformulated  as field theoritical dualities. A well known example is the particle-vortex duality \cite{peskin_madelstam_1978, dasgupta_phase_1981}. Recently, many new dualities have been found in $2+1$ dimensional gauge theories (earlier examples of this resurgence can be found in Refs.~\cite{son_is_2015, seiberg_a-duality_2016, karch_particle_2016}). Studying critical properties of topological disorder operators provides useful data which may serve to verify conjectured dualities. Confinement-deconfinement transitions are also important in particle physics where the confinement of quarks into hadrons at low energy is a long-standing issue. In fact, the original motivation of Polyakov to study  compact $\mQED_3$ was to obtain a toy model of confinement of quantum chromodynamics.  These relations to deep advancements in quantum phases of matter and quantum field theories motivate further our study of monopole operators.

The objective of this paper is to provide a field theoretical characterization of the confinement-deconfinement transition from a DSL to an antiferromagnetic phase.  We will study the properties  of a quantum critical point (QCP) separating these phases, which is in fact a conformal field theory. The transition will be described with a Gross-Neveu like deformation of $\mQED_3$, where a fermion mass is condensed by tuning a fermion interaction. In turn, the gapped fermions no longer screen the monopole operators which can proliferate. Special attention is given to these topological operators. The central result of our work is the scaling dimension of monopole operators at the QCP.    The field theory used to describe this confinement-deconfinement transition with the condensation of a spin-Hall mass   was proposed in Ref.~\cite{lu_unification_2017}. The idea was later generalized to include the condensation of any monopole operator following the condensation of an appropriate fermion bilinear \cite{song_unifying_2018}.

The paper is structured as follows. In Sec.~\ref{sec:prel}, we give the theoretical background for the monopole operators and the confinement-deconfinement transition driven by the condensation of a spin-Hall mass. In Sec.~\ref{sec:scaling}, we compute the lowest scaling dimension of monopole operators  at the QCP using the state-operator correspondence. We find that the monopole scaling dimension is lower at the QCP than at the $\mQED_3$ fixed point. We also obtain an analytical approximation of the scaling dimension in the limit of large magnetic charge. In Sec.~\ref{sec:monopole}, we consider distinct fermionic dressings that define  monopole operators with various quantum numbers, and show there is a hierarchy in the related scaling dimensions. We also discuss the role of these distinct monopole flavors when constructing perturbations allowed by lattice symmetries.  In Sec.~\ref{sec:comparison}, we do the same analysis in a transition to a chiral spin liquid, where a mass respecting the full flavor group is condensed instead of a spin-Hall mass. Extrapolating our results to $2N_f=2$ allows us to test the duality between the QED$_3$-Gross-Neveu QCP and the bosonic CP$^1$ theory. In Sec.~\ref{sec:RG}, we do a one-loop perturbative renormalization group analysis of the non-compact field theory describing the confinement-deconfinement transition. We find an infrared fixed point corresponding to the QCP and we compute various critical exponents. In Sec.~\ref{sec:kago}, we discuss the implications of our results for the phase transition in the particular case of the Kagome Heisenberg lattice model.  We summarize our results and discuss directions for future research in  Sec.~\ref{sec:conclusion}.

\section{Preliminaries \label{sec:prel}}

\subsection{Monopole operators in $\mQED_3$  \label{sec:mon_QED3}}
Let us consider $\mQED_3$ with $2 \nf$ flavors of massless two-component Dirac fermions, $\psi_A$ where $A = 1,2, \dots 2 \nf$. The flavors could correspond to magnetic spin and valley degrees of freedom, see  Sec.~\ref{sec:kago} for a discussion of how they arise in the Kagome Heisenberg model. These fermions can be organized as a  spinor in flavor space, ${\Psi=\pmatr{\psi_1,&\psi_2,&\dots&\psi_{2\nf}}^\intercal}$.  In Euclidean signature, the bare action reads 
\eqn{
S_{\mQED_3} =  \int d^3 x \lc  - \Psib \sl{D}_a \Psi +  \frac{1}{2e^2} \lb \epsilon_{\mu \nu \rho} \pa_\nu a_{\rho} \rb^2 \rc \,
\label{eq:QED3}
}
where $a_{\mu}$ is the $\U(1)$ gauge field, $\Psib =  \Psi^\dag \gamma_0$  and $\sl{D}_a$ is the gauge covariant derivative 
\eqn{
\sl{D}_a \Psi = \gamma_\mu \lb \pa_\mu - i  a_\mu \rb \Psi \,.
} 
The Dirac matrices $\gamma_\mu$  act on Lorentz spinor components and realize a two-dimensional representation of the Clifford algebra, $\{\gamma_\mu, \gamma_\nu\} = 2 \delta_{\mu \nu} \Id_2$. They can be  chosen as ${\gamma_\mu=(\tau_3, \tau_2, - \tau_1)}$ where the $\tau_i$ are the Pauli matrices.

 As it is written in Eq.~\eqref{eq:QED3}, $\mQED_3$ has a global symmetry, $\U(1)_{\rm top}$, which is related to the conservation of the magnetic current $j_{\rm top}^\mu(x) = \frac{1}{2 \pi} \epsilon^{\mu \nu \rho} \pa_\nu a_\rho(x)$. In the lattice regularization of this theory, it may no longer be the case that this current is conserved. Indeed, in the compact version of  $\mQED_3$, $a_\mu$ is a periodic gauge field  which takes values in the compact $\U(1)$ gauge group. This implies $2\pi$ quantization of the magnetic flux and the existence of instantons  called monopole operators in this context.  These operators insert integer multiples of the flux quantum and break the $\U(1)_{\rm top}$ symmetry.  Non-compact $\mQED_3$ may still describe correctly the infrared (IR) limit of compact $\mQED_3$ if monopole operators are irrelevant. The theory is then said to exhibit an emergent $\U(1)_{\rm top}$ global symmetry in the infrared. Unless stated otherwise, we mean compact $\mQED_3$ when we simply write $\mQED_3$ throughout the paper. 

Let $\man^\dagger_\g(x)$ be a monopole operator with  a $\U(1)_{\rm top}$ charge $\g$ at spacetime point $x$ such that $2 \g \in \Z$. This disorder operator inserts a $4 \pi \g$ magnetic flux. More precisely, the Operator Product Expansion (OPE) of  the  magnetic current operator and the monopole operator  yields the expected magnetic field for a magnetic monopole with charge $\g$ \cite{borokhov_topological_2003}
\eqn{
j_{\rm top}^\mu(x) \man^\dagger_\g(0) \sim \frac{\g}{4 \pi} \frac{x^\mu}{|x|^3}  \man^\dagger_\g(0) + \cdots \,,
\label{eq:flux-op}
}
where the ellipsis denotes less singular terms as ${|x| \to 0}$. Apart from the magnetic flux they insert,  another important property defining monopole operators is their gauge invariance. In particular, these operators must have a vanishing fermionic number. Among $\U(1)$ gauge invariant $4 \pi \g$-flux inserting operators, monopole operators are the most relevant, that is, they have the lowest scaling dimension. Only certain fermionic occupations can produce such operators: Among the $4 \g \nf$ fermion zero modes existing in the monopole background \cite{atiyah_index_1963},  half of them must be filled. There are many ways to satisfy this condition, and all the distinct zero modes dressings define monopole operators with different quantum numbers but with equal scaling dimensions. In particular, for a minimal magnetic charge $\g=1/2$, there are precisely $\binom{2\nf}{\nf}$ monopole operators \cite{borokhov_topological_2003}.

\subsection{Confinement-deconfinement transition to an antiferromagnet \label{sec:GN}}

We mentioned in the last section that non-compact $\mQED_3$ provides an incomplete IR description of compact $\mQED_3$ if monopoles are relevant excitations. In fact, the theory is very different in this case. In pure $\U(1)$ compact gauge theory, monopole operators are relevant and condense. This leads to  confinement and to the emergence of a mass gap \cite{polyakov_compact_1975, polyakov_quark_1977}. This effect can be prevented if there are enough massless fermion flavors to screen the monopoles. Indeed, at leading order in $1/\nf$, the monopole scaling dimension is proportional to $\nf$ \cite{borokhov_topological_2003, hermele_stability_2004}: The operator becomes irrelevant for a sufficiently large number of massless fermion flavors $2\nf$.  Otherwise, the fermions confine.  Even if monopoles turn out to be irrelevant and do not destabilize QSL phases in magnets described by emergent $\mQED_3$, they may play an important role elsewhere  in the phase diagram. In particular, as new fermion interactions are tuned, fermion masses can be generated. In this case, the screening effect by fermions is lost and monopoles are free to proliferate.

For the rest of this section,  we examine the aforementioned  monopole proliferation subsequent to a fermion mass condensation. We study the deformation of compact $\mQED_3$ with a chiral Heisenberg Gross-Neveu ($\cHGN$)  interaction with coupling strength $h$
\eqn{
S_{\mQED_3-\cHGN} =  \int d^3 x \lc - \Psib \sl{D}_a \Psi  -  \frac{h^2}{2} \lb \Psib \bm{\sigma} \Psi \rb^2     \rc   + \dotsb  \,,
 \label{eq:QFT_transition}
}
where the ellipsis denotes the Maxwell free action and the contribution from monopole operators. Here, $\bm{\sigma}$ is a Pauli matrix vector acting on a $\SU(2)$ subspace of flavors. For definiteness, we introduce right now the language natural for quantum magnets. The $\SU(2)$ subspace in question consists of two magnetic spin degrees of freedom $\{\u, \d\}$. The other $\SU(\nf)$ subspace consists of valley degrees of freedom, i.e.\ locations of Dirac point in the Brillouin zone.  $\mQED_3$ has the full flavor symmetry $\SU(2\nf)$.\footnote{The center of $\SU(2\nf)$ coincides with $\U(1)$ gauge transformations and we should quotient the symmetry group \cite{song_unifying_2018}. For simplicity, we keep the redundancy and refer to $\SU(2\nf)$ as the flavor symmetry group.}  The $\cHGN$ interaction breaks down the  global flavor symmetry, ${\SU(2\nf) \to \SU(2) \times \SU(\nf)}$. This is broken further to $\SU(\nf)$ when a spin-Hall mass $\moye{\Psib \bm{\sigma} \Psi}$ is condensed following the tuning of the coupling constant $h$. The condensate direction spontaneously chosen sets  a preferred axis of quantization for the magnetic spin. Monopole operators which then condense have, accordingly, spin quantum numbers. We shall examine this point more thoroughly when we discuss the distinct flavors of monopole operators in Sec.~\ref{sec:monopole}.  

We just described how an AFM order appears when, following the tuning of a spin-dependent fermion interaction,  monopole operators proliferate. This mechanism  was described in Ref.~\cite{lu_unification_2017} in the contexts of Kagome antiferromagnets.  It was also considered to describe a transition on the square lattice \cite{ghaemi_neel_2006} where a $\SU(\nf)_{\rm valley}$ breaking interaction, $\delta \lag \sim ( \Psib \mu_z \bm{\sigma} \Psi)^2 $, is considered instead. This confinement of the DSL on the square lattice has also been studied numerically with quantum Monte Carlo \cite{yang-xu_monte_2018}. An extended version of this mechanism involving  general fermion bilinears was also considered in  Ref.~\cite{song_unifying_2018}. 

\subsubsection{Spin-Hall mass condensation in the $1/\nf$ expansion}

In what follows, we demonstrate, using a $1/\nf$ expansion, that a spin-Hall mass does condense when a sufficiently strong Gross-Neveu interaction is present. Performing a Hubbard-Stratonovich transformation on the action \eqref{eq:QFT_transition}, we obtain 
\eqn{	
S'_{\mQED_3-\cHGN} = \int d^3 x  \left [- \Psib \lb \sl{D}_a + \bm{\phi} \cdot \bm{\sigma}  \rb \Psi +  \frac{2 \nf}{2 h^2} \bm{\phi}^2  \right ]  \,,
\label{eq:action_hubbard}
}
where $\bm{\phi}$ is a three-component auxiliary bosonic field and we rescaled $h^2$ with the number of fermion flavors $2 \nf$.  The fermions can be integrated to a determinant operator. Tracing out the valley subspace,   the effective action becomes  
\eqn{\label{eq:Seff}
S_{\rm eff} = 2 N_f \lb -    \half \log \det\lb \sl{D}_a + \bm \phi \cdot \bm\sigma \rb  +   \int d^3 x    \frac{1}{2 h^2} \bm{\phi}^2 \rb  \,,
} 
where $\det$ is the determinant over the magnetic spin and the Dirac spaces. The saddle point solution for the gauge field is $a_\mu=0$. We take a homogeneous ansatz  for the saddle point value of the bosonic field $\moye{\bm \phi} = M \hat n $, where $\hat n$ is a unit vector.  Eigenstates of the resulting determinant operator are plane waves and are used to obtain the saddle point equation for $M$ in a diagonalized form 
\eqn{\label{eq:gap0}
2 M \lb  \frac{1}{2h^2} - \int \frac{d^3 p}{(2 \pi)^3} \frac{1}{p^2 + M^2}   \rb = 0  \,.
}
There is a trivial solution $M = 0$ which represents the symmetric phase. A critical coupling $h_c^{-2}$ defines  the transition to the ordered phase $M > 0$ through the relation
\eqn{
\frac{1}{2 h_c^2} = \int \frac{d^3 p}{(2 \pi)^3} \frac{1}{p^2}   = 0\, , \label{eq:crit_coupling}
}
where we used a zeta function regularization to evaluate the divergent integral. 
%, we obtain the following critical coupling
%\eqn{
%h_c^{-2} = 0 \,.\label{eq:crit_coupling}
%}
For future reference, we evaluate the effective action \eqref{eq:Seff} at the critical point \eqref{eq:crit_coupling}
\eqn{
S_{\rm eff}^c = - \nf \log \det \lb \sl{D}_a + \bm{\phi} \cdot \bm  \sigma \rb  \,.
\label{eq:S_eff_c}
}
In the  ordered phase, the expectation value of the bosonic field can be found by rewriting the saddle point equation as   
\eqn{
\frac{1}{2h_c^2} - \frac{1}{2h^2}  = \int \frac{d^3 p}{(2 \pi)^3} \lc \frac{1}{p^2 } - \frac{1}{p^2 + M^2} \rc   \,.
}
Using \eqref{eq:crit_coupling}, we find $\m = -2 \pi h^{-2}$ for $h^{-2} < 0$. More generally, the condensed mass is given by
\eqn{
\moye{|\bm{\phi}|} \equiv \m     
=  
\left \{
\begin{alignedat}{2}
& 0 \,, \qquad  &&  h^{-2} > h_{c}^{-2}\,, \\
& 2 \pi \lb h_c^{-2} - h^{-2} \rb  \,, \qquad && h^{-2} < h_{c}^{-2}\,.
 \end{alignedat}
\right . 
}
Note that in momentum regularization we would have obtained a non-zero value for the critical coupling, $h_c^{-2}$.

\section{Scaling dimensions of monopole operators \label{sec:scaling}}

We established in Sec~.\ref{sec:GN} the existence of the large$-\nf$  $\mQED_3- \mathrm{cHGN}$ critical fixed point in the non-compact theory which leads to a spin-Hall mass  condensation. In turn, this implies the proliferation of monopoles  in the compact theory.  Given the primordial role that monopole operators play in the quantum phase transition, we compute their scaling dimensions at the QCP. We shall restrict our computation to leading order in $1/\nf$.

Monopole operators are usually defined as operators with the lowest scaling dimension among $4 \pi \g$ flux-creating operators. In $\mQED_3$, there are many monopole operators due to the presence of fermion zero modes \cite{borokhov_topological_2003}. One important result we shall show in the next section is that the analogous operators in $\mQED_3-\cHGN$ develop a non-trivial hierarchy in their scaling dimensions. Nevertheless, we keep referring to those operators as monopole operators. In the  present section, we will compute the lowest scaling  dimension among these monopole operators.

\subsection{State-operator correspondence and $1/\nf$ expansion}
A monopole operator $\man^\dagger_q$ is characterized by a scaling dimension $\Delta_{\man_q}$ which determines the power law decay of its two-point correlation function. The scaling dimension can be determined through the state-operator correspondence. This correspondence implies that the insertion of a local operator at the origin of flat spacetime $\R^{1,2}$ can be mapped to a state of the conformal field theory (CFT) on $S^2 \times \R$ (see \cite{rychkov_epfl_2017} for a clear and concise explanation of this correspondence). Specifically, the monopole operator with the lowest scaling dimension corresponds to the ground state of fermions in $\mQED_3-\cHGN$  living on $S^2$ in a background magnetic flux $4 \pi \g$. The relation also implies that the energy $F_\g$ of this ground state and the scaling dimension  of this monopole operator {$\Delta_\g = \min (\Delta_{\man_q})$}  are equal
\eqn{
\Delta_\g = F_\g \equiv - \log Z_{S^{2} \times \R}[A^\g]  \,,
}
where $A^\g$ is an external gauge field yielding the magnetic flux $\int_{S^2} \dd A^\g = 4 \pi \g$.
The notation $F_\g$ stands for free energy, which, in the present non-thermal setup, is the same as the ground state energy.\footnote{The free energy should be understood as a zero-temperature limit, $\lim_{\beta \to \infty} (- \log Z_{S^{2} \times S^1_{\beta}}[A^\g] / \beta ) $ \cite{dyer_monopole_2013}. This definition is considered when needed later on.}
%$^,$\footnote{The scaling dimension of the identity operator has to vanish $\Delta_{g=0} =0$. Changing the normalization can guarantee it, but here it will  turn out unnecessary.} . 
Our strategy to obtain the scaling dimension $\Delta_\g$  will be to perform  a $1/\nf$ expansion  of the free energy 
\eqn{
 F_\g =  \nf  F^{(0)}_\g +  F^{(1)}_\g + \dots 
 \label{eq:Fn}
 } 
 We restrict our study to leading order in $1/\nf$ \footnote{The appropriate relation is actually $\Delta_{\g}^{(0)} = F_\g^{(0)}  - F_{\g=0}^{(0)}$, but we find later on that for the case we study, $F_{\g=0}^{(0)} = 0$. }
 \eqn{
 \Delta_\g^{(0)} = F_\g^{(0)}  \,.
 \label{eq:del}
 }

The state-operator correspondence was first used to compute the scaling dimension of a topological disorder operator in the context of $\mQED_3$ \cite{borokhov_topological_2003}. A similar computation was made for the bosonic theory $\CP^{\nb-1}$   \cite{metlitski_monopoles_2008}.  The path integral formalism was also used to obtain $1/N$ corrections  for $\mQED_3$ \cite{pufu_anomalous_2014} and $\CP^{\nb-1}$ \cite{dyer_scaling_2015}. The ungauged version of $\CP^{\nb-1}$ was also investigated  using these techniques   \cite{pufu_monopoles_2013}. Monopole operators were also studied in non-abelian gauge theories, in presence of supersymmetries and in presence of a Chern-Simons term \cite{borokhov_monopole_2002, dyer_monopole_2013, radicevic_disorder_2016, chester_monopole_2018}. 

\subsection{Spectrum of the Dirac operator with a spin-Hall mass}
In order to obtain the free energy $F_\g$, we study the effective action obtained after integrating out the fermions. The analysis is similar to the one in Sec.~\ref{sec:GN}, but we must now work on a sphere with a background magnetic flux. This latter consideration  is incorporated through an  external gauge field 
\eqn{
A^\g(x) = \g (1 - \cos \theta ) \dd \phi  \,,
}
whose flux integral is  ${\int \dd A^\g = 4 \pi \g}$. The singular part at the south pole $\theta =\pi$ can be compensated by a Dirac string. The requirement that the Dirac string should be invisible imposes the Dirac condition $2 \g \in \Z$.   On the other hand, the spacetime $S^2 \times \R$ is encoded in a non-trivial metric $g_{\mu \nu}(x)$ which we parameterize with $(\theta, \phi, \tau)$ 
\eqn{
g_{\mu \nu} \dd x^\mu \dd x^\nu = \dd \tau^2 + R^2 \lb \dd \theta^2 + \sin^2 \theta \dd \phi^2 \rb  \,,
\label{eq:gmunu}
}
where $R$ is the radius of $S^2$. The metric can be decomposed as ${ g_{\mu \nu} = e_{\mu}^a e_{\nu}^b \eta_{ab}}$, where $\eta_{ab}$ is the flat spacetime metric and $ e_{\mu}^a(x)$ are the  tetrad fields. A spin connection $\Omega_{\mu}$ transporting the fermion fields on the curved spacetime can be found from the tetrad fields. Taking into account both $\U(1)$ and spacetime connections, the Dirac operator in the critical action \eqref{eq:S_eff_c} now reads \cite{weinberg_quantum_2013}
\eqn{
\sl{D}^{S^2 \times \R}_{a, A^\g} = e\indices{_b^\mu} \gamma^b \lc  \pa_\mu + \Omega_\mu - i\lb A^\g_\mu + a_\mu \rb   \rc  \,.
\label{eq:Dirac}
}
The critical effective action $S_{\rm eff}^c$ \eqref{eq:S_eff_c} with the modified Dirac operator, $\sl{D}_a\to \sl{D}_{a,A^\g}^{S^2 \times \R}$, becomes
\eqn{
S_{\rm eff}^{\prime c} = - \nf \log \det \lb \sl{D}_{a,A^\g}^{S^2 \times \R}+ \bm{\phi}\cdot \bm \sigma \rb  \,.
\label{eq:S_eff_c_p}
} 
The saddle point condition for this modified effective action still implies a vanishing expectation value for the gauge field $\moye{a_\mu}  = 0$.  We take a homogeneous ansatz for the saddle point value of the bosonic  field on the sphere with flux $4 \pi \g$,  $\moye{\bm \phi} = \mg \hat n$, where $\hat n$ is a unit vector. Without loss of generality, we can orient the condensate such that $\hat n  = \hat z$.   Inserting this ansatz in the effective action \eqref{eq:S_eff_c_p}, we find the leading order free energy
\eqn{
F_\g^{(0)} = - \log \det  \lb \sl{D}_{A^\g} + \mg \sigma_z \rb  \,,
\label{eq:F0_curv}
}
where  $\sl{D}_{A^\g} \equiv \sl{D}^{S^2 \times \R}_{a, A^\g}\big |_{a=0}$.\footnote{In this section, we simply write $\sl{D}_{A^\g}$ as we assume a curved spacetime $S^2 \times \R$ whenever the external gauge field $A^q$ is present.}  The spectrum of the operator appearing 
inside the determinant in Eq.~\eqref{eq:F0_curv} must be found to obtain the leading order free energy $F_\g^{(0)}$. 

We first review how the spectrum of the Dirac operator $\sl{D}_{A^\g}$ was found   in Refs.~\cite{borokhov_topological_2003, pufu_anomalous_2014} by using analogs of spherical harmonics \cite{wu_dirac_1976} appropriate for describing spin-$1/2$ particles in the monopole background. A first step in the generalization is to define the generalized angular momentum  $L_\g^i = - i \epsilon_{ijk} x_j (\pa_k - A^\g_{k}) - r^2  \epsilon_{ijk} \pa_j A^\g_{k} $   which includes the effect of the magnetic charge.  The $SU(2)$ algebra remains  after the generalization,  $[L^i_\g, L^j_\g] = i \epsilon_{ijk}  L_\g^k$, so there exists eigenfunctions  $Y_{\g,\ell,m}(\theta, \phi)$, called monopole harmonics, which simultaneously diagonalize $L^2_\g$ and  $L_\g^z$ \cite{wu_dirac_1976}
\eqna{2}{
L^2_\g Y_{\g,\ell,m} 
&= \ell \lb \ell + 1 \rb Y_{\g,\ell,m}  \,, &\quad \ell &= |\g|, |\g|+1, \dots 
\label{eq:l2} \\
L_\g^z Y_{\g,\ell,m}  \,,
&= m  Y_{\g,\ell,m}  \,, &\quad m &= -\ell, -\ell+1, \dots, \ell  \,.
\label{eq:lz}
}
As the Dirac operator acts on spinors,   one must consider the  total angular momentum  $\bm J_\g = \bm L_\g + \bm \tau/2$ as well. Two-component spinors  $S^\pm_{\g,\ell,m}$ that  diagonalize the operators $\{L_\g^2, J_\g^{z}, J_\g^2 \}$ are thus introduced
\eqn{
J^2_\g S^\pm_{\g,\ell,m} &= j_\pm \lb j_\pm +1 \rb S^\pm_{\g,\ell,m}  \,, \quad j_{\pm} = \ell \pm 1/2  \,,\\
L_\g^2 S^\pm_{\g,\ell,m} &= \ell \lb  \ell + 1 \rb  S^\pm_{\g,\ell,m}  \,,\\
J^z_\g S^\pm_{\g,\ell,m} &= (m + 1/2) S^\pm_{\g,\ell,m}  \,. 
}
Such spinors $S^\pm_{\g,\ell,m}$, dubbed spinor monopole harmonics, are built  using  monopole harmonics as components  \cite{borokhov_topological_2003, pufu_anomalous_2014}
\eqn{
S^+_{\g,\ell,m} &= \mqty({\sqrt{\frac{\ell+m+1}{2\ell+1}}  Y_{\g,\ell,m} \\ \sqrt{\frac{\ell-m}{2\ell+1}}  Y_{\g,\ell,m+1} })  \,, \\
S^-_{\g,\ell,m} &= \mqty({-\sqrt{\frac{\ell-m}{2\ell+1}}  Y_{\g,\ell,m} \\ \sqrt{\frac{\ell+m+1}{2\ell+1}}  Y_{\g,\ell,m+1} })  \,.
}
These spinors can be organized as doublets $[S^+_{\g,\ell-1,m}, \quad S^-_{\g,\ell,m}]^\intercal $ with total angular momentum $j = \ell - 1/2$. 
 Adding plane waves $e^{- i \omega \tau}$ to describe the ``time" direction\footnote{We emphasize that this ``time" dimension on $S^2 \times \R$ does not correspond to the original time dimension on $\R^{1,2}$.},  the action of the Dirac operator on this basis is \cite{borokhov_topological_2003, pufu_anomalous_2014}
\eqn{
\sl{D}_{A^\g}\bmatrs{e^{-i \omega \tau} S^+_{\g,\ell-1,m}  \\ e^{-i \omega \tau} S^-_{\g,\ell,m}}  = -i \bm{O}_{\g, \ell} \lb \omega + i \bm{P}_{\g, \ell} \rb
 \bmatrs{
 e^{-i \omega \tau} S^+_{\g,\ell-1,m}
 \\
e^{-i \omega \tau}  S^-_{\g,\ell,m}
 }
 \label{eq:DTS}
}
where the matrices $\bm{O}_{\g, \ell}$ and $\bm{P}_{\g, \ell}$  are given by
\eqn{
\bm{O}_{\g, \ell} = \frac{1}{\ell} \bmatr{-\g  & - R \veps_\ell^0 \\ - R \veps_\ell^0 &  \g } \,, \quad 
\bm{P}_{\g, \ell} = \frac{\veps_\ell^0}{\ell} \bmatr{
R \veps_\ell^0 	& -\g  \\
-\g  	& - R \veps_\ell^0
}
}
and where $\veps_\ell^0 \equiv R^{-1} \sqrt{\ell^2 - \g^2}$. For the minimal total angular momentum $j=|\g|-1/2$, only the spinor $S^-_{\g,|\g|,m}$ is defined\footnote{The other would-be spinor $S^+_{\g, |\g|-1,m}$ with $j=|\g|-1/2$ does not exist  since $\ell=|\g|-1$ is smaller that the minimal angular momentum allowed   for monopole harmonics \eqref{eq:l2}.} and the action of the Dirac operator on this mode reduces to 
\eqn{
\sl{D}_{A^\g} \bmatr{0 \\ e^{-i \omega \tau} S^-_{\g,|\g|,m}}   = -i \omega \sgn(\g) \bmatr{0 \\ e^{-i \omega \tau} S^-_{\g,|\g|,m}}  \,.
 \label{eq:DS}
}
This mode has a vanishing energy and thus corresponds to  a fermion zero mode in the monopole background.

We now study the complete determinant operator with the contribution of the spin-Hall mass term appearing in Eq.~\eqref{eq:F0_curv}. The additional term is diagonal in the spinor monopole harmonics basis. Therefore, we can still use this basis to compute the determinant operator
\eqn{
&  \log \det \lb \sl{D}_{A^\g}  + \mg \sigma_z \rb = \sum_{\sigma = \pm 1} \int_{-\infty}^{\infty} \frac{d \omega}{2 \pi} \nn \\
 & \times\Bigg [ 
 d_{|\g|} \log \lc  -i  \omega \sgn(\g)  + \mg \sigma  \rc \nn \\
& + \sum_{\ell=|\g|+1}^\infty d_\ell \log \det \lc -i \bm{O}_{\g, \ell} \lb \omega  + i \bm{P}_{\g, \ell} \rb +  \mg \sigma  	\rc   \Bigg]   \,,
\label{eq:det}
}
where $d_{\ell} = 2 \ell$ is the degeneracy coming from azimuthal quantum numbers. Simplifying further, we obtain
\eqn{
 &\log \det \lb \sl{D}_{A^\g}  + \mg \sigma_z  \rb  = \sum_{\sigma = \pm 1} \int_{-\infty}^{\infty} \frac{d \omega}{2 \pi} \nn\\
 & \times  \Bigg[ d_{|\g|}\log \lc  \omega  + i \mg \sigma \sgn(\g)  \rc \nn \\
  &
  +\sum_{\ell=|\g|+1}^\infty d_\ell \log \lc  \omega^2  + (\veps_\ell^0)^2 + \mg^2  \rc \Bigg] \nn \\
=& 
\int_{-\infty}^{\infty} \frac{d \omega}{2 \pi} 
  \Bigg[
 d_{|\g|}\log ( \omega^2 +  \mg^2  )
  +   \sum_{\ell=|\g|+1}^\infty 2 d_\ell \log \lc  \omega^2  +  \veps_{\ell}^2\rc  \Bigg]  \,,
  \label{eq:det_op}
}
where  we removed inessential constants and we defined the mass-deformed eigenvalues 
\eqn{
\veps_\ell = R^{-1} \sqrt{\ell^2 -\g^2   + \mg^2 R^2}  \,.
\label{eq:veps_l}
}

 We now explicitly write the spectrum of the Dirac operator with a  spin-Hall mass on the magnetically charged sphere found from \eqref{eq:det}
\eqn{
\omega  + i \sigma \sgn (\g) \veps_{|\g|} & \,,    	\quad \ell=|\g|  \,, 
\label{eq:spectrum-a}
\\
\pm \sqrt{\omega^2 + \veps_\ell^2} & \,,	\quad	\ell= \{|\g| +1, |\g| +2, \dots\}  \,,
\label{eq:spectrum-b}
}
where ${\sigma \in \{-1, +1\}}$. The $\pm$ modes for $\ell \geq  |\g| + 1$ in \eqref{eq:spectrum-b} are referred to as conduction $(+)$ and valence $(-)$ modes. There is no such  doubling of the $\ell = |\g|$ modes in \eqref{eq:spectrum-a} which are descendants of the  $\mQED_3$ zero modes. This is why  we refer to these modes as \zero modes even though they have non-vanishing energy $\pm \veps_{|\g|} = \pm \mg $  with the inclusion of the spin-Hall mass. We restate that the $\sigma$ eigenvalue refers to magnetic spin orientation relative to the quantization axis defined by  $\moye{\bm{\phi}}$.  For positive magnetic charge $\g>0$, a \zero mode with spin up,  $\sigma=1$, has an energy $\veps_{|\g|}= \mg$, whereas a spin down \zero mode, $\sigma=-1$ has an opposite energy $-\mg$.  These \zero modes are responsible for the first term  in Eq.~\eqref{eq:det_op}

\subsection{Scaling dimension computation \label{sec:Fscal}}
\begin{figure}[ht!]
\centering
{\includegraphics[width=\linewidth]{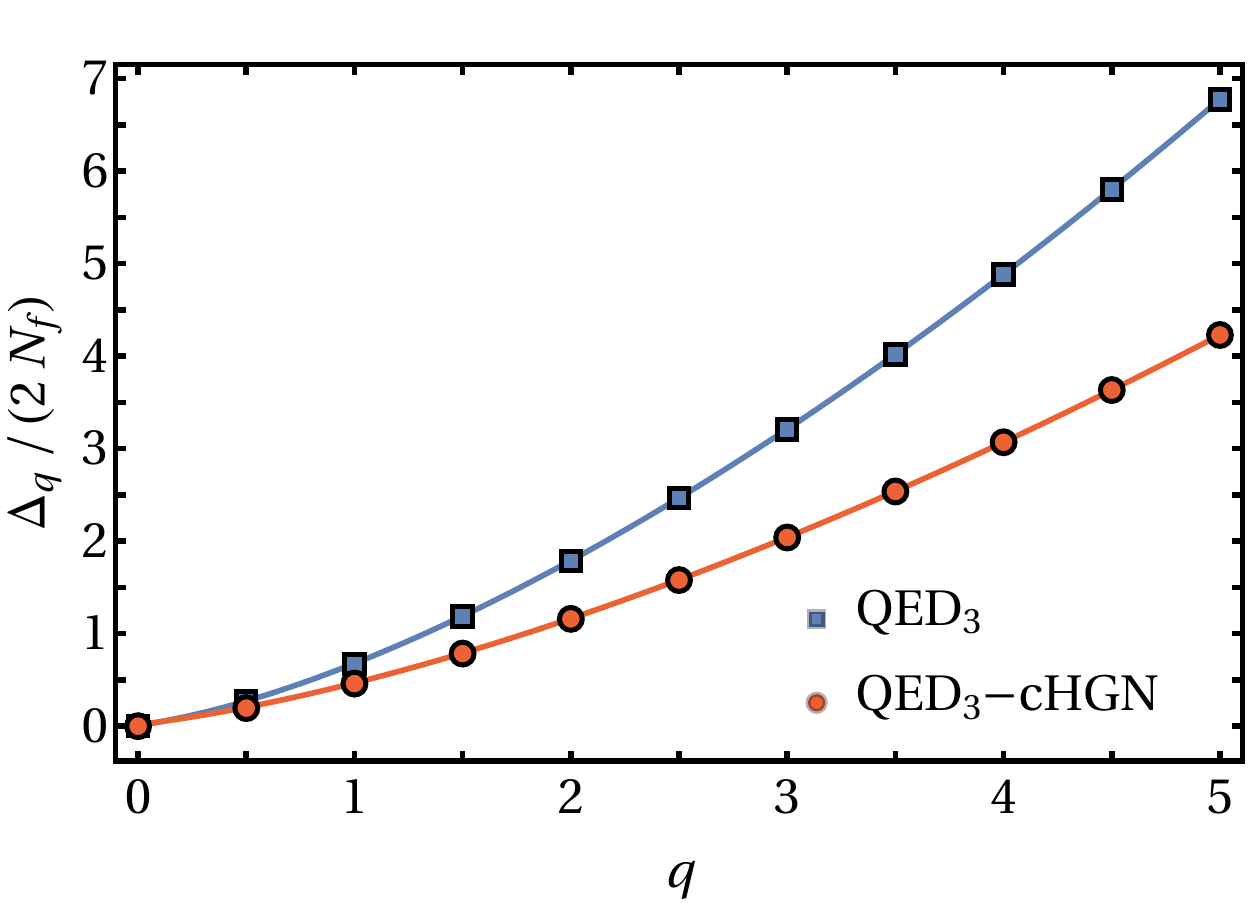}}
\caption{Lowest scaling dimension of monopole operators $\Delta_\g$ per number of fermion flavors $2 \nf$ as a function of the magnetic charge $\g$. 
Analytical approximations in the large $\g$ limit of the scaling dimension in $\mQED_3-\cHGN$  and $\mQED_3$, respectively  \eqref{eq:F0g_2} and \eqref{eq:F0g_1}, are plotted in solid lines. These are compared to their respective numerical values shown in Tab.~\ref{tab:scaling}. \label{fig:scaling} }
\end{figure}

\begin{figure}[ht!]
\centering
{\includegraphics[width=\linewidth]{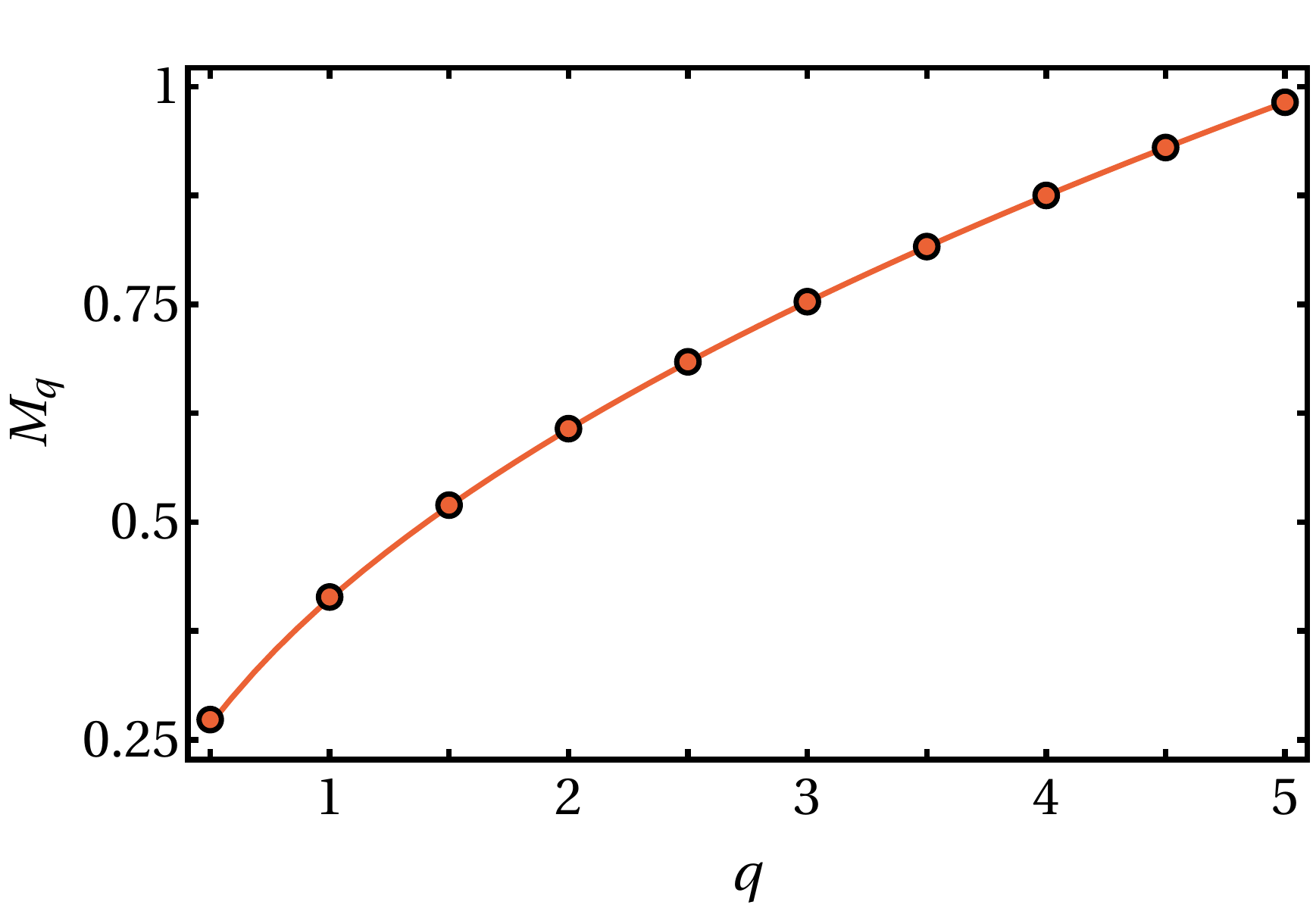}}
\caption{Mass $\mg \equiv \moye{|\bm \phi|}$ as a function of the magnetic charge $\g$ at leading order in $1/\nf$. The solid line corresponds to the large $\g$ analytical approximation of $M_{\g}$ \eqref{eq:m2_g}. The circles are the numerical values of $M_{\g}$ shown in Tab.~\ref{tab:scaling}. 
 \label{fig:scaling2} }
\end{figure}

The free energy at leading order  \eqref{eq:F0_curv} is rewritten using the result \eqref{eq:det_op} (from now on, we assume a positive magnetic charge $\g>0$), 
\eqna{2}{
F_\g^{(0)} 
&= - \!\int\! \dfrac{d \omega}{2 \pi} 
&
 \bigg[  &    d_\g   \log \lc \omega^2 + \mg^2\rc      +\!\!\! \sum_{\ell = \g+ 1}^\infty  2 d_\ell \log \lc \omega^2 + \veps_\ell^2 \rc \bigg ] 
\label{eq:f0} \,,
} 
where the radius $R$ of the sphere was eliminated  by changing the integration variable $\omega \to \omega / R$, by  rescaling the parameters $\{\veps_\ell, F_\g^{(0)}, \mg\} \to\{\veps_\ell, F_\g^{(0)}, \mg\} /R $ and by removing an  inessential constant. The free energy \eqref{eq:f0} needs regularization. We first treat the diverging integral over frequencies by  rewriting the integrand using the identity $\log A = - \dd A^{-s}/ \dd s |_{s=0}$ and doing an analytic continuation to $s = 0$. This procedure is presented in App.~\ref{app:reg}. The resulting free energy is
\eqn{
F_\g^{(0)}  
&= -  d_\g \mg - \sum_{\ell = \g  +1}^\infty 2 d_\ell \veps_\ell  \,.
\label{eq:F0_unreg}
}
By setting  $\mg=0$ in this free energy, we obtain the $\mQED_3$ results shown in Ref.~\cite{pufu_anomalous_2014}.\footnote{Our definition of $F^{(0)}_\g$ has an extra factor of $2$  because we defined the total number of fermion flavors as $2\nf$ but we expanded the free energy in powers of $1/\nf$ \eqref{eq:Fn}. This procedure is more natural since the spin degeneracy does not factor out like the valley degeneracy because of the \zero modes.} This free energy \eqref{eq:F0_unreg} still needs regularization. The divergent sum is  rewritten by adding and subtracting its diverging part
\eqn{
 \sum_{\ell = \g +  1}^\infty d_\ell \veps_\ell =&\sum_{\ell = \g +  1}^\infty \left [  2 \ell \sqrt{\ell^2 +  \mg^2- \g^2} - 2\ell^2 - \lb \mg^2- \g^2 \rb \right ] \nn \\
 +&2 \sum_{\ell = \g +  1}^\infty \left [ \ell^{2(1-s)} + \lb \half - s \rb \lb \mg^2- \g^2 \rb \ell^{-2s} \right ] \!\! \bigg |_{s=0} \hspace{-1em}  \,.
}
Now, only the last sum is divergent and we treat it  with a zeta function regularization by using ${\sum_{n=0}^\infty \lb n + a \rb^{-s} = \zeta(s,a)}$. 
This sum then becomes $ 2 \zeta(-2,\g+1) +  \lb \mg^2- \g^2 \rb \zeta(0,\g+1)$,  an expression for which a polynomial form may be found using Ref.~\cite{NIST:DLMF}. The resulting finite expression is then inserted  in \eqref{eq:F0_unreg} to obtain the regularized free energy  
\eqn{
\begin{split}
F_\g^{(0)}  
=&- d_\g \mg - \!\! \sum_{\ell  = \g + 1}^\infty \Big[ 2 d_\ell \veps_\ell -  d_\ell^2 - 2 \lb \mg^2- \g^2 \rb \Big]
\\
&  +(2 \g +1) \lb \mg^2 -  \g (\g-2) / 3 \rb  \,.
\end{split}
\label{eq:F0_reg}
}
 We can then find the regularized  gap equation, i.e. the saddle-point equation $\pa F_\g^{(0)} / \pa \mg = 0$
\eqn{
-  d_\g +  2 \mg (2 \g + 1)   -   2 \mg \sum_{\ell = \g + 1}^\infty   \left [  d_\ell \veps_\ell^{-1} -   2 \right ] = 0  \,.
\label{eq:gap_reg}
}
For a vanishing magnetic charge $\g=0$, the contribution from \zero modes vanishes since $d_{\g=0} =0$. The saddle point equation then only has a trivial solution $M_{\g=0}=0$. This case coincides with  $\mQED_3$ where there is no mass $\mg$ to start with and the free energy vanishes, $F^{(0)}_{\g=0} = 0$. For $\g \neq 0$, the saddle point equation only has a non-trivial solution  $\mg>0$ which must be determined numerically.\footnote{If the ``time" direction $\R$ is compactified on a circle $S^1_\beta$, a trivial solution does exist for $\g>0$. It persists as the radius $\beta$ is taken to infinity to retrieve the real line. However, this solution is a \textit{maximum} of free energy and does not determine $\Delta_q$. See App.~\ref{app:thermal}} 
The resulting mass $\mg$ is then inserted in \eqref{eq:F0_reg} to obtain the scaling dimension at leading order in $1/\nf$,  $\Delta_\g = \nf F_\g^{(0)} + \O(\nf^0)$. The mass $\mg$ and the  scaling dimension of monopole operators in $\mQED_3-\cHGN$ and $\mQED_3$ are obtained for multiple values of $\g$ and are shown in Tab.~\ref{tab:scaling}. 
\begin{table}[ht!]
\caption{Numerical results for the mass $\mg$ and the lowest scaling dimension of monopole operators  in $\mQED_3-\cHGN$ and $\mQED_3$, respectively $\Delta_\g$ and $\Delta^{\mQED_3}_\g = \Delta_\g|_{\mg=0}$. These results are at leading order in $1/\nf$. We show the scaling dimensions per number of fermion flavors $2 \nf$. These quantities are shown for the first few allowed values of the magnetic charge $\g$. 
\label{tab:scaling}
}
\centering
\begin{ruledtabular}
\begin{tabular}{rccc}
$\g$ &$\mg$ &  $\frac{1}{2\nf}\Delta_\g$   &  $\frac{1}{2\nf}\Delta_\g^{\mQED_3}$ \\ \hline 
$0$ & $0$ & $0$ & $0$\\
$0.5$ & $0.27318$ & $0.19539$ & $0.26510$\\
$1.0$ & $0.41395$ & $0.46039$ & $0.67315$\\
$1.5$ & $0.51946$ & $0.78471$ & $1.18643$\\
$2.0$ & $0.60728$ & $1.15964$ & $1.78690$\\
$2.5$ & $0.68406$ & $1.57928$ & $2.46345$\\
$3.0$ & $0.75311$ & $2.03939$ & $3.20837$\\
$3.5$ & $0.81638$ & $2.53671$ & $4.01591$\\
$4.0$ & $0.87510$ & $3.06867$ & $4.88154$\\
$4.5$ & $0.93014$ & $3.63315$ & $5.80162$\\
$5.0$ & $0.98211$ & $4.22839$ & $6.77309$\\
\end{tabular}
\end{ruledtabular}
\end{table}
These numerical results are also plotted in Fig.~\ref{fig:scaling} and Fig.~\ref{fig:scaling2} along with corresponding analytical approximations obtained in Sec.~\ref{sec:large_g}. The numerical and analytical results agree very well even for small values of $\g$.  We also note that the monopole operator scaling dimension is always smaller in  $\mQED_3-\cHGN$ than in $\mQED_3$. The fact that $\Delta_\g \leq \Delta_\g^{\mQED_3}$ is expected since the case $M_q = 0$ implies $\Delta_\g = \Delta_\g^{\mQED_3}$ and sets an upper bound  for the scaling dimension $\Delta_q$.  
Setting $2 \nf =4$ gives an estimate on the scaling dimension of monopole operators for certain quantum magnets. One should be careful with these results as the expansion parameter is not small and corrections to the leading order may be important. Nevertheless, if we  consider the monopole operator with a minimal magnetic charge $\g = 1/2$,  the lowest scaling dimension  is
\eqn{
\Delta_{\g=1/2} = 2 \nf \cdot 0.19539 + \O(\nf^0)  \,.  \label{eq:min_mon}
} 
In the case $2 \nf =4$, which is interesting for application to quantum magnets, we find a strongly relevant operator
\eqn{
\Delta_{\g=1/2}\Big|_{2\nf=4} \approx 0.78156  < 3  \,. \label{eq:min_mon_2}
}
However, the monopole operator with the minimal magnetic charge is not allowed in many contexts, we discuss this matter in Sec.~\ref{sec:kago}. We also note that the unitarity bound is violated for $2 \nf < 2.56$ as $\Delta_{q=1/2}<1/2$~\cite{rychkov_epfl_2017}. It would be interesting to see if the violation persist with higher order corrections. This would shed light on the phase diagram of QED$_3$ at
$2N_f=2$.

\begin{figure}[ht!]
\centering
{\includegraphics[width=\linewidth]{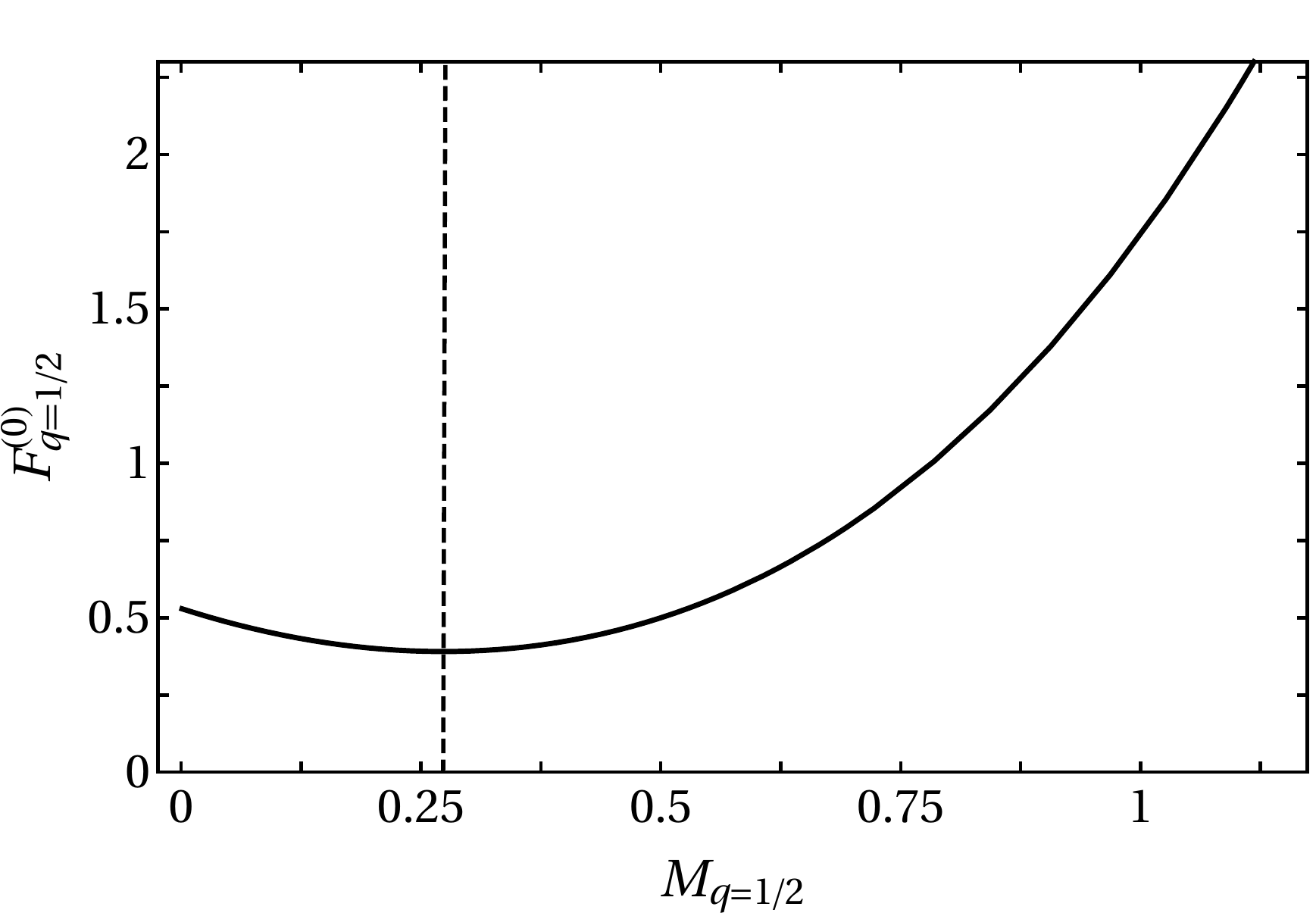}}
\caption{Leading order in $1/\nf$ of the free energy with minimal magnetic charge $F^{(0)}_{q=1/2}$ \eqref{eq:F0_reg}  as a function of the mass $M_{q=1/2}$. The appropriate value of the mass $M_{q=1/2}$ found with the saddle point equation \eqref{eq:gap_reg} corresponds to the minimum of this function. For large mass, the free energy behaves as $F^{(0)}_{q=1/2} \sim 4 (M_{q=1/2})^3 /3$.\label{fig:F0_as_func_m}}
\end{figure}

For demonstration purpose, we plot in  Fig. \ref{fig:F0_as_func_m}  the free energy for the minimal magnetic charge as a function of the mass $M_{\g=1/2}$. The free energy minimum is identified and corresponds to the solution to the saddle point equation, $M_{q=1/2} = 0.27318$, as shown in Tab. \ref{tab:scaling}. For large values of $M_{q=1/2}$, the sum in \eqref{eq:F0_reg} can be approximated as an integral, and we find that the free energy grows as $F^{(0)}_{q=1/2} \sim 4 (M_{q=1/2})^3 / 3$ at leading order in $M_{q=1/2}$. 

Let us restate one important result. For $\g \neq 0$, we found that the monopole operator with the lowest scaling dimension is described  with a mass condensate, ${ \mg \equiv \moye{|\bm{\phi}|} > 0}$. We stress out that this does not imply a non-vanishing spin-Hall mass expectation value at the critical point of the phase transition. Our computation is done on a compact space using the state-operator correspondence and simply serves as way to compute the scaling dimension. The condensate is natural in this context since, once $R$ is reintroduced by undoing our previous rescaling \eqref{eq:f0}, there is a characteristic length to build a non-vanishing mass  $\mg \propto R^{-1}$. 

\subsection{Monopole scaling dimensions for large $\g$  \label{sec:large_g}}
We now obtain an analytical approximation of the condensed mass $\mg$ and the lowest scaling dimension of monopole operators $\Delta_\g$ by studying the free energy $F_\g^{(0)}$ in  a large $\g$ limit. It is simpler to work with the unregularized free energy \eqref{eq:F0_unreg}.  We first change the summation index   $\ell \to \ell + \g + 1$ so that only the summand depends on $\g$. The free energy then becomes
\eqn{
F_\g^{(0)}  
&= -  2\g \mg - 4\sum_{\ell = 0}^\infty (\ell + \g +1) \sqrt{(\ell + \g +1)^2 -\g^2 + \mg^2}  \,.
\label{eq:F0_unreg_2}
}
The saddle point equation $\pa F_\g^{(0)}/ \pa \mg=0$ defining the mass $\mg$ is, up to multiplicative factors, 
\eqn{
 \g + 2 \mg \sum_{\ell = 0}^\infty \frac{ \ell + \g + 1  }{\sqrt{\lb \ell + \g + 1 \rb^2 + \mg^2- \g^2}} = 0  \,.
\label{eq:cond}
}
We introduce an ansatz for the mass squared that is inspired by an analog computation in  the bosonic theory  $\mathbb CP^{N-1}$  \cite{dyer_scaling_2015}
\eqn{
\mg^2 = 2 \chi_0 \g + \chi_1 + \O(\g^{-1})  \,.
\label{eq:ansatz_m2}
}
With this ansatz, we expand  Eq.~\eqref{eq:cond} in powers of $1/\g$. Using once again the zeta function regularization, the condition becomes
\eqn{
0 =& \; 4 \g^{1/2} \lb 2 \zeta_{1/2} + \chi_0^{-1/2} \rb 
 - \g^{-1/2} \Big( 2 \lb   \chi_1  +  \chi _0^2\rb \zeta_{3/2} \nn \\
& + 4  \chi_0 \zeta_{1/2} -6  \zeta_{-1/2}  + \chi_1 \chi _0^{-3/2} \Big)  +\O(\g^{-3/2})  \,,
\label{eq:gapg}
}
where we defined $\zeta_s \equiv \zeta(s, 1 + \chi_0)$.\footnote{Neglecting powers of $\ell$ compared to powers of $\g$ in the large$-\ell$ portion of the sum in condition \eqref{eq:cond} may seem problematic. These terms do appear to higher order in $1/\g$ but cause no problem once properly regularized.}  Solving  \eqref{eq:gapg} order by order, we find a transcendental condition defining $\chi_0$ and a linear condition for $\chi_1$ 
\eqn{	
2 \zeta_{1/2} + \chi_0^{-1/2}  &= 0  \,,
\label{eq:chi0}\\
\chi_1 + \frac{ 2  \chi_0^{3/2} \lb
 \chi_0^2 \zeta_{3/2}
 +2  \chi_0 \zeta_{1/2} 
 -3   \zeta_{-1/2} 
 \rb }{1 + 2 \chi_0^{3/2} \zeta_{3/2}} &=0  \,.
\label{eq:chi1}
}
Inserting the solution of (\ref{eq:chi0}, \ref{eq:chi1}) into the mass ansatz \eqref{eq:ansatz_m2}, we find the mass squared $\mg^2 = 0.199 \g- 0.030 + \O(\g^{-1})$ from which the mass is found to be
\eqn{	
\mg = 0.446 \g^{1/2}- 0.0341 \g^{-1/2} + \O(\g^{-3/2})  \,.
\label{eq:m2_g}
}
As shown in Fig.~\ref{fig:scaling2}, this asymptotic expansion is found to agree extremely well with the exact mass $\mg$, even at small magnetic charge $\g$. 

We repeat this procedure  for the free energy. We insert the mass ansatz \eqref{eq:ansatz_m2} in the free energy \eqref{eq:F0_unreg_2} and we perform a $1/\g$ expansion 
\eqn{
\begin{split}
\half F_\g^{(0)}
=& - \sqrt{2} \g^{3/2} \lb 2 \zeta_{-1/2} + \chi _0^{1/2}  \rb  - \frac{\g^{1/2}}{\sqrt{2}} \bigg( \!   \left(\chi _0^2+\chi _1\right) \zeta_{1/2} \\
&- 6  \chi_0 \zeta_{-1/2} + 5 \zeta_{-3/2} + \frac{1}{2} \chi_1 \chi_0^{-1/2} \bigg)  + \O(\g^{-1/2}) \,. \label{eq:F0g_0}
\end{split}
}
Inserting the solution of (\ref{eq:chi0}, \ref{eq:chi1}) in this result, we obtain the leading order scaling dimension \eqref{eq:del}
\eqn{
\Delta_\g = 2 \nf \lb 0.356 \g^{3/2} + 0.111 \g^{1/2} + \O(\g^{-1/2})\rb + \O(\nf^{0})  \,.
\label{eq:F0g_2}
}
The  scaling dimension  for $\mQED_3$ is found by doing the same expansion but starting with $\mg=0$
\eqn{	
\Delta^{\mQED_3}_\g = 2 \nf \lb 0.588 \g^{3/2}+0.090 \g^{1/2} + \O(\g^{-1/2}) \rb  + \O(\nf^{0})  \,.
\label{eq:F0g_1}
}
Once again, these asymptotic expansions are in good agreement with the corresponding numerical results as shown in Fig. \ref{fig:scaling}. Note that there is no term at order $q^0$ in the scaling dimensions at leading order in $1/\nf$~(\ref{eq:F0g_2}, \ref{eq:F0g_1}). This is expected since all CFTs in $d=2+1$ dimensions with a $\U(1)$ global charge  should have the same $q^0$ term \cite{Hellerman_on_2015, delaFuente_large_2018}. As such, it can't depend on $\nf$. It thus has to vanish at order $\nf$.

We take a step back to appreciate the leading order relation $\Delta_\g^{(0)} \sim \g^{3/2}$. We recall that the theory is set on $S^2 \times \R$. The background magnetic flux on the sphere is $B  R^2 = \g$, where $B$ is the magnetic field and $R$ is the radius of the sphere. Taking  $\g \to \infty$ and $R \to \infty$ while keeping $B$ finite, the theory is reduced to Dirac fermions on a plane in a uniform magnetic field. The eigenvalues are then given by relativistic Landau levels
\eqn{
E_n = \sqrt{2 B \lb n + 1/2 \rb + \mg^2} \approx R^{-1} \g^{1/2} \sqrt{2  \lb n + 1/2 \rb + 2 \chi_0}  \,.
}
In terms of a free energy density $[\mathcal{F}_\g] = [E]^3$,  this means we have $\mathcal{F}_\g \sim R^{-3} \g^{3/2}$. The free energy should  then scale as $F^{(0)}_\g \sim R^2 \mathcal{F}_\g \sim R^{-1} \g^{3/2}$. Reintroducing a missing power of $R$ in \eqref{eq:F0g_0} that was rescaled away in \eqref{eq:f0}, this is indeed the relation we get. The behaviour $\Delta_\g^{(0)} \sim \g^{3/2}$ is then coherent with the large-$\g$ interpretation in terms of Landau levels. 

\section{Monopole dressing \label{sec:monopole}}
In the previous section where we computed the lowest scaling dimension of monopole operators $\Delta_\g$,  the fermionic occupation of the corresponding ground state was not explicited.  In this section, we specify the \zero modes dressing of this ground state and consider other possible  \zero modes dressings  defining other monopole operators. We  show a  non-trivial   hierarchy in the scaling dimensions of  monopole operators in $\mQED_3-\cHGN$ that is not present in $\mQED_3$.  

Monopole operators correspond to  $\U(1)$  gauge invariant states which means they have a vanishing fermion number $\moye{\hat N}$. In  a $\C\T$ quantized theory where ${\{CT, \hat N \} = 0}$, this conditions enforces half filling of the fermion modes. The monopole operators  correspond to states where the Dirac sea is filled as well as half of the \zero modes. This condition that was discussed in the case of $\mQED_3$ \cite{borokhov_topological_2003}. It is also valid in $\mQED_3-\cHGN$ where the spectrum has the same structure, as was shown in Sec. \ref{sec:scaling}. One other consideration is that monopole operators should be Lorentz scalars. We shall focus on $\g = 1/2$ monopoles where this is not an issue as the unique \zero mode for each fermion flavor corresponds to a $j = 0$ Lorentz $\SU(2)_{\rm rot}$  singlet \cite{borokhov_topological_2003}.

\begin{figure}[ht!]
\centering
\vspace{1em}
{\includegraphics[height=5.5cm]{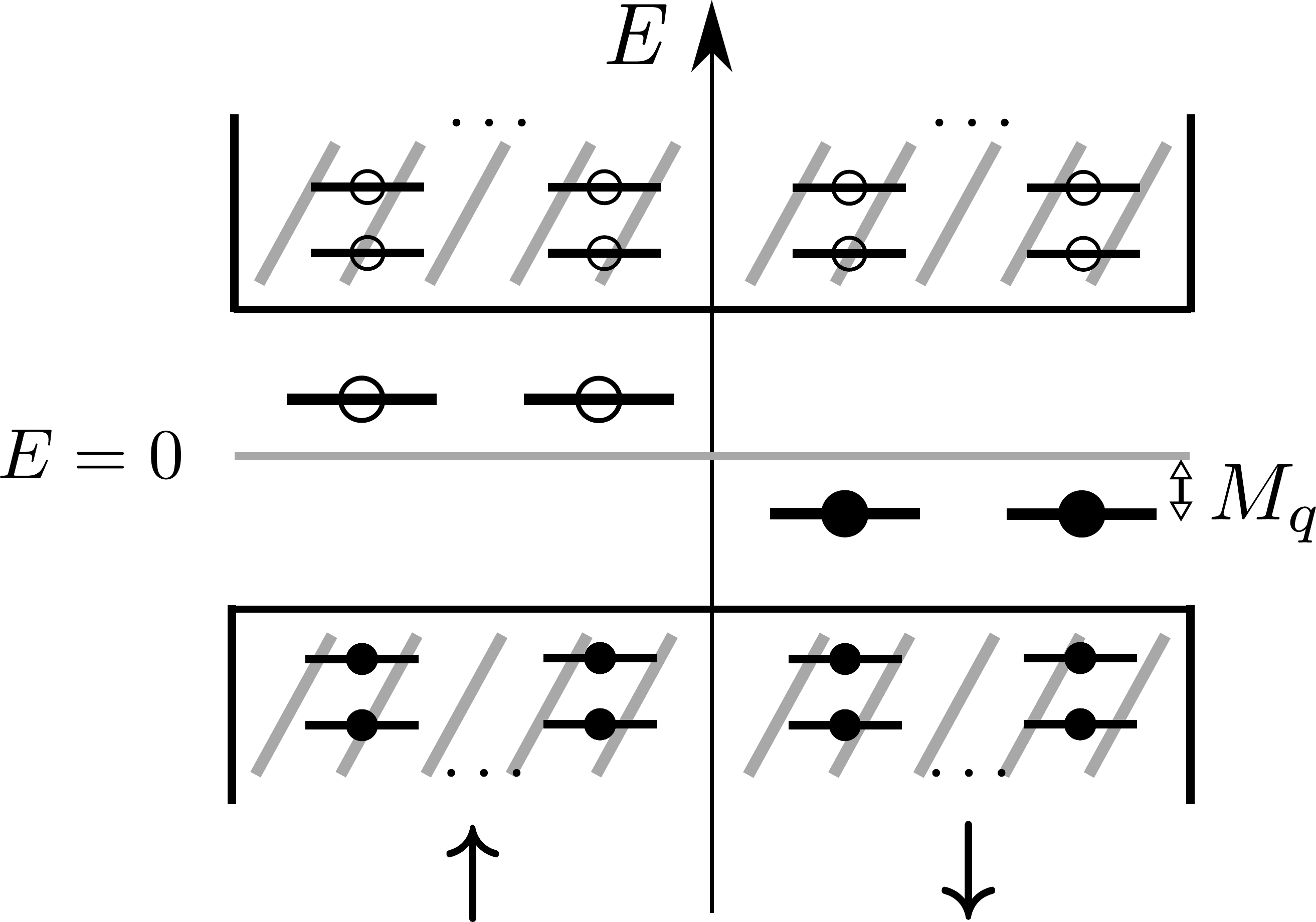}}
\caption{Schematic representation of the energy spectrum and fermionic occupation for the ground state monopole  in presence of a spin-Hall mass $\mg \sigma^z$. The spectrum is shown for $2 \nf = 4$ fermion flavors and minimal magnetic charge $\g=1/2$. Modes with spin up (down) are shown on left (right). A state of minimal energy is achieved by occupying all the Dirac sea as well as all the spin down zero modes which have energy $-\mg$. \label{fig:spectrum_sigz}}
\end{figure}

The energy spectrum of fermions in the monopole background (\ref{eq:spectrum-a},~\ref{eq:spectrum-b}) shows that spin down \zero modes have a lower energy than spin up \zero modes. The monopole operator with the lowest scaling dimension $\Delta_\g$ thus corresponds to the state with all the spin down \zero modes occupied and all the spin up \zero modes empty, as shown  in Fig. \ref{fig:spectrum_sigz}.  We refer to this operator as the ground state monopole.
This fermionic configuration can also be read off the free energy \eqref{eq:F0_unreg} by rewritting it suggestively as
\eqn{
\begin{split}
F_\g^{(0)} 
=&  \; d_\g \lb -\mg \rb  \lb \half \rb
+ d_\g \lb \mg \rb  \lb -\half \rb \\
 & +\sum_{\ell = \g  +1}^\infty \lc  2 d_\ell ( - \veps_\ell ) \lb \half \rb +   2 d_\ell (  \veps_\ell ) \lb - \half \rb \rc  \,.
\end{split}
}
This form puts emphasis on the fact that modes with positive (negative) energy are empty (filled), corresponding to an occupation factor $\moye{c_{n}^\dag c_n -  1/2} = \mp 1/2 $, where $c_n^\dagger$ are the creation operators for the fermion modes in the monopole background.

\begin{figure}[ht!]
\centering
\subfigure[\label{fig:mon^+}]
{\includegraphics[height=3.5cm]{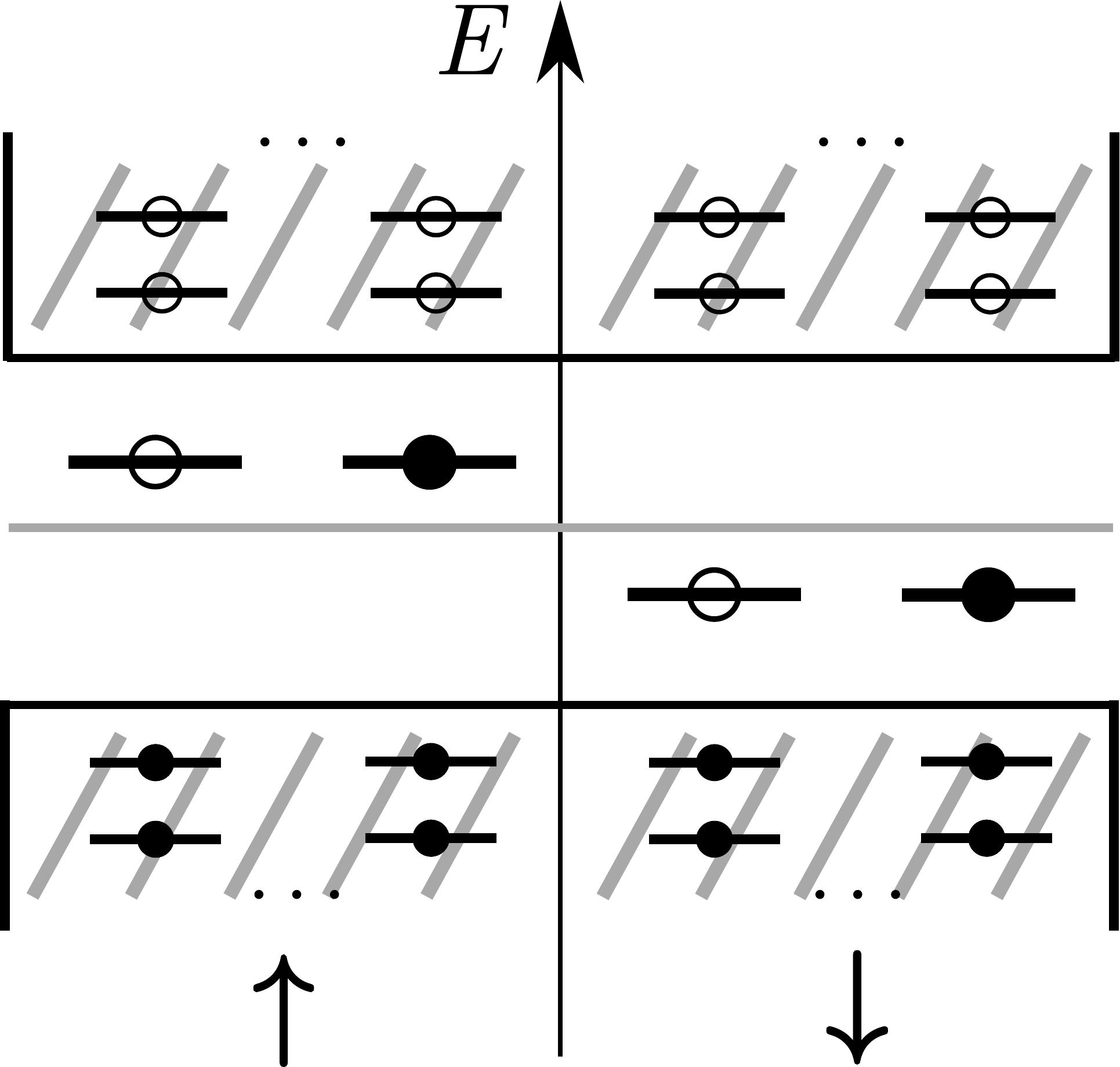}}
\qquad 
\subfigure[\label{fig:mon^up}]
{\includegraphics[height=3.5cm]{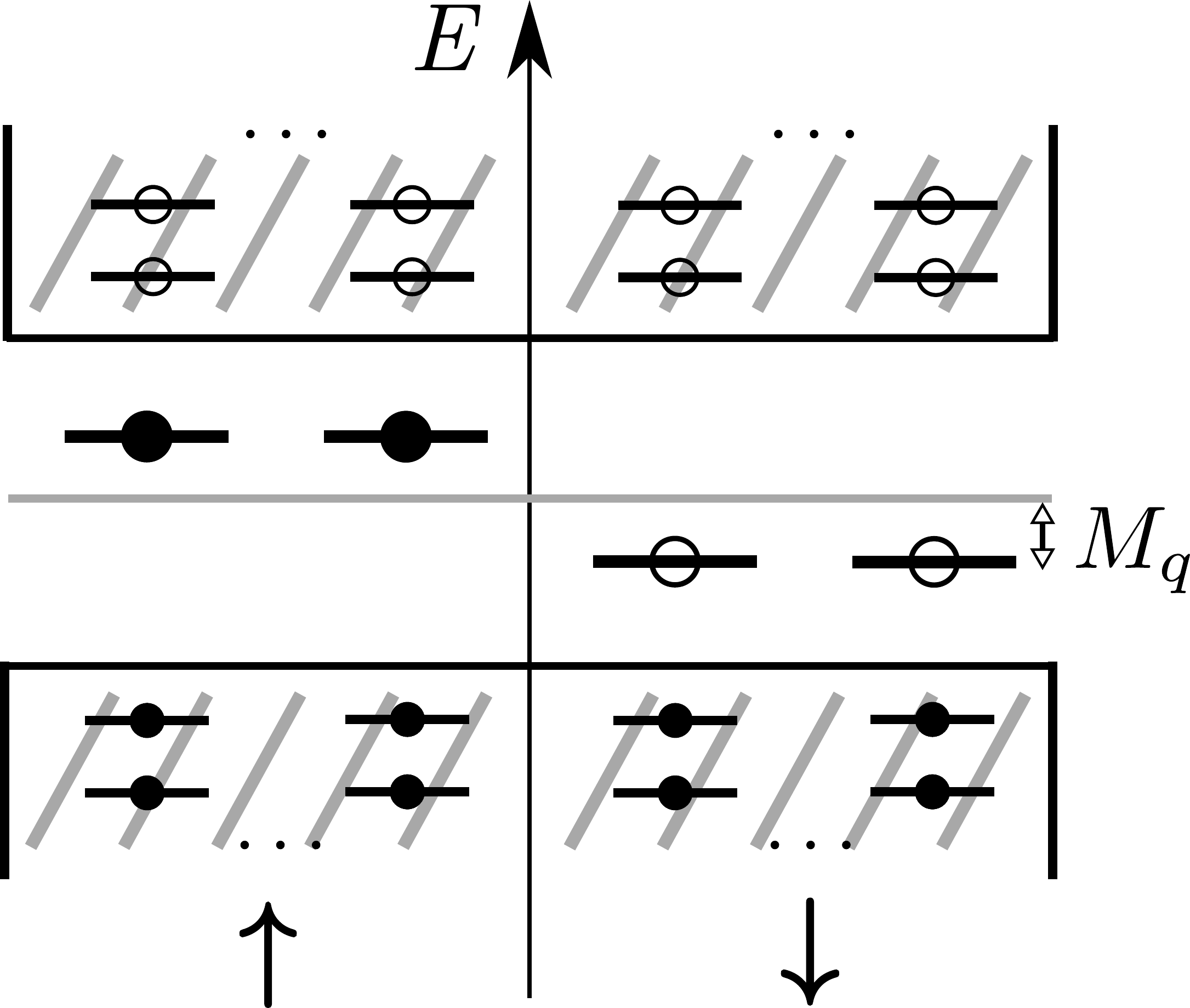}}
\caption{Schematic representation of the fermionic occupation of two excited monopole operators in presence of a spin-Hall mass $\mg \sigma^z$. Modes with spin up (down) are shown on left (right).  The spectrum is shown for $2 \nf = 4$ fermion flavors and minimal magnetic charge $\g=1/2$.  a)~First excited monopole operator; b)~Monopole operator dressed only with spin up \zero modes. \label{fig:mon_exc}}
\end{figure}

We now evaluate the scaling dimensions of monopole operators, which are  defined by the various \zero modes dressings. These operators can be built by annihilating some or all the spin down \zero modes of the ground state monopole and creating  an equal amount of spin up \zero modes. Each such change increases the energy by $\mg - (-\mg) = 2 \mg$.  We study explicitly excited monopole operators with minimal magnetic charge  ${\g = 1/2}$. For example, the first excited monopole operator, whose fermionic occupation is represented in Fig.~\ref{fig:mon^+}, has a scaling dimension $\Delta_{q=1/2}^{+} = \Delta_{q=1/2} + 2 M_{q=1/2} $.  The monopole operator dressed with all the $\nf$ spin up \zero modes, represented in Fig.\ref{fig:mon^up}, has the largest scaling dimension among monopole operators which is ${\Delta_{q=1/2}^{\u} = \Delta_{q=1/2} + 2 \nf M_{q=1/2}}$. 
Using the numerical results for the mass $M_{q=1/2}$ in Tab. \ref{tab:scaling}, we find leading order scaling dimensions of excited monopole operators
\eqn{
\Delta_{q=1/2}^{+} -  \Delta_{q=1/2} &\approx  0.546 + \O(\nf^{-1}) \,, \quad\\
\Delta_{q=1/2}^{\u} -  \Delta_{q=1/2} &\approx 2 \nf (0.273) + \O(\nf^{0})  \,.
}
The  first excited monopole scaling dimension is order $(1/\nf)^0$ larger than the ground state. This difference becomes relatively less important as $\nf$ becomes large.  On the other hand, the scaling dimension of the monopole operator dressed with all spin up \zero modes is order $\nf$ and the difference with the lowest scaling dimension grows larger as $\nf$ is increased.

We can also find the range of the scaling dimensions analytically  for large $\g$.  We consider the scaling dimension of the monopole operator dressed with all the $ 2 \nf \g $ spin up \zero modes
\eqn{
\Delta_{\g}^{\u} = \Delta_{\g} + 4 \nf \g \mg \,.
}  
Using large $\g$ results (\ref{eq:m2_g}, \ref{eq:F0g_2}), we find the leading order results in $1/\nf$ 
\eqn{
\Delta_{\g}^{\u} &= 2\nf \lb 1.248 \g^{3/2}+ 0.0426 \g^{1/2}  + \O(\g^{-1/2})\rb + \O(\nf^{0}) \,.
}

The hierarchy observed  in the scaling dimensions of monopole operators in $\mQED_3-\cHGN$ represents a symmetry lifting of the degenerate monopole multiplet in compact $\mQED_3$. The global $\SU(2 \nf)$ flavor symmetry of $\mQED_3$ implies that monopole operators are organized in a completely antisymmetric representation  of $\SU(2 \nf)$ and consequently must have equal scaling dimensions \cite{borokhov_topological_2003}. Indeed, the symmetry is responsible for the degeneracy of the distinct zero modes dressings defining the monopole operators.  When the spin-dependent interaction is added,  the global symmetry is broken down   $\SU(2 \nf) \to \SU(2)_{\rm spin} \times \SU(\nf)_{\rm valley}$, and there is a symmetry lifting. This lifting is only partial since, for example, monopole operators with trivial spin quantum numbers remain degenerate, having the same scaling dimensions.  
 
\section{Comparison with $\SU(2 \nf)$ symmetric critical point \label{sec:comparison}}
Other Gross-Neveu deformations of $\mQED_3$  yield distinct CFTs and different monopole operators. In this section, we study the monopole operators at the QCP between a DSL and a chiral spin liquid. In the latter phase, the fermions acquire the same mass, leading to a Chern-Simons term for the dynamical gauge field.  This phase transition is driven by the  interaction $( \Psib \Psi )^2$,
\eqn{
S_{\mQED_3-\GN} =  \int d^3 x \lc - \Psib \sl{D}_{a} \Psi  - \frac{h^2}{2} \lb \Psib  \Psi \rb^2     \rc \,,
 \label{eq:QFT_sym_transition}
}
where a $\SU(2\nf)$ symmetric mass is condensed for sufficiently strong coupling strength $h>h_c$.

The procedure to obtain the scaling dimensions of monopole operators must be modified. For this model, there is only a single pseudo-scalar boson $\phi$ entering the Hubbard-Stratonovich transform. The effective action at the critical point can be obtained in the same way we derived the analogous quantity for the spin-dependent case \eqref{eq:S_eff_c_p}
\eqn{
S_{\rm eff}^{\prime \prime c} = - \nf \log \det \lb \sl{D}_{a,A^\g}^{S^2 \times \R}+ \phi  \rb \,,
\label{eq:S_eff_c_pp}
}
where we now work on $S^2 \times \R$ with the sphere pierced by a $4\pi q$ flux. 
Here, there is no constraint on the sign of the mass given by the expectation value of the bosonic field $\mg = \moye{\phi}$.  

Another important difference when computing the lowest scaling dimension of monopole operators in this model is that   a chemical potential $\mu$ must be introduced. This is used to enforce half-filling of the \zero modes.\footnote{We will find  that these modes have vanishing energy and are truly zero modes. However, it is simpler for the discussion that follows to assume that these modes might  have non-zero energy, so we refer to them as \zero modes at this point of the analysis.}  This was not necessary for the model with a spin-dependent interaction. The reason is that the chemical potential is by default set to  zero. Thus, half of the \zero modes are below this level and get filled up. In $\mQED_3$ \cite{borokhov_topological_2003, pufu_anomalous_2014}, the zero modes are also half filled  as a chemical potential set to zero sits at the level of all the zero modes. This is not the case when a $\SU(2 \nf)$ symmetric interaction is activated since the \zero modes all get shifted below the chemical potential if  the boson condensate is non-zero. Thus, directly setting $\mu=0$ only yields the correct fermionic occupation when there is an equal number of modes above and below zero energy.

\begin{figure}[ht!]
\centering
{\includegraphics[height=5.5cm]{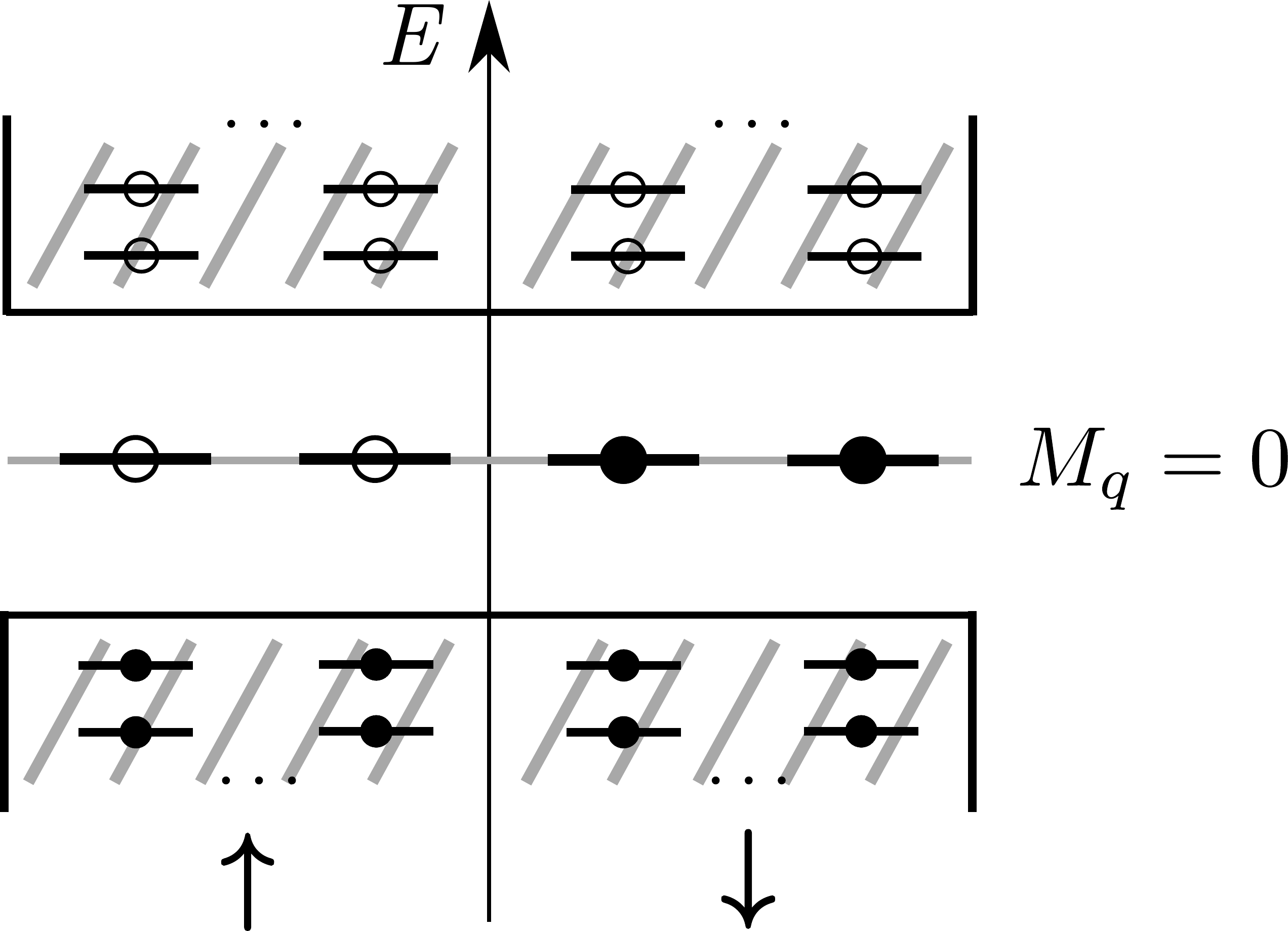}}
\caption{Schematic representation of the energy spectrum and fermionic occupation for one of  the monopole ground state in the $\SU(2\nf)$ symmetric QCP. The spectrum is shown for $2 \nf = 4$ fermion flavors and minimal magnetic charge ${\g = 1/2}$. The zero modes could be shifted by an energy $\mg$, but the saddle point equations force this quantity to vanish.  \label{fig:M_symmetric}}
\end{figure}

The chemical potential can be incorporated within the path integral formalism. We first compactify the ``time" direction to a circle $S^1_{\beta}$ with a radius $\beta$. This radius is taken to infinity, $\beta \to \infty$, at the end of our computations. When working on this ``thermal" circle, the modified relation between the monopole operator scaling dimension $\Delta_\g$  and the free energy $F_\g$  for $\beta \gg 1$ is \cite{chester_monopole_2018}
\eqn{
 \Delta_\g - \frac{1}{\beta} \log(\Omega_\g) + \O \lb e^{-c\beta} \rb = F_\g \equiv - \frac{1}{\beta}\log Z_{S^2 \times S^1_\beta}[A^\g] \,,
 } 
 where $\Omega_\g$  is the ground state degeneracy and $c$ gives the energy spacing between the ground state and the first excited state.\footnote{The identity operator $(\g=0)$ should have a vanishing scaling dimension $\Delta_{\g=0} =0$ \textit{and} no degeneracy, $\Omega_{\g=0} = 1$. This can be guaranteed by a proper normalization, but just as in the spin-dependent case, it turns out to be unnecessary.}  On this space, the  chemical potential can be defined as the homogeneous saddle point value of the imaginary gauge field $\moye{a_{\tau}} = - i \mu$. This approach was used  in Ref.~\cite{chester_monopole_2018}. The  chemical potential is a source for the fermion number $\Psi^\dag \Psi$. Requiring a vanishing fermion number, we have the  condition    
\eqn{
\frac{1}{\beta} \fdv{\ln Z_{S^2 \times S^1_{\beta}}[A^q]}{\mu} = 
 \moy{N_{\rm fermions}} = 0 \,.
 }
This can also be written in terms of a saddle point equation. Once again, the spatial part of the gauge field has a vanishing expectation value $\moye{a_i} =0$. We are left with our homogeneous ansatz $\moye{a_\tau} = - i \mu$ and $\moye{\phi} = \mg$, 
\eqn{
\pdv{F^{(0)}_\g}{\mu} 
&= 0 \,, \label{eq:sp1} \\
 \pdv{F^{(0)}_\g}{\mg} 
&= 0 \,,
\label{eq:sp2}
}
where the leading order in $1/\nf$ of the free energy is 
\eqn{
F^{(0)}_\g = - \log \det (\sl{D}_{-i \mu, A^q}^{S^2 \times S^1_{\beta}} + M_q ) \,.
\label{eq:Fg_sym}
} 
As the ``time" direction is compact, the spectrum of the Dirac operator in Eq.~\eqref{eq:Fg_sym} is now defined by Matsubara fermionic frequencies $\omega_n = (2 \pi / \beta) (n + 1/2)$ where $n \in \Z$.  In contrast to our previous computations, the mass term is now spin symmetric in contrast to our previous computation. 
With these considerations, the previous result shown in Eq.~\eqref{eq:det_op} can be adapted so that  \eqref{eq:Fg_sym} becomes
\eqn{
\begin{split}
F_\g^{(0)} 
=& - \frac{2}{\beta} \sum_{n=-\infty}^\infty
 \Bigg[     d_\g   \log \lc \omega_n - i \mu +  i \mg \rc    \\
 & + \sum_{\ell = \g  +1}^\infty   d_\ell \log \lc (\omega_n - i \mu)^2 + \veps_\ell^2 \rc \Bigg ] \,,
\end{split}
\label{eq:f0_s1}
} 
where as before $\varepsilon_\ell$ is given by Eq.~\eqref{eq:veps_l}
Regularizing the sum over Matsubara frequencies, we obtain
\eqn{
\begin{split}
F_\g^{(0)} 
=& - \frac{2}{\beta} 
 \Bigg[     d_\g  \log \lc 2 \cosh \lb \frac{\beta (\mu-\mg)}{2} \rb \rc \\
 & + \sum_{\ell = \g  +1}^\infty  d_\ell \log \lc 2 ( \cosh{}(\beta \veps_\ell) + \cosh{}(\beta \mu) ) \rc \Bigg ] \,.
\end{split}
}
The saddle point equations (\ref{eq:sp1}, \ref{eq:sp2}) become
\eqna{2}{
-d_{\g} \tanh \lb \frac{\beta (\mu - \mg)}{2}  \rb 
&-
\sum_{\ell = q+ 1}^\infty  \frac{2 d_\ell  \sinh{}(\beta  \mu )}{\cosh{}(\beta \veps_\ell) + \cosh(\beta \mu)} &&= 0 \,, \\
d_{\g} \tanh \lb \frac{\beta (\mu - \mg)}{2}  \rb  
&-
 \sum_{\ell = \g+ 1}^\infty   \frac{ 2 d_\ell \veps_\ell^{-1}   \mg  \sinh(\beta  \veps_\ell)}{\cosh(\beta \veps_\ell) + \cosh(\beta \mu)}    &&=0 \,.
}
Taking $\mu = \mg$ eliminates the first term in both equations. These equations can be further simplified by taking  the large $\beta$ limit. The sum in the first equation vanishes to leading order in $1/\beta$ and the first saddle point equation is then satisfied for $\beta \to \infty$. In the same way, the second saddle point equation to leading order in $1/\beta$ with $\mu = \mg$ is given by
\eqn{
2 \mg \sum_{\ell = q+ 1}^\infty d_\ell \veps_\ell^{-1} = 0 \,.
}
In this limit, this saddle point equation is  only satisfied for $\mg = 0$, which  implies a vanishing chemical potential $\mu=0$. More directly, this means that the expectation value of the bosonic field vanishes,  $\moye{\phi} =0$. Thus, monopole operators at the symmetric QCP are dressed with true zero modes. One of the monopole ground states is shown in Fig.   \ref{fig:M_symmetric}
Since there is no boson condensate, the fermion energies are unchanged when compared to those in $\mQED_3$. This means that  we are left, at leading order in $1/\nf$, with the same scaling dimensions as in $\mQED_3$, $\Delta_q^{\mQED_3-\GN} = \Delta_q^{\mQED_3} + O(1/\nf^0) $. A similar result was obtained for the QCP between a DSL and a ${\Z_{2}- \text{spin liquid}}$~\cite{boyack_transition_2018}.

We emphasize that this is a result at leading order in $1/\nf$. The presence of the bosonic field includes more quantum fluctuations that may change corrections at non-leading orders. We also note that the first condition $\mu = \mg$ is indeed the condition enforcing   half-filling of the zero modes and is needed even if in the end, we find $\mu = \mg  = 0$. Had we not included the chemical potential, we would minimize the free energy that has all the zero modes empty and we would find $\mg \neq 0$. In the case of the spin-dependent interaction we studied before, the saddle point solution for the chemical potential is $\mu = 0$, independently of the value of the mass $\mg$. We show this explicitly in App.~\ref{app:thermal}. This is why the inclusion of a chemical potential was not necessary in this case.

\subsection{Testing a duality with bosonic CP$^1$ model}

The case of compact $\mQED_3-\GN$ theory with $2\nf=2$ fermion flavors is of particular interest since it is conjectured to be dual to the $\CP^1$ theory when both theories are tuned at their critical point, assuming the fixed points exist \cite{wang_deconfined_2018}. This latter theory describes $\nb =2$ flavors of complex bosonic fields forming a $\SU(2)$ doublet $z = \pmatr{z_1, z_2}^\intercal$ which satisfy a length constraint
$z^\dag z= 1$ and interact with a compact gauge field. This model notably describes a deconfined quantum critical point between N\'eel and VBS phases, which is relevant for quantum magnets on various lattices \cite{metlitski_intrinsic_2018},  with the case of square lattice being the prototype example of deconfined criticality \cite{senthil_quantum_2004}. Operators on one side of the conjectured duality have the same scaling dimension as their dual on the other side of the duality. More specifically, the duality relates a set of operators in the fermionic theory 
\eqn{
\begin{split}
\Big[&\Re  \left( \psi_1^\dag \widetilde{\man}_{q=1/2}\right), -\Im  \left(\psi_1^\dag  \widetilde{\man}_{q=1/2}\right), \\ 
&\Re  \left( \psi_2^\dag \widetilde{\man}_{q=1/2}\right), \Im \left( \psi_2^\dag  \widetilde{\man}_{q=1/2}\right), \phi \Big]\,,
\end{split}
\label{eq:op_qed3-GN}
}
to a set of operators in the bosonic theory
\eqn{
\Big[ 2 \Re   \man_{q=1/2}^{\CP^1}, 2 \Im \man_{q=1/2}^{\CP^1}, z^{\dagger} \sigma_{1} z, z^{\dagger} \sigma_{2} z, z^{\dagger} \sigma_{3} z \Big ]\,.
}   
Here,  $\psi_{I=1,2}^\dag \widetilde{\man}_{q=1/2}$ is a monopole with a minimal magnetic charge ${q=1/2}$ dressed with one of the two fermion zero modes. On the bosonic side,   $\man_{q=1/2}^{\CP^1}$ is the unique monopole operator with a minimal magnetic charge ${q=1/2}$. Additionally, both theories have global symmetries relating certain operators of these sets,  i.e. some operators within a set  have the same scaling dimension. Global $\U(1)$ symmetry relates the real and imaginary parts of monopole operators in both theories. Flavor symmetry on the fermionic side relates the two types of monopole operators while a $\SU(2)$ symmetry on the bosonic side implies that all three components of the bilinear $z^\dag \bm \sigma z$ share the same scaling dimension.   Taking into account these global symmetries and the conjectured duality, it is deduced that all operators above should have the same scaling dimension. Said otherwise, the duality predicts  an emergent $\SO(5)$ symmetry at the fixed point of these theories. The duality can thus be tested by comparing the monopole scaling dimension obtained above  to other scaling dimensions conjectured to be the same. 

 The scaling dimension of the monopole operator with $q\!=\! 1/2$ in $\mQED_3-\GN$ for $2\nf \!=\! 2$ fermion flavors is ${\Delta^{\mQED_3-\GN}_{\man_{q=1/2}} =  2 \nf (0.26510) \sim 0.53}$. We compare this value to results in the literature for the other scaling dimensions.   For the monopole operator with minimal magnetic charge in the bosonic theory $\CP^{\nb-1}$ with $\nb=2$, it was found in a similar computation using the state-operator correspondence that ${\Delta^{\CP^{\nb-1}}_{\man_{q=1/2}} = 0.1245922 \nb + 0.05992 \sim 0.31}$ \cite{dyer_scaling_2015}. Note that this result violates the unitary bound; in addition, it is small when compared to other results in the literature.  Using a large $\nb$ expansion in a functional renormalization group analysis~\cite{bartosch_corrections_2013}, it was found that $\Delta_{\text{N\'eel}} \equiv \Delta_{z^\dag \bm \sigma z} =0.61$,
 which is closer to our $0.53$. On the numerical front,  the scaling dimensions of the VBS order parameter $\Delta_{\rm VBS} \equiv \Delta^{\CP^1}_{\man_{q=1/2}}$ and the N\'eel order parameter $\Delta_{\text{N\'eel}}$ were found to be in the range $\Delta_{\text{VBS, N\'eel}} \in [0.60, 0.68]$~\cite{sandvik_evidence_2007, melko_scaling_2008, kaul_lattice_2012, pujari_neel_2013, nahum_deconfined_2015}.   On the fermionic side, a large $\nf$ expansion suggest that the scaling dimension of the boson might be in the range $\Delta_{\phi} \in [0.59, 0.65]$~\cite{boyack_deconfined_2018}. While these last results do not seem far off from the scaling dimension of the monopole operator that we obtain, a more precise computation could help clarify the situation.

\section{Renormalization group analysis of the critical fixed point \label{sec:RG}}
We study more thoroughly the critical fixed point by considering the Yukawa theory that is the UV completion of the $\mQED_3-\cHGN$ model \eqref{eq:action_hubbard}. This model is called  $\mQED_3-\cHGNY$ and  is defined by the following bare Euclidean lagrangian
 \eqn{
\begin{split}
\lag%_{\mQED_3-\cHGNY} 
=& 
- \Psib \sl{\pa} \Psi + \frac{1}{2}  \lb \epsilon_{\mu \nu \rho} \pa_\nu a_\rho\rb^2 + \half  ( \pa_\mu \bm{\phi} )^2  \\
&+ i e    \Psib \sl{a} \Psi +  h  \bm{\phi} \cdot  \Psib \bm{\sigma} \Psi 
+  \half m_{\phi}^2 \bm{\phi}^2  +  \lambda  (\bm{\phi}^2)^2 \,.
\end{split}
\label{eq:lag_QED3-cHGNY}
}
In this section, we consider the non-compact version of $\mQED_3$.  As before, $\Psi$ denotes the spinor with $2\nf$ flavors of two-component Dirac fermion field   and  $\bm \phi$ is  a boson field with $\nb =3$ components. We perform the renormalization group (RG) analysis for general $\nf$ and $\nb$, and then we specialize to $\nb =3$ and $2\nf=4$, which are relevant parameters for certain quantum magnets. For simplicity, we keep referring to this model as $\mQED_3-\cHGNY$ although we don't fix $\nb=3$ from the outset.  RG studies for similar quantum field theories have been considered before. For example, a gauged theory with a  valley-dependent $\cHGNY$-like interaction, $\bm \phi  \Psib \mu_z \bm \sigma \Psi$,  and with $2 \nf =4$ fermion flavors was studied to leading order in $\epsilon = 4 -d$ expansion \cite{ghaemi_neel_2006}. Here, $d$ is the spacetime dimension. 

 The analysis was also done in the ungauged theory, i.e. the $\cHGNY$ model, with general $\nf$ at four loops in the $\epsilon = 4 - d $ expansion \cite{zerf_four-loop_2017} and to order $1/\nf^2$ \cite{gracey_large_0_2018}. The $\mQED_3- \mathrm{GNY}$ was also considered for general $\nf$ using dimensional regularization at one-loop \cite{janssen_critical_2017}, three-loop \cite{bernhard_deconfined_2018} and four-loop \cite{zerf_critical_2018} and to order $1/\nf^2$ \cite{gracey_fermion_2018, boyack_deconfined_2018, benvenuti_easy-plane_2019}.  In the last reference, a spin-dependent Yukawa interaction term with $\nb=1$,  $\phi \Psib \sigma_z \Psi$, has also been considered.  See \cite{gracey_large_2018} for a comprehensive review on large$-\nf$ methods. 

\subsection{Setup}
The first step in our RG study is to write the renormalized Euclidean lagrangian
\eqn{
\lag =& -Z_\psi \Psib \sl{\pa} \Psi + \frac{1}{2} Z_a \lb \epsilon_{\mu \nu \rho} \pa_\nu a_\rho\rb^2  + \half Z_\phi ( \pa_\mu \bm{\phi} )^2 \nn \\
&+  Z_e \lb i e \rb  \mu^{\frac{4-d}{2}} \Psib \sl{a} \Psi + Z_h h \mu^{\frac{4-d}{2}} \bm{\phi} \cdot  \Psib \bm{\sigma} \Psi  \\
&+  \half Z_{m_\phi^2} m_\phi^2 \mu^{2} \bm{\phi}^2  + Z_\lambda \lambda \mu^{4-d} (\bm{\phi}^2)^2 \,,\nn
}
where we introduced wave function renormalization constants $Z_\psi$,  $Z_\phi$ and $Z_a$ as well as vertex renormalization constants $Z_e$, $Z_h$, $Z_{m_\phi^2}$ and $Z_\lambda$. Coupling constants are also rescaled by powers of energy $\mu$ factoring out their naive scaling dimension. The renormalized fields are obtained by a rescaling of the bare fields 
\eqn{
\Psi_0 = \sqrt{Z_\psi} \Psi, \quad \bm{\phi}_0 = \sqrt{Z_\phi} \bm{\phi}, \quad  (a_\mu)_0  = \sqrt{Z_a} a_\mu \,. \label{eq:wave_normalization}
}
From \eqref{eq:wave_normalization}, analog relations between  the renormaliz
d coupling constants $(e^2, h^2, \lambda) \equiv (c_1, c_2, c_3) = \bm c$ the bare coupling constants are obtained
\eqn{
e^2		&= e_0^2 \mu^{d-4}  Z_a, \label{eq:coupling0}  \\
h^2		&= h_0^2 \mu^{d-4} Z_\psi^2 Z_\phi Z_h^{-2}, \\
\lambda &= \lambda_0 \mu^{d-4} Z_\phi^2 Z_\lambda^{-1} \,,
\label{eq:coupling}
}
where the Ward identity, $Z_\psi^2 Z_e^{-2} =1$,  was used to  simplify the  renormalization of the gauge charge $e^2$.    We also add a gauge fixing term $\lag_{\rm g.f.} = \lb \pa_\mu a_\mu \rb^2 / \lb 2 \xi \rb $  to ensure physical quantities are gauge independent.  We have  explicitly written a quadratic boson term since, as we discussed in Sec.\ref{sec:GN}, the mass is the tuning parameter for the  phase transition.  We first study the RG flow equations at the critical value of the boson mass, $m_{\phi}^c =0$. Later on,  we incorporate the boson mass term, along with fermion bilinears, as perturbations away from the QCP.  
 
 By rescaling the energy $\mu \to \mu e^{-l}$ in (\ref{eq:coupling0} - \ref{eq:coupling}), we can analyze how the coupling constants vary with the scale factor \cite{sachdev2011quantum}. The RG flow equation are found by differentiating the renormalized coupling constants with respect to the scale factor, i.e. by obtaining the beta functions $ \beta_{c_I} = \dd c_I / \dd l$ with $I \in \{1,2,3\}$
\eqna{2}{	
\beta_{e^2} &\equiv \dv{e^2}{l} 
&&=  	\lb  4-d   - \gamma_a \rb e^2 \,, \label{eq:beta_e2} \\
\beta_{h^2} &\equiv \dv{h^2}{l} 
&&=  \lb  4-d  - 2 \gamma_\psi - \gamma_\phi + 2 \gamma_h \rb h^2 \,,  
 \\
\beta_{\lambda^2} &\equiv \dv{\lambda}{l} 
&&=  \lb 4-d  - 2 \gamma_\phi + \gamma_\lambda \rb \lambda \,,
}
where the coefficients $\gamma_{x_i}$ with $x_{i} \in \{ \psi, \phi, a, h, \lambda \}$ are defined as 
\eqn{
\gamma_{x_i}  = - \dv{\ln Z_{x_i}}{l} \,.
\label{eq:gamma}
}
These coefficients $\gamma_{x_i}$ are obtained in App.~\ref{app:dia} and allow to find the following RG flow equations
\eqn{	
\dv{e^2}{l} 
=& 	\lb 4-d \rb e^2 - \frac{4\nf}{3} e^4 \,,
\\
\dv{h^2}{l} 
=& \lb 4-d \rb h^2+ 8 \lb 1 -\frac{1}{d} \rb e^2 h^2 \nn \\
& - 2\lb \nf + 2  - \frac{2\nb}{d}   \rb  h^4 \,,\\
\dv{\lambda}{l} 
=& \lb 4-d \rb \lambda - 4 \nf h^2 \lambda + \nf h^4 - 4  \lb \nb + 8 \rb \lambda^2 \,,
}
where the coupling constants have been rescaled to eliminate a loop integral factor.  We will first find the fixed points of the RG flow, that is the  critical coupling constants  $\lb e^2_*, h^2_*, \lambda_* \rb \equiv \bm c_*$ for which the beta functions vanish, $\beta_{c_I} = 0$. The $\mQED_3-\cHGNY$ infrared fixed point corresponding to the QCP will be found, and we will evaluate critical exponents at this point.

\subsection{Fixed points in the $1/\nf$ expansion}
We first control the convergence of the flow by a $1/\nf$ expansion. Setting $d=3$ and assuming that coupling constants  are order $1/\nf$,  the flow equations become
\eqn{
\dv{e^2}{l} 
&=  e^2 - \frac{4 \nf}{3} e^4 \,,  \label{eq:RG_N_0}  \\
\dv{h^2}{l} 
&= h^2  -  2 \nf h^4 \,,\\
\dv{\lambda}{l}  
&=\lambda -4 \nf h^2 \lambda +  \nf h^4 \,.
\label{eq:RG_N}
}
This set of equations leads to the subset of fixed points shown in Tab.~\ref{tab:fptsN}. This excludes Wilson-Fisher type fixed points for which $\lambda$ is not controlled by the $1/\nf$ expansion.
\begin{table*}[ht!]
\caption{Fixed points obtained at leading order in the $1/\nf$  expansion of RG flow equations  (\ref{eq:RG_N_0} - \ref{eq:RG_N}).  There are four such fixed points, where  $G$ stands for ``Gaussian". This excludes Wilson-Fisher type fixed points which are not controlled by the $1/\nf$ expansion. \label{tab:fptsN}}
\centering
\begin{ruledtabular}
\begin{tabular}{cccc}
	Fixed points 
 &$e_*^2 $
 & $h_*^2$ 
 & $\lambda _*$ \\
\hline
 $\mathrm{G}$
 & $0$ 
 & $0$ 
 & $0$ \\
$\mQED_3$
 & $\dfrac{3}{4 N_f}$ 
 & $0$ 
 & $0$ \\
 $\cHGNY$
 & $0$ 
 & $\dfrac{1}{2 N_f}$ 
 & $\dfrac{1}{4 N_f}$ \\
 $\mQED_3-\cHGNY$
 & $\dfrac{3}{4 N_f}$ 
 & $\dfrac{1}{2 N_f}$ 
 & $\dfrac{1}{4 N_f}$ 
\end{tabular}
\end{ruledtabular}
\end{table*}
The linearized RG flow equations around a fixed point $\bm c^*$ yields a matrix equation for the coupling constant perturbations
\eqn{
\dv{}{l} (c_I - c_I^*) = \sum_{J=1,2,3} M_{IJ} (c_J - c_J^*)\,,
}
where $M_{IJ}$ is the stability matrix 
\eqn{
M_{IJ} = \pdv{\beta_{c_I}}{c_J}\bigg|_{\bm c= \bm c^*}\,.
}
Eigenvectors of this matrix equation yield proper directions in the parameter space of couplings $\{e^2, h^2, \lambda\}$. The related eigenvalues $\lambda_I$ indicate relevant (${\lambda_I > 0}$), marginal (${\lambda_I= 0}$), or irrelevant (${\lambda_I < 0}$) perturbations. An infrared fixed point is characterized by irrelevant perturbations in all proper directions. For the theory we study, there is a unique infrared fixed point which, at order $1/\nf$, is given by
\eqn{
e^2_* = \frac{3}{4\nf} \,, \quad h^2_* = \frac{1}{2 \nf} \,, \quad \lambda_* = \frac{1}{4 \nf} \,.
\label{eq:IFP_N}
}
At this infrared fixed point, called $\mQED_3-\cHGNY$, all  critical coupling constants are non-zero.

\subsection{RG study in the $\epsilon  = 4-d$ expansion} 
We now use a dimensional regularization  to control the RG flow. We study the theory at finite $\nf$ by working  near the  upper critical number of spacetime dimensions   ${d = 4-\epsilon}$, where $\epsilon$ is treated as a small expansion parameter.  Assuming the coupling constants are $\order{\epsilon}$, we obtain the flow equations in the $\epsilon$  expansion
\eqn{	
\dv{e^2}{l} 
&=  \epsilon e^2  - \frac{4 \nf}{3} e^4 \,,   \label{eq:rge}  \\
\dv{h^2}{l} 
&=  \epsilon h^2 + 6 e^2 h^2 -  \lb 2\nf + 4 - \nb \rb h^4 \,, \label{eq:rgh} \\
\dv{\lambda}{l} 
&=  \epsilon \lambda +  \nf h^4  - 4   \nf h^2 \lambda   - 4\lb \nb + 8 \rb \lambda^2 \,. \label{eq:rgl}
}
The corresponding physical fixed points are shown in Tab.~\ref{tab:fptseps}, where we have defined 
\eqn{
\K_{\nf, \nb} 	
=& \big[4 N_f^4 +  4 \left(5 N_b+46\right) N_f^3  +324 \left(N_b+8\right)N_f  \nn \\ 
&+\left(N_b^2+172 N_b+1348\right) N_f^2 \big]^{1/2} \,, \label{eq:k_nfnb}\\
\K'_{\nf, \nb} 	
=& \left [4 N_f^2  + 4 \left(5 N_b+28\right) N_f+\left(N_b-4\right){}^2 \right ]^{1/2} \,.
\label{eq:kp_nfnb}
}
By studying the linearized RG flow around the fixed points, one can again show that the $\mQED_3-\cHGNY$ is an infrared fixed point. We note in this case that the infrared fixed point is in the physical region $h^2>0$ only if $2 \nf  - \nb + 4 >  0$  . 
\begin{table*}[ht!]
\caption[]{Fixed points obtained at leading order in the $\epsilon = 4 -d$  expansion of RG flow equations  (\ref{eq:rge} - \ref{eq:rgl}).  $\K_{\nf, \nb}$ and $\K'_{\nf, \nb}$ are defined in Eqs. (\ref{eq:k_nfnb}, \ref{eq:kp_nfnb}), respectively. There are six fixed points, where  $G$ stands for ``Gaussian" and $\WF$ for ``Wilson-Fisher". 
\label{tab:fptseps}
}
\centering
\begin{ruledtabular}
\begin{tabular}{cccc}
 Fixed points
&  $e_*^2$
& $h_*^2$ 
&$\lambda_*$ \\
\hline
G&
 $0$ 
 & $0$ 
 & $0$ \\
 $\mQED_3$&
 $\dfrac{3}{4  N_f} \epsilon $
  & $0$
 & $0$ \\
 $\WF$&
 $0$ 
 & $0$
 & $\dfrac{1 }{4 \left(N_b+8\right)}\epsilon $\\
 $\mQED_3 -\WF$&
 $\dfrac{3}{4  N_f} \epsilon $
 & $0$ 
 & $\dfrac{1 }{4 \left(N_b+8\right)}\epsilon $\\
 $\cHGNY$&
 $0 $
 & $\dfrac{1}{2 \nf - \nb + 4  } \epsilon $
 & $\dfrac{\K'_{N_f,N_b}-2 N_f -N_b +4}{8 \left(N_b+8\right) \lb 2 N_f - \nb +4  \rb} \epsilon$ \\
 $\mQED_3-\cHGNY$&
 $\dfrac{3}{4  N_f} \epsilon$ 
 & $\dfrac{2 N_f+9  }{2 \nf \lb 2 \nf - \nb +4  \rb} \epsilon$
 & $\dfrac{\K_{N_f,N_b} -2 N_f^2 -\left( N_b + 14 \right) N_f}{8 N_f \left( \nb +8\right)  \lb 2 \nf - \nb +4  \rb} \epsilon$ \\
\end{tabular}
\end{ruledtabular} 
\normalsize
\end{table*}

\begin{figure*}[ht!]
\centering
\subfigure[\label{fig:flow-a}]
{\includegraphics[width=0.45\linewidth]{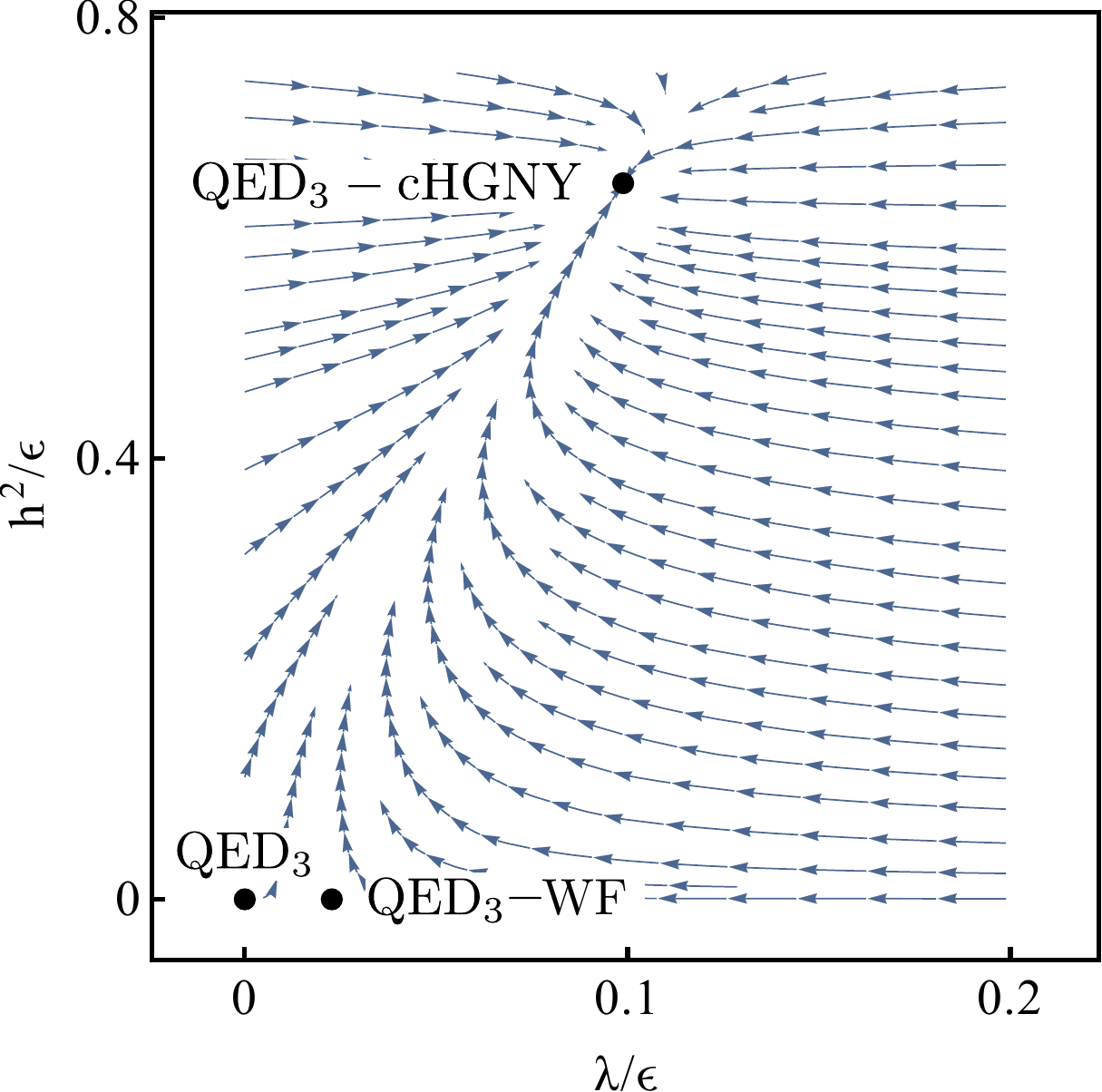}}
\quad
\subfigure[]
{\includegraphics[width=0.45\linewidth]{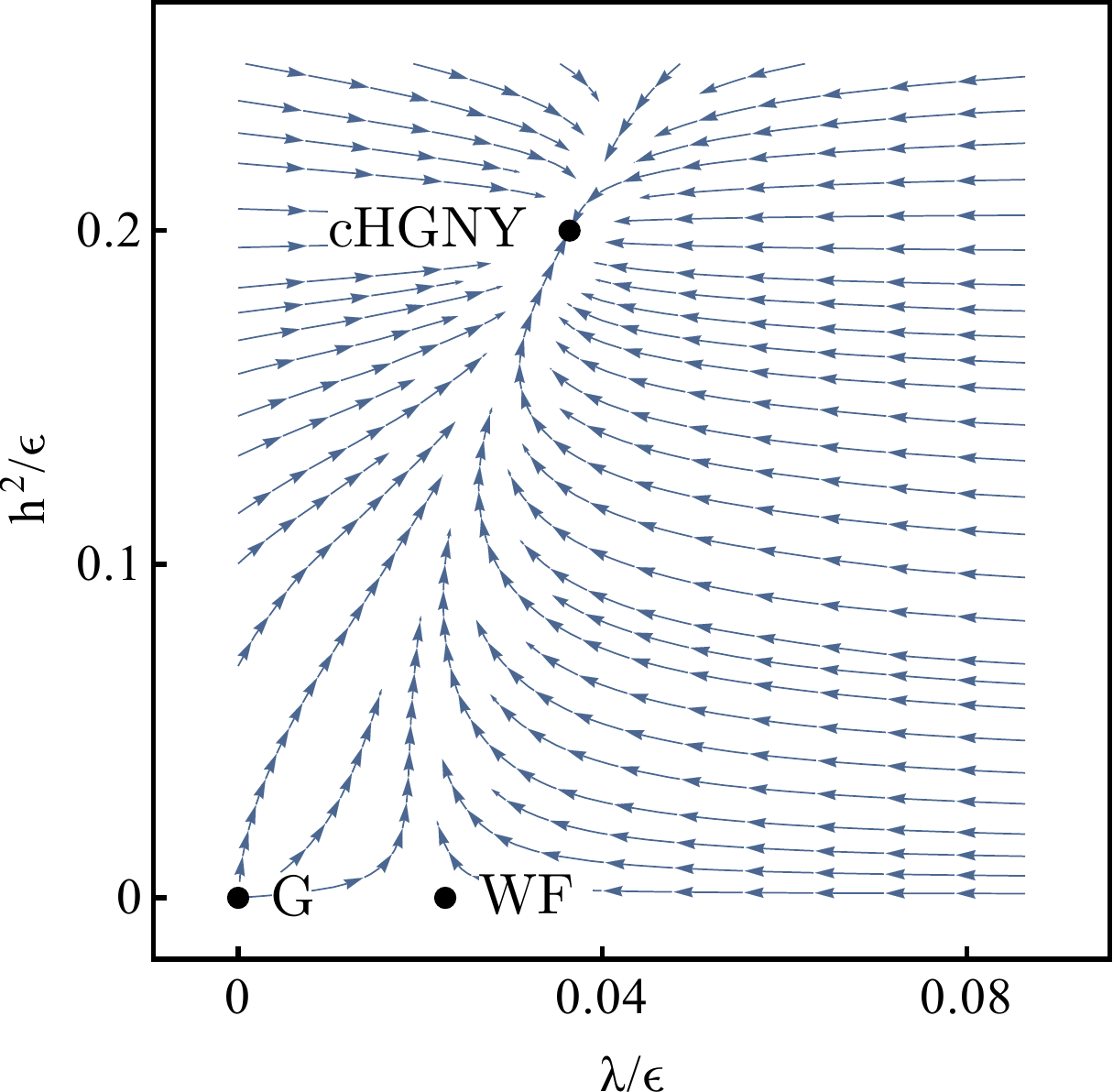}}
\caption{RG flow for $2\nf=4$ and $\nb=3$ in the $\lb \lambda/\epsilon, h^2/\epsilon \rb$ plane to leading order in $\epsilon = 4 - d$. 
a) Flow for $e^2=3 \epsilon / 4 \nf$;
b) Flow for $e^2=0$.
\label{fig:flow}}
\end{figure*}

The flow in the $\lb \lambda/\epsilon, h^2/\epsilon \rb $ plane with $e^2$ fixed to its two possible critical values   is shown Fig.~\ref{fig:flow}. In particular, the  flow from the $\mQED_3$ fixed point ${(h^2 =\lambda=0)}$ to the $\mQED_3-\cHGNY$ fixed point is shown in  Fig.~\ref{fig:flow-a}.
 Setting $\nb=3$, this infrared fixed point is given by  following critical coupling constants
\eqn{
e^2_* 		&=  \frac{3}{4\nf}\epsilon  \,, \label{eq:ec}\\
h^2_* 		&= \frac{2 \nf + 9}{2 \nf \lb 2 \nf +1\rb} \epsilon \,, \label{eq:hc} \\
\lambda_* 	&= \frac{-2 \nf^2 - 17 \nf + \K_{\nf, 3}}{88 \nf  \lb 2 \nf + 1\rb} \epsilon \,, \label{eq:lc}
}
where 
\eqn{
\K_{\nf, 3} = [\nf (4 \nf^3+244 \nf^2+1873 \nf+3564)]^{1/2} \,. \label{eq:knf3}
}
Expanding in powers of $1 / \nf$ and setting $\epsilon =1$, the infrared fixed point corresponds to the previous $1/\nf$ result \eqref{eq:IFP_N}.

\subsubsection{Critical exponents at the quantum critical point}
We now compute  the critical exponents at the infrared fixed point. The scaling dimension of an operator $\O(x)$ at the QCP is defined  by the following expectation value
\eqn{
\moye{\O(x_1) \O^\dagger(x_2)} \propto\frac{1}{|x_1-x_2|^{2 \Delta_\O}}\,,
}
 Using scaling arguments, one can write a naive dimension $\Delta_{\O}^0$. This value is corrected by an anomalous dimension $\eta_{\O}$  once interactions are taken into account
\eqn{
\Delta_{\O} = \Delta_{\O}^0 + \frac{\eta_\O}{2}\,.
}

We start by studying the  anomalous dimensions of the fields $\eta_{\varPhi}$ with $\varPhi \in \{\phi, \psi, a\}$. These quantities are found by evaluating  the corresponding coefficients $\gamma_{\Phi}$ \eqref{eq:gamma} at the QCP, that is $\eta_{\varPhi} = \gamma_{\varPhi}|_{\bm c = \bm c_*^{\rm QCP}}$. First, we write the general expressions for these coefficients  obtained in App.~\ref{app:dia}
\eqn{
\gamma_\phi &= 2 \nf h^2 \,, \quad \label{eq:gamma_phi}  \\
\gamma_\psi &= \lb d + \xi - 5 + \frac{4}{d} \rb e^2 + \lb 1 -\frac{2}{d} \rb \nb h^2 \,, \quad  \\
\gamma_a &= \frac{4 \nf}{3} e^2 \,.  \label{eq:gamma_a}
}
Replacing the coupling constants in Eqs.~(\ref{eq:gamma_phi} - \ref{eq:gamma_a}) by their critical value at the $\mQED_3 - \cHGNY$  fixed point in Tab.~\ref{tab:fptseps}, we obtain the anomalous dimensions
\eqn{
\eta_\phi 
&=   
\frac{2 \nf + 9}{2 \nf - \nb + 4 } \epsilon \,,  \label{eq:eta_phi} \\
\eta_\psi
&= 	\frac{\nb \lb 2 \nf - 3 \xi + 9 \rb + 6 \xi \lb \nf + 2 \rb }{4 \nf \lb 2 \nf  - \nb + 4 \rb} \epsilon \,, \label{eq:eta_psi} \\
\eta_a	
&=  	\epsilon   \,.
}
Note that the anomalous dimension of the fermion depends on the gauge fixing parameter $\xi$, but this is expected since it is not a gauge invariant quantity.  The gauge field anomalous dimension is $\epsilon$ which implies that the one-loop corrected gauge field propagator is $\moye{a_{\mu}(p) a_{\nu}(-p)} \sim |p|^{-1}$ once we set $\epsilon=1$. By setting $e^2_* \neq 0$ in the gauge charge flow equation \eqref{eq:beta_e2}, it is seen that the one-loop result we found, $ \eta_a = \epsilon$, is actually valid to all orders. This is again a consequence of the Ward identity. Now, setting $\nb =3$ in the anomalous dimensions (\ref{eq:eta_phi} - \ref{eq:eta_psi}), we obtain
\eqn{
\eta_\phi		= \frac{ 2 N_f+9}{2 N_f+1} \epsilon \,, \,\;
\eta_\psi 	= \frac{3   \left(2 (\xi+1) N_f+\xi+9\right)}{4 N_f \left(2 N_f+1\right)}\epsilon \,.
}
Setting $2 \nf =4$, we obtain
\eqn{
\eta_\phi		= \frac{13}{5}\epsilon \,,
\quad
\eta_\psi 	= \frac{3 \lb 5 \xi + 13 \rb}{40}\epsilon \,.
}
Setting $\epsilon=1$, we find $\Delta_{\phi}  = (d-2)/2 + \eta_\phi/2 \approx 1.8$.

We now study the scaling dimension of mass operators by introducing mass perturbations at the QCP
\eqn{
\Delta \lag  = \frac{1}{2} Z_{m_{\phi}^2} m_\phi^2 \bm{\phi}^2 +  Z_{m_\psi} m_\psi \Psib \Psi + Z_{\tilde m_\psi} \tilde m_\psi (\bm{\hat n} \cdot  \Psib \bm \sigma \Psi) \,,
}
where $\bm{\hat n}$ is a unit vector indicating the direction of the spin-Hall bilinear perturbation.  We do not include a valley-Hall bilinear $\Psib \mu_A\Psi$, where  $A \in \{1,2, \dots \nf\}$, since its scaling dimension is the same as the $\SU(2 \nf)$ symmetric bilinear $\Psib \Psi$ at leading order in $\epsilon  = 4-d$. The masses we introduced can be related to bare masses like we did with  the other coupling constants
\eqn{
m_\phi^2 &= \lb m_\phi^2 \rb_0 \mu^2 Z_\phi  Z_{m_\phi^2}^{-1} \,,\\
m_\psi &= \lb m_\psi \rb_0 \mu Z_{\psi} Z_{m_\psi}^{-1} \,, \\
\tilde{m} _{\psi}  &= \lb \tilde{m} _{\psi} \rb_0 \mu Z_{\psi} Z_{\tilde{m} _{\psi}}^{-1} \,.
\label{eq:mcouplings}
}
The RG flow equations follow from these relations are
\eqna{2}{
\dv{m_\phi^2}{l} 
&= \lb 2 - \gamma_\phi + \gamma_{m_\phi^2} \rb m_\phi^2 \,,
\\
\dv{m_\psi}{l} 
&= \lb 1 - \gamma_\psi + \gamma_{m_\psi} \rb m_\psi  \,,
\\
\dv{\tilde{m} _{\psi}}{l}
&= \lb 1 - \gamma_\psi + \gamma_{\tilde{m} _{\psi}} \rb \tilde{m} _{\psi} \,, 
\label{eq:flowmass}
}
where the $\gamma_{x_i}$ with  $x_{i} \in \{\phi, \psi, m_{\phi}^2, m_{\psi}, \tilde{m}_{\psi}\}$ are defined in Eq.~\eqref{eq:gamma}. Using results of App.~\ref{app:dia}, the RG flow equations become
\eqn{
\dv{m_\phi^2}{l} 
&= \lb  2  -2  \nf h^2 -4   \left(N_b+2\right) \lambda \rb m_{\phi}^2 \,,  \label{eq:mphi} \\
\dv{m_\psi}{l} 
&= \lb  1 + 4 \lb \frac{d-1}{d}  \rb e^2 + 2 \nb \lb \frac{1-d}{d}  \rb h^2 \rb m_{\psi} \,, \label{eq:mpsi} \\
\dv{\tilde{m} _{\psi}}{l}
&= \lb 1 + 4 \lb \frac{d-1}{d}  \rb e^2 + 2 \lb \frac{ \nb -  d }{d} \rb  h^2 \rb \tilde{m}_{\psi} \,.
\label{eq:mtpsi}
}
When the coupling constants are evaluated at their $\mQED_3-\cHGN$ critical value, the RG flow of the masses is controlled by the scaling dimension of the related mass operators at the QCP 
\eqn{
\dv{m_\phi^2}{l}  			&=	(d- \Delta_{\bm{\phi}^2}) m_{\phi}^2 \,,\\
\dv{m_\psi}{l}  			&=   (d- \Delta_{\Psib \Psi}) m_{\psi}\,,\\
\dv{\tilde{m}_\psi}{l}  	&=   (d- \Delta_{\Psib \sigma_a \Psi}) \tilde{m}_{\psi}\,.
}

We first study the $\bm \phi^2$ perturbation.  The phase transition is controlled by the mass $m_{\phi}^2$ auxiliary boson and the correlation length exponent ${\nu^{-1} = d - \Delta_{\bm{\phi}^2}}$ is obtained by evaluating  \eqref{eq:mphi} at the QCP  
\eqn{
\nu^{-1}
=& \, 2 
-\frac{2 N_f+9}{2 \nf  - \nb + 4 }\epsilon \nn \\
&- \frac{ \left(N_b+2\right) \left(\K_{\nf, \nb}-N_f \left(2 \nf + \nb +  14\right)\right)}{2 N_f  \left(N_b+8\right) \lb 2 \nf  - \nb + 4  \rb}\epsilon \,.
}
Setting $\nb=3$, we obtain
\eqn{
\nu^{-1} 
&=2- \frac{34 N_f^2+113 N_f + 5 \K_{\nf, 3}}{22 N_f \left(2 N_f+1\right)} \epsilon \,,
}
which gives the correct $\nf \to \infty$ limit: $\nu = 1$.
Setting $2\nf=4$, we obtain
\eqn{
\nu^{-1} 
= 2 -4.577 \epsilon \,.
}
Setting $\epsilon=1$, our one loop result yields a negative correlation length exponent.  This is was also observed for  the $\mQED_3-\GNY$ model in Ref.~\cite{janssen_critical_2017} where a dimensional regularization around $d=2+\epsilon$ was also performed to do an interpolation and obtain a positive correlation length exponent. One could also go further in the loop expansion to  obtain a physical result in the $\epsilon \to 1$ limit.  However,  a physical estimate for $\nu$ can be obtained by inverting  $\nu^{-1}$
\eqn{
\nu = \frac{1}{2} + \frac{34 N_f^2+113 N_f + 5 \K_{\nf, 3}}{88 N_f \left(2 N_f+1\right)} \epsilon \,\bigg|_{2\nf = 4} =  \frac{1}{2} + 1.144 \epsilon\,.
}
Setting $\epsilon=1$ now yields the physical result $\nu = 1.644$. Inverting this exponent once again, we obtain the scaling dimension of $\phi^2$ which is given by $\Delta_{\bm{\phi}^2} = d - \nu^{-1} \approx 2.392$. 

We now turn our attention to the fermion bilinears perturbations. Evaluating (\ref{eq:mpsi}, \ref{eq:mtpsi}) with critical couplings of the $\mQED_3-\cHGN$ fixed point shown at leading order  in $\epsilon$ in Tab.~\ref{tab:fptseps}, we find the scaling dimensions at the QCP are given by
\eqn{
\Delta_{\Psib \Psi} 
&= 3-\frac{4 \nf^2 -  (5\nb - 17) \nf - 18(\nb-1)  }{2 \nf \lb 2 \nf  - \nb + 4  \rb}\epsilon  \,, \\
\Delta_{\Psib \sigma_a \Psi}
&= 3 -  \frac{4 \nf - \nb + 13}{2 \lb 2 \nf  - \nb + 4  \rb}\epsilon  \,.
  }
Setting $\nb = 3$, we obtain
  \eqn{
 \Delta_{\Psib \Psi}		&= 3 - \frac{2 \nf^2 + \nf - 18}{ \nf \lb 2 \nf+1 \rb}\epsilon  \,,  \label{eq:scal_psibpsi}  \\
\Delta_{\Psib \sigma_a \Psi} 	&=3 - \frac{2 N_f + 5}{2 N_f+1} \epsilon \,.
}
Setting $2\nf = 4$, this becomes
\eqn{
\Delta_{\Psib \Psi}				&= 3 + \frac{4}{5}\epsilon \,, \quad
\Delta_{\Psib \sigma_a \Psi}    	= 3 - \frac{9}{5}\epsilon \,.
}
Once we set $\epsilon=1$, the spin-Hall bilinear is relevant at the QCP, ${\Delta_{\Psib \sigma_a \Psi} =1.2}$, but the  symmetric bilinear is not,   ${\Delta_{\Psib \Psi} = 3.8}$.  This contradicts what we obtain by taking the large $\nf$ limit in \eqref{eq:scal_psibpsi} since the scaling dimension $\Delta_{\Psib \Psi}|_{\nf \to \infty} =  3 - \epsilon$ then implies a relevant operator for $\epsilon=1$. It is expected that higher order corrections in $\epsilon = 4 - d$ would render $\Psib \Psi$ relevant.

The critical exponents we found are compiled in Tab.~\ref{tab:critical_exponents}. In principle, many of our scaling dimensions should agree with the results in Ref.~\cite{ghaemi_neel_2006} for $2\nf=4$ and $\nb=3$ since the theory considered in this case is almost the same. We find small discrepancies attributable the RG flow equation for the Yukawa coupling, i.e. using their normalization, our Eq.~\eqref{eq:rgh} doesn't match  Eq.~(27) in Ref.~\cite{ghaemi_neel_2006}. Fortunately, the mismatch comes from diagrams which are independent of the number of boson components, thus we can compare with studies of the $\mQED_3-\GNY$ model (see for example Ref.~\cite{janssen_critical_2017}) which confirm our result.  We did other verifications for different regions of the parameter space of our theory. First, we considered the ungauged theory $e^2=0$  where the QCP point is given by the $\cHGNY$ fixed point. Setting $\nb=3$, the fixed points, the RG flow equations and the critical exponents match those of the $\cHGNY$ model presented in Ref.~\cite{zerf_four-loop_2017}. The fermion bilinear scaling dimensions, which were not computed in this last reference,  match the leading order results in $1/\nf$ of \cite{gracey_large_0_2018}. We also verified the gauged theory when $\nb = 1$. In this model,  the spin-Hall bilinear scaling dimension $\Delta_{\Psib \sigma_a \Psi}$ is equal, at one-loop order, to the symmetric bilinear scaling dimension $\Delta_{\Psib \Psi}$ in $\mQED_3-\GNY$. We find agreement with the results obtained in $\epsilon=4-d$ expansions presented in Refs.~\cite{janssen_critical_2017, zerf_four-loop_2017}, and by taking the large $\nf$ limit in our $\epsilon$ expansion and comparing to results obtained with large $\nf$ expansions in Refs.~\cite{iliesiu_bootstrapping_2016, gracey_fermion_2018, benvenuti_easy-plane_2019}. As noted in Ref.~\cite{zerf_critical_2018}, the result for $\Delta_{\Psib \Psi}$ disagrees with the one  presented in \cite{tarnopolsky_yukawa_2016}. We find that the latter result would be obtained if the renormalization of the fermion mass included a  Hartree diagram. This contribution is not generated in Wilsonian RG.

\begin{table}[ht!]
\caption[]{Critical exponents at the $\mQED_3-\cHGNY$ fixed point with $\nb = 3$ at leading order in $\epsilon = 4 -d$. $\K_{\nf,3}$ is defined in \eqref{eq:knf3}. The scaling dimension of valley-Hall bilinears is  $\Delta_{\Psib \mu_A \Psi} = \Delta_{\Psib \Psi}$ where $A \in \{1,2,\dots \nf\}$. 
\label{tab:critical_exponents}
}
\centering
\begin{ruledtabular}
\begin{tabular}{cccc}
& & $2 \nf =4$ & $2 \nf =4$\\
&&&
 $\epsilon=1$
\\
\hline
$\eta_\phi$ 				
& 	$\dfrac{2 \nf + 9}{2 \nf + 1} \epsilon$ 	
&		$2.6\epsilon$ 
& 2.6
\\
$\nu  $ 					
& $\dfrac{1}{2} + \dfrac{34 N_f^2+113 N_f + 5 \K_{\nf, 3}}{88 N_f \left(2 N_f+1\right)} \epsilon$  
& $\dfrac{1}{2} +1.144\epsilon$ 
&1.644
	\\
$\Delta_{\Psib \Psi}$ 				
&	$3 - \dfrac{2 \nf^2+ \nf-18}{ \nf \lb 2 \nf+1 \rb}\epsilon$ 	& $3 + 0.8 \epsilon$
& 3.8
\\
$\Delta_{\Psib  \sigma_a \Psi}$ 	& $3 - \dfrac{2 N_f + 5}{2 N_f+1} \epsilon$						& $3 - 1.8\epsilon$
& 1.2
\end{tabular}
\end{ruledtabular} 
\normalsize

\end{table}

\section{Quantum phase transition in the Kagome magnet \label{sec:kago}}
The $\mQED_3-\cHGN$ model considered in the previous sections finds a natural application in quantum magnets where the underlying lattice  implies  the compactness of the emergent gauge field and the  existence of the monopole operators. We specialize our analysis to the quantum magnet on the Kagome lattice. We first review how a DSL emerges as a possible ground state of the Kagome Heisenberg Antiferromagnet (KHAFM) model. This simple Hamiltonian serves as a starting point to describe  the magnetic $\chem{Cu}$ atoms in Hebertsmithite $\chem{ZnCu_3(OH)_6 Cl_2}$ \cite{shores_a-structurally_2005}. We  also review the confinement-deconfinement transition  from this DSL to a ${\bm q=0}$ coplanar antiferromagnetic phase.  We then examine  the  properties of the monopole operators perturbations which drive this quantum phase transition.

\subsection{Emergent $\mQED_3$}
 The Hamiltonian of the KHAFM is 
\eqn{
 H_{H} = J_{1} \sum_{\expval*{ij}} \bm{S}_i \cdot \bm{S}_j \,,
 \label{eq:HH}
 }
 where $J_1>0$ gives the coupling strength of AFM interactions between nearest neighbors of the Kagome lattice. 
 The emergent fractional spin excitations and gauge field in this model arise due to fractionalization. This phenomenon is studied using a parton construction. The spin operator on site $i$  is decomposed as 
 \eqn{
 \bm{S}_i = \half f_{i, \s}^\dag \bm{\sigma}_{\s \s'} f_{i,\s'}  \,,  \label{eq:spinon_decomposition}
 }
 where $f_{i, \s}$ is a slave-fermion (spinon) with spin ${\s \in \{ \u, \d\}}$ and $\bm{\sigma}$ is a vector of  Pauli matrices  acting on this spin space. The spinon  variables introduce a $\U(1)$ gauge redundancy\footnote{There is a larger $\SU(2)$ gauge symmetry, but the $\U(1)$ subgroup is sufficient for our discussion.} through the symmetry transformation ${f_{i, \s} \to e^{i \theta_{i}} f_{i, \s}}$.  The new Hilbert space is doubled compared to the original spin model, therefore an occupation constraint, $f^\dag_{i, \s} f_{i, \s}=1$, must be imposed. A QSL arises when spinon and gauge degrees of freedom are deconfined. The ground state of the KHAFM is not yet well established. Many numerical studies indicate  a $\U(1)$ spin liquid for the ground state \cite{ran_projected-wave-function_2007, sindzingre_low-energy_2009,  iqbal_gapless_2013, iqbal_spin_2015,  he_signatures_2017, zhu_entanglement_2018} while other investigations point towards a $\Z_2$ spin liquid \cite{jiang_density_2008, gotze_heisenberg_2011, yan_spin-liquid_2011, jiang_identifying_2012, depenbrock_nature_2012, mei_gapped_2017}. In the latter class of spin liquids, the $\U(1)$ gauge symmetry is broken due to a  non-vanishing expectation value  of spinon pairs $\expval*{f_{i}^\dag f^\dag_{j}}$. 
 
 We focus our attention on $\U(1)$ spin liquids.    Using the spinon decomposition \eqref{eq:spinon_decomposition} and applying the occupation constraint, the Hamiltonian developed around the hopping expectation value $\expval*{f^\dag_{i \s} f_{j \s}} \neq 0$ becomes 
 \eqn{
  	\tilde{H}_{\rm H} = - \sum_{\expval*{ij}} t_{ij} e^{i a_{ij}} f_{i}^\dag f_{j} + \hc  \,,    \label{eq:HMF}
  } 
  where the sum on spin indices is now implicit,  ${t_{ij} = J_1 \expval*{f^\dag_{i} f_{j}} / 2}$ and $a_{ij}$  are the phase fluctuations around the expectation value $\moye{f_i^\dagger f_j}$.   The $\U(1)$ gauge symmetry is preserved if the phase fluctuation transforms as $a_{ij} \to a_{ij} +\theta_{i} - \theta_{j}$. This degree of freedom is thus a dynamical $\U(1)$ gauge field. 
  
\begin{figure}[ht!]
\vspace{1em}
\centering
{\includegraphics[height=5cm, width=0.75\linewidth, keepaspectratio]{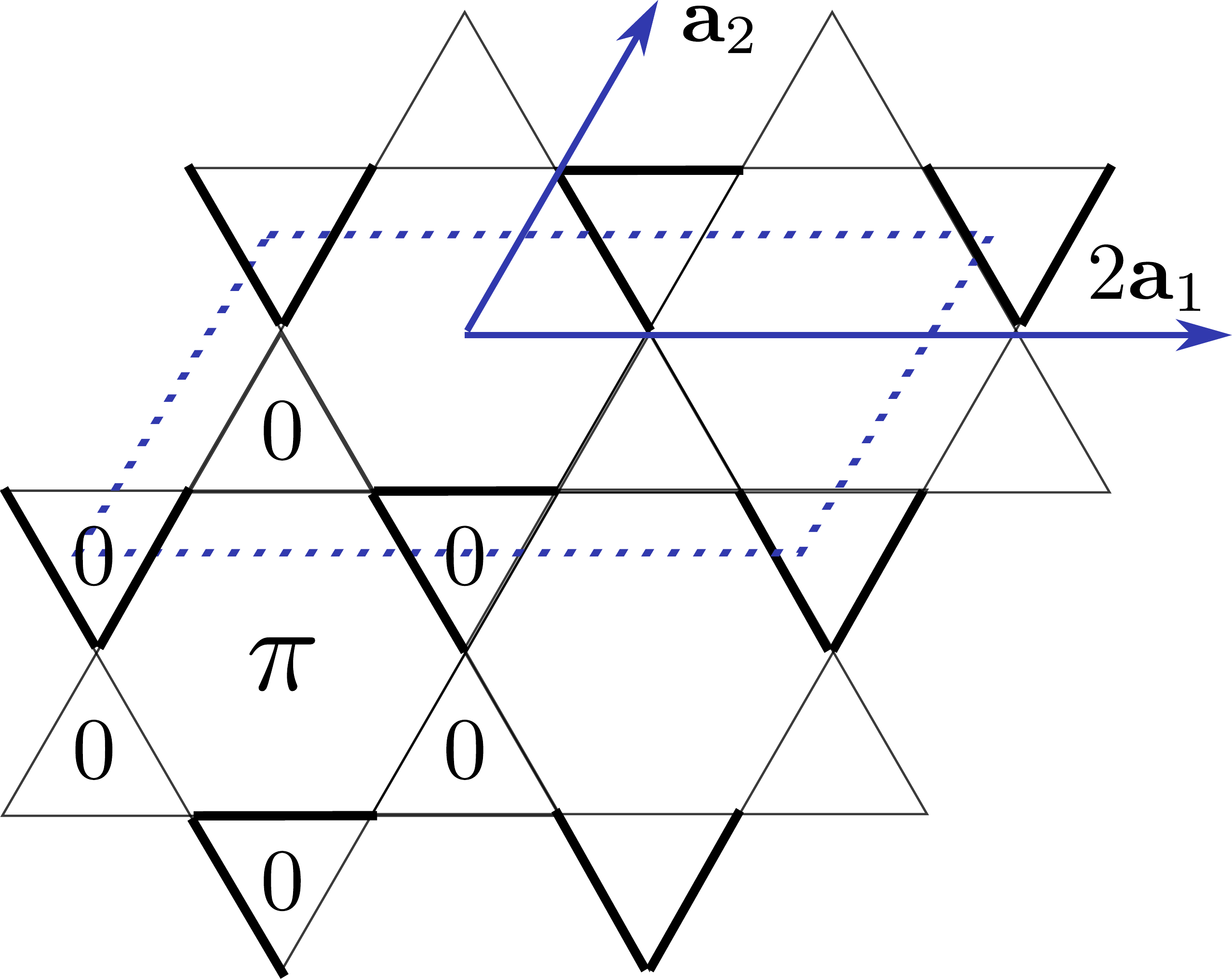}}
\caption{$\pi$-flux pattern on the Kagome lattice. Bold bonds and regulard bonds have opposite signs for their corresponding hopping parameters.   \label{fig:pi-flux}}
\end{figure}

   Among the possible realizations  of a $\U(1)$ spin liquid, the candidate ground state is obtained with the {$\pi$-flux} pattern of the bond orders which is depicted in Fig.~\ref{fig:pi-flux}. 
This pattern defines the ground-state called $\U(1)$ Dirac spin liquid (DSL) and which has $4$ Dirac cones at the Fermi level \cite{hastings_dirac_2000}. The low energy limit is described by $\mQED_3$ with $2\nf =4$ flavors of massless two-component Dirac fermions,  two spin components and two nodes $\pm \bm Q$ in momentum  space \cite{hastings_dirac_2000}. An eight-component spinor regrouping all degrees of freedom can be written as $\Psi = (\psi_{\u, \uff}, \psi_{\u, \dff},  \psi_{\d, \uff}, \psi_{\d, \dff})^\intercal$. Vectors of Pauli matrices acting on $\SU(2)_{\rm spin}$ and $\SU(2)_{\rm valley}$ subspaces, respectively labeled as $\bm{\sigma}$ and $\bm{\mu}$, allow to form spin and valley vectors, $\Psib \bm{\sigma} \Psi$ and $\Psib \bm{\mu} \Psi$. Specifically, the third Pauli matrices in each subspace act as $\sigma_z = \ket{\u}\bra{\u} - \ket{\d}\bra{\d}$ and $\mu_z = \ket{+\bm Q}\bra{+\bm Q} - \ket{-\bm Q}\bra{-\bm Q}$. In similar fashion, Dirac matrices acting on the two-dimension spinor space are represented by Pauli matrices, $\gamma_\mu = (\tau_3, \tau_2, -\tau_1)$. The transformations of these fermions under Kagome lattice symmetries and time reversal are shown in Tab.~\ref{tab:transf_fermions}~\cite{hermele_properties_2008}.
\begin{table*}[ht!]
 \caption{Transformation properties of spinons under discrete symmetries of the KHAFM~\cite{hermele_properties_2008}  where $\mu_{C_6} 			
= ( \mu_1 + \mu_2 - \mu_3)/\sqrt 3$ and  $\mu_{\mathcal{R}_y}  = - (\mu_1 + \mu_3)/\sqrt 2$.}
\centering
\begin{ruledtabular}
\begin{tabular}{r ccccc}
&$T_{\a_1} $
&$T_{\a_2} $
&$\mathcal{R}_y$
&$C_6$ 
&$\T$ \\
\hline\\[-1em]
$\Psi \to$  
&$i \mu_2 \Psi$  
&$i \mu_3  \Psi$
&$\exp \lb \dfrac{i \pi}{2} \mu_{\mathcal{R}_y} \rb \lb i \tau_1 \rb \Psi$
&  $\exp \lb \dfrac{2 \pi i}{3} \mu_{C_6} \rb \exp \lb \dfrac{i \pi}{6} \tau_3 \rb \Psi $
& $\lb i \sigma_2 \rb  \lb -i \mu_2 \rb \lb i \tau_2 \rb  \Psi$ 
\end{tabular}
\end{ruledtabular}
 \label{tab:transf_fermions}
\end{table*}

\subsection{Antiferromagnetic order parameter}

\begin{figure}[ht!]
\centering
\subfigure[]
{\includegraphics[height=2.5 cm]{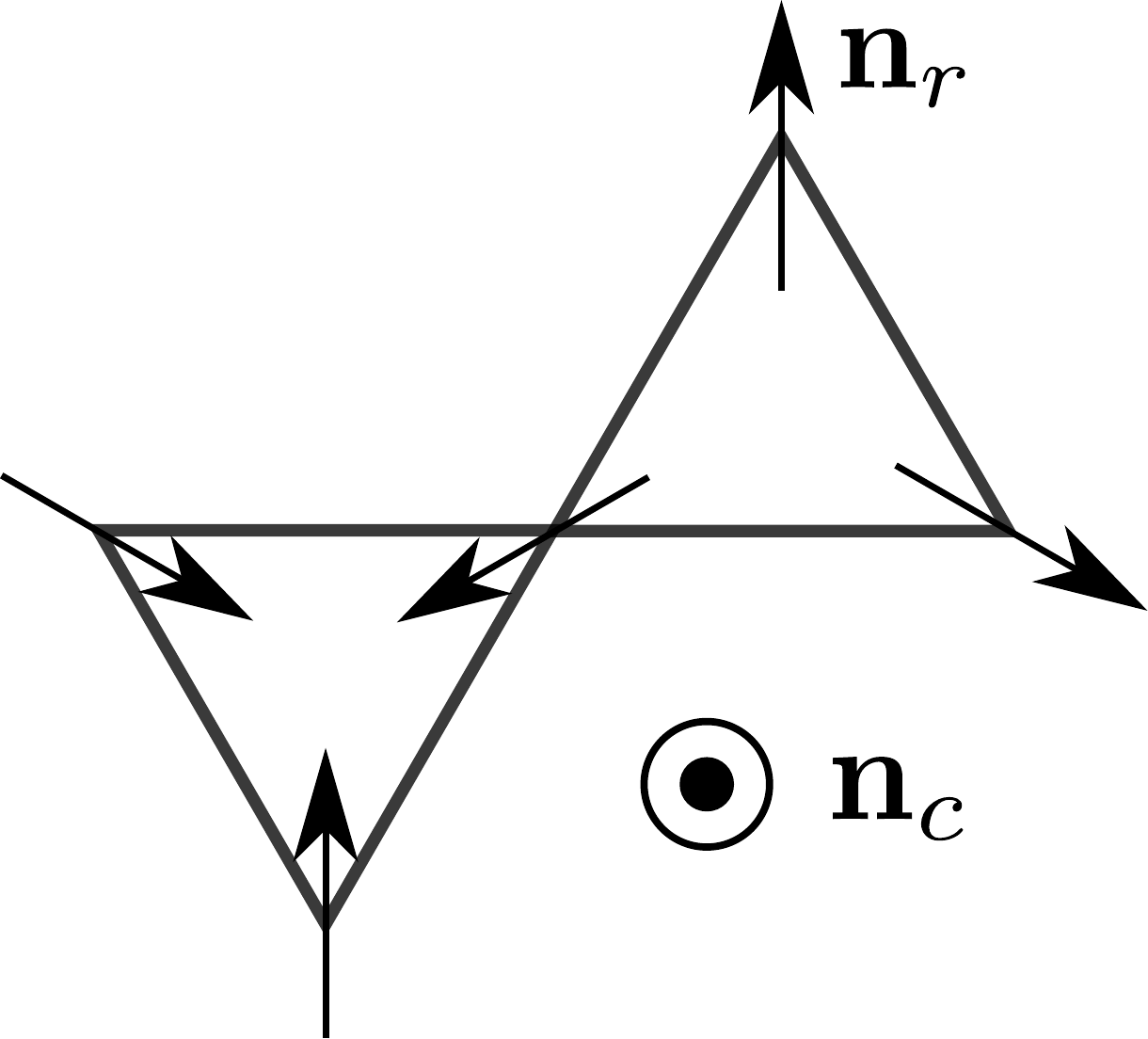}}
\quad
\subfigure[]
{\includegraphics[height=2.5 cm]{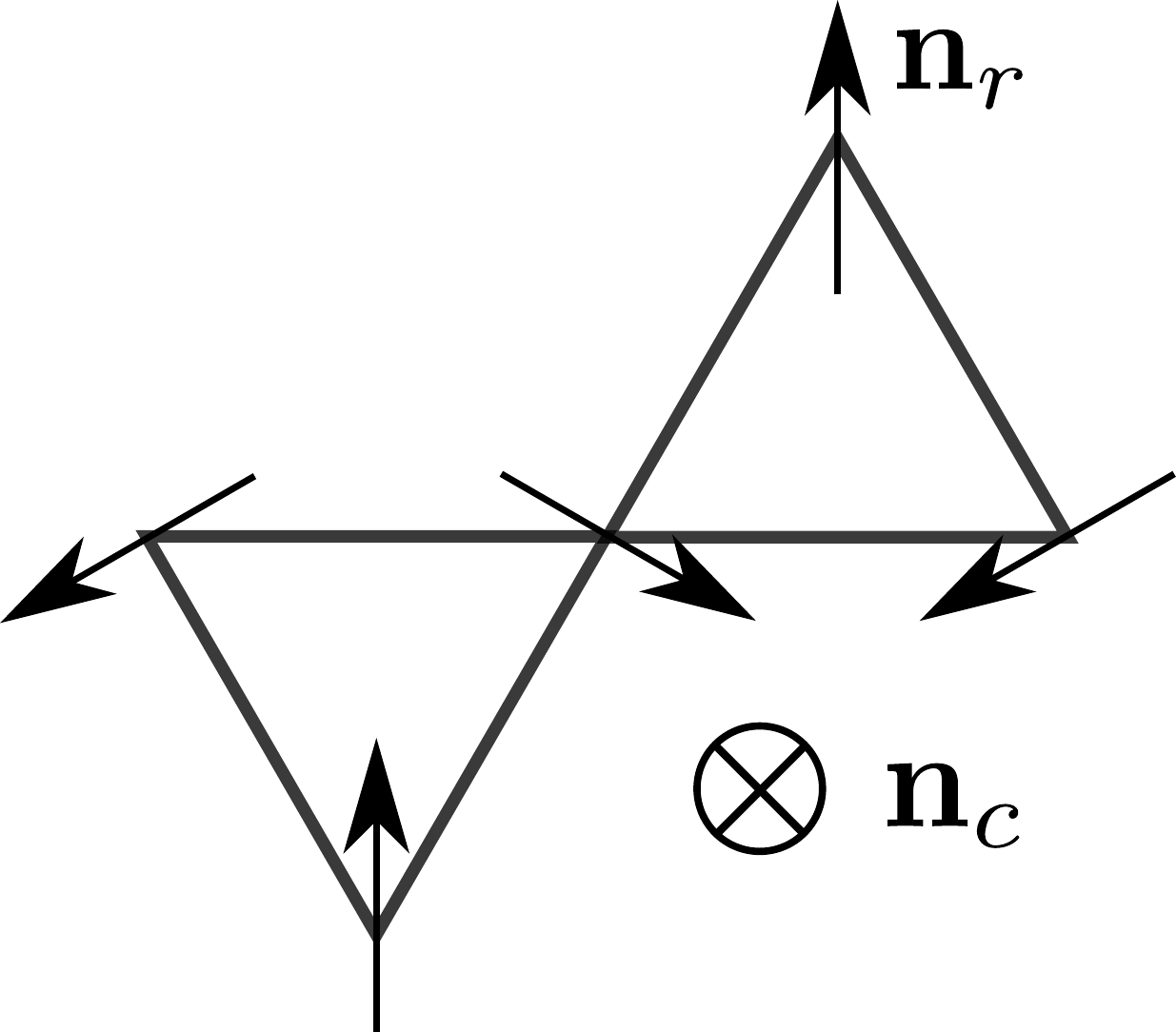}}
\caption{Antiferromagnetic $\bm{q} = 0$ non-collinear phase with a complex order parameter $\bm{n} = \bm{n}_r + i \lb \bm{n}_r \cross \bm{n}_c \rb$ with a) positive chirality;  b) negative chirality. 
\label{fig:q0}}
\end{figure}

We now modify the lattice model to include a next-nearest neighbor AFM coupling $J_2$. The resulting Hamiltonian describes the spin$-1/2$ $J_1-J_2$ Heisenberg model
\eqn{
 H' = J_{1} \sum_{\expval*{ij}} \bm{S}_i \cdot \bm{S}_j +  J_{2} \sum_{\expval*{\expval*{ij}}} \bm{S}_i \cdot \bm{S}_j \,.
 } 
When the ratio $J_2/J_1$ is sufficiently large, the Kagome frustrated magnet orders to a  $\bm{q} = 0$ AFM coplanar phase \cite{gong_global_2015, iqbal_spin_2015, kolley_phase_2015} shown in Fig. \ref{fig:q0}.
The order parameter can be described as a complex vector $\bm{n} = \bm{n}_r + i \lb \bm{n}_r \cross \bm{n}_c \rb$ whose real part encodes the orientation of the spin on one of Kagome's three sub-lattices. On each triangle, the two remaining spins are separated by $120^\circ$ angles with chirality determined by $\bm{n}_c$. The transformation properties of this vector are shown in Tab. \ref{tab:n}.
\begin{table}[h]
\caption{\label{tab:n} Transformation properties of the $\bm{q}=0$ coplanar AFM order  $\bm{n}$ shown in Fig. \ref{fig:q0} under Kagome lattice discrete symmetries and time reversal\cite{hermele_properties_2008}. }
\centering
\begin{ruledtabular}
\begin{tabular}{r ccccc} 
  & $T_{\a_1}$ 
  & $T_{\a_2}$ 
  & $R_y$ 
  & $C_6$ 
  & $\mathcal{T}$ 
  \\ \hline
  $ \bm{n} \rightarrow$ 
  & $\bm{n}$ 
  & $\bm{n}$ 
  & $\bm{n}^*$ 
  & $e^{\frac{2 \pi i}{3}} \bm{n} $ 
  & $- \bm{n}^*$ 
\end{tabular}
\end{ruledtabular} 
\end{table}  

We now show that monopole operators with spin quantum numbers have the same transformation properties as the AFM order parameter shown in Tab.~\ref{tab:n}. This was first argued in Ref.~\cite{hermele_properties_2008}. A comprehensive study of the quantum numbers of monopole operators on the square, triangular honeycomb and Kagome lattices can be found in Ref.~\cite{song_unifying_2018, song_spinon_2018}. One important contribution of this work was to verify numerically the contribution of the $\U(1)_{\rm top}$ charge for the space rotation. This amounts to deducing how  the topologically charged Dirac sea transforms under the  system symmetries. This result can be combined with the transformation properties of the zero modes creation operators to obtain the transformation properties of flux operators. This approach is well explained in Ref.~\cite{alicea_monopole_2008}. We apply this procedure to the case of the Kagome lattice. 

 Here, we  restrict the discussion to monopoles with minimal magnetic charge $\g=1/2$. This means there are four zero modes, two of which must be filled. The $\g=1/2$ flux operators take the form 
\eqn{
\Phi_+^\dag \sim f^\dag_{+;\s, \v} f^\dag_{+;\s', \v'} \widetilde{\man}^\dagger_{+}  \,,
}
where the label $+$ gives the sign of the magnetic charge, $f^\dag_{+;\s, \v}$ is a zero mode creation operator and $\widetilde{\man}^\dagger_{+}$ defines a \textit{bare} $2 \pi$-flux creating operator, an operator  similar to a monopole operator but with all the zero modes empty (in what follows, we refer to it as the bare monopole operator). The six different zero modes filling can be organized as a triplet of $\SU(2)_{\rm valley}$ and a triplet of $\SU(2)_{\rm spin}$, yielding three valley-type and three spin-type flux operators:
\eqn{
\Phi_{+;1,2,3}^\dag 
&=
\frac{1}{2} f^\dag_{+; \s, \v} 
\lc
\lb  i \sigma_2  \rb_{\s \s'} \lb  i \mu_2 \mu_{1,2,3} \rb_{\v \v'}
\rc
f^\dag_{+; \s', \v'}  \widetilde{\man}^\dagger_{+} \,,\\
\Phi_{+;4,5,6}^\dag 
&=
\frac{i}{2} f^\dag_{+; \s, \v} 
\lc
\lb  i \sigma_2 \sigma_{1,2,3} \rb_{\s \s'} \lb  i \mu_2 \rb_{\v \v'}
\rc
f^\dag_{+; \s', \v'}  \widetilde{\man}^\dagger_{+} \,,
}
where the global phases are chosen to reproduce the flux operators introduced in \cite{song_unifying_2018}.
We focus on spin-type flux operator for which we introduce the following short-hand notation
\eqn{
(\Phi_{+;4}^\dag, \Phi_{+;5}^\dag, \Phi_{+;6}^\dag)  \equiv
\bm \Phi_{+;S}^\dag   \equiv  \bm F_{+;S}^\dag \widetilde{\man}^\dagger_{+} \,.
}
The flux operator  with spin down quantum number is
\eqn{
\frac{1}{2} \lb \Phi_{+,4}^\dag - i \Phi_{+;5}^\dag \rb = i f^\dag_{+;\d , -\bm Q} f^\dag_{+;\d , \bm Q} \widetilde{\man}^\dagger_{+} \,. \label{eq:spin_down_flux}
}
We define monopole operators, similar to Ref.~\cite{alicea_monopole_2008}, as combinations of flux creation operators and anti-flux destruction operators, e.g. spin-type monopole operators are given by 
\eqn{
  \bm \S^\dag
&\equiv (\S_1^\dag \,, \S_2^\dag \,, \S_3^\dag)   \nn
  \\
   &= \lb \Phi_{+;4}^\dag  + \Phi_{-;4}\, , \, \Phi_{+;5}^\dag + \Phi_{-;5}\, , \, \Phi_{+;6}^\dag  + \Phi_{-;6}    \rb\,. \label{eq:spin_monopole}
  }
 
 The transformations properties of these operators are obtained as follows.  The transformation properties of the zero modes creation operators $ \bm F_{+;S}^\dag$ can be found using Tab.~\ref{tab:transf_fermions}. As for the bare monopole $\widetilde{\man}^\dagger_{+}$, its transformations are partially constrained by symmetries and the requirement that flux operators $\Phi^\dag_{+}$ and anti-flux operators $\Phi^\dag_{-}$ transform between themselves. This does not completely fix $\U(1)_{\rm top}$ phases which were determined numerically in Refs.~\cite{hermele_properties_2008, song_unifying_2018} and analytically \cite{song_spinon_2018}. From these transformations, one can then find how flux operators and monopole operators transform under symmetries. The transformation properties of all the operators mentioned above are shown in Tab. \ref{tab:sym}. The definition of spin-type monopole operators was chosen \eqref{eq:spin_monopole} such that these operators are odd under time reversal.  Note that the $C_6$ transformation induces an additional phase common to all the monopole operators which can be dropped.  
\begin{table}[ht!]
\caption{\label{tab:sym} 
Transformation properties under Kagome lattice symmetries and time reversal of a bare monopole $\widetilde{\man}^\dagger_{+}$, of a spin-type combination of zero modes creation operators  $\bm{F}_{+;S}^\dagger$ and the corresponding flux operators $\bm{\Phi}_{+;S}^\dagger$ and monopole operators $\bm{\S}^\dagger$. }
\centering
\begin{ruledtabular}
\begin{tabular}{r rrrrr} 	
  & $T_{\a_1}$ 
  & $T_{\a_2}$ 
  & $R_y$ 
  & $C_6$ 
  & $\mathcal{T}$ 
  \\ \hline \\[-1em]
  $ \bm F^\dag_{+;S} \rightarrow$ 
  & $\bm F^\dag_{+;S}$ 
  & $\bm F^\dag_{+;S}$ 
  & $-\bm F^\dag_{-;S}$ 
  & $\bm F^\dag_{+;S}$ 
  & $\bm F^\dag_{-;S}$
  \\
  $\widetilde{\man}^\dagger_{+} \to $
  & $\widetilde{\man}^\dagger_{+}$
  & $\widetilde{\man}^\dagger_{+}$
  & $\widetilde{\man}^\dagger_{-}$
  & $e^{\frac{2 \pi i}{3}} \widetilde{\man}^\dagger_{+}$
  & $- \widetilde{\man}^\dagger_{-}$
\\
    $ \bm{\Phi_{+;S}}^\dag\rightarrow$ 
  & $\bm{\Phi_{+;S}}^\dag$ 
  & $\bm{\Phi_{+;S}}^\dag$ 
  & $-\bm{\Phi_{-;S}}^\dag$ 
  & $e^{\frac{2 \pi i}{3}} \bm{\Phi_{+;S}}^\dag $ 
  & $- \bm{\Phi_{-;S}}^\dag$   
    \\
    $ \bm{\S}^\dag\rightarrow$ 
  & $\bm{\S}^\dag$ 
  & $\bm{\S}^\dag$ 
  & $\bm{\S}$ 
  & $e^{\frac{2 \pi i}{3}} \bm{\S}^\dag $ 
  & $- \bm{\S}$ 
\end{tabular} 
\end{ruledtabular}
\end{table} 
By comparing Tab.~\ref{tab:n} and Tab.~\ref{tab:sym}, we see that the spin triplet monopole $\bm{\S}^\dag$ is the right operator to produce the $\bm{q}=0$ AFM order. 

\subsection{Quantum phase transition}
The confinement-deconfinement mechanism  introduced earlier is thus appropriate to describe the transition from the DSL to the $\bm q = 0$ coplanar AFM on the Kagome lattice. Following the condensation of a spin-Hall mass driven by the $\cHGN$ interaction, spin-type monopole operators proliferate \cite{lu_unification_2017} and condense the AFM order. In terms of lattice operators, the spin-Hall bilinear $\Psib \bm \sigma \Psi$, or equivalently the auxiliary boson $\bm \phi$, corresponds to a vector spin chirality ${V^{a} \sim \sum_{\moye{ij}\in \hexagon}\left(\vec{S}_{i} \times \vec{S}_{j}\right)^{a}}$ \cite{hermele_properties_2008}, where $\hexagon$ denotes an hexagonal plaquette on the Kagome lattice. In this language, the transition is driven by the second neighbor antiferromagnetic interaction which condenses the vector spin chirality, which in turn allows the monopole operators to proliferate on the lattice.

We have just seen that lattice quantum numbers are important to identify the spin down monopole operator as the right operator to induce the AFM. They also determine which combinations of monopole operators transform trivially under all the symmetries of the DSL and thus constitute allowed perturbations in this phase. By inspection of Tab.~\ref{tab:sym}, the sextupled spin down monopole operator, ${ \O = (\S_{1}^\dag - i \S_{2}^\dag)^6 + \hc}$ is identified as a symmetry-allowed perturbation. This is reminiscent of the role that $n-$tupled monopole operators play for the Neel-VBS transition described by $\CP^{\nb-1}$ bosonic theory \cite{read_spin_1990, senthil_quantum_2004,  lee_wess_2014,  metlitski_intrinsic_2018}.    The perturbation ${\O' = \S_1^\dag \S_2 + \hc}$ also respects the symmetries of the Kagome lattice. Among those symmetric perturbations built from spin down monopole operators $\O_{\S}$, the one with the lowest scaling dimension $\Delta_{\O_\S}$ controls the scale $\xi_{\S}$ of the AFM order. The spin-spin connected correlation function  is controlled by this length scale  $\xi_\S$ and the scaling dimension $\Delta_q$  of the spin down monopole operator (see Tab.~\ref{tab:scaling}), scaling as ${\moye{\bm n(\bm r) \cdot \bm n^\dagger(0)}_c \sim 1/ r^{2 \Delta_q}}$ for $r \ll \xi_\S$. 

Determining the scaling dimension of these monopole perturbations $\O_{\S}$ for  $2 \nf  = 4$ is important, as the quantum phase transition works out very differently whether these operators are relevant  at the QCP or not~\cite{lee_signature_2019}. If all allowed monopole perturbations $\O_\S$ turn out irrelevant at the QCP --- as is the case for all monopole operators  in the large$-\nf$ limit --- then these perturbations are dangerously irrelevant, and  monopole operators only proliferate once the spin-Hall mass is condensed. Said otherwise, the QCP is a simple fixed point with one relevant direction being controlled by the boson mass $m_{\phi}^2$. The length scale $\xi$ controlling the spin-Hall mass condensation is then  determined by the critical exponent ${\nu^{-1} = d - \Delta_{\phi^2}}$. The vector spin chirality connected correlation function   depends on this length scale $\xi$ and on the scaling dimension  ${\Delta_V}$ given by ${\min(\Delta_\phi, \Delta_{\Psib \sigma_a \Psi})}$  (see Tab.~\ref{tab:critical_exponents}) at leading order in the loop-expansion \cite{ghaemi_neel_2006}, scaling as  $\moye{\bm V(\bm r) \cdot \bm V(0)}_c \sim 1 / r^{2 \Delta_V}$ for $r \ll \xi$.   In this case where the monopole perturbation is a dangerously irrelevant operator, the two length scales $\xi_{\S}$ and $\xi$ are interdependent~\cite{ghaemi_neel_2006}.  At intermediate scales ${\xi \ll L \ll \xi_{\S}}$, the system is described by a spin liquid where the spinons are gapped since a spin-Hall mass is condensed. At longer scale ${L \gg \xi_{\S}}$, the spinons are confined and the system is well described by the AFM phase.

\section{Conclusion \label{sec:conclusion}}

We have computed the scaling dimension of monopole operators at the QCP of a confinement-deconfinement transition between a DSL and an AFM phase in the large $\nf$ expansion where $2 \nf$ is the number of fermion flavors.  We find that the lowest scaling dimension of monopole operators at the QCP is always smaller than at the $\mQED_3$ point. For the minimal magnetic charge, this scaling dimension is ${\Delta_{\g=1/2} = (2 \nf) 0.19539 + \mathcal{O}(\nf^0)}$. We have considered other possible fermion \zero modes dressings and found a hierachy in the scaling dimension of different monopole operators. We also computed the lowest scaling dimension and the range of the monopole operators scaling dimensions using a large $\g$ limit.  In contrast, the case of the transition to a chiral spin liquid, where a $\SU(2 \nf)$ symmetric mass is condensed, was shown to have the same leading order scaling dimension as $\mQED_3$. To complement our large $\nf$ analysis, we also studied the RG flow of $\mQED_3$ with a spin-dependent Yukawa coupling. We found the existence of the QCP and computed  critical exponents at one loop using the $d= 4 - \epsilon$ expansion. We characterized the QCP in the case of a Kagome quantum magnet.

 We mainly focused on the transition from a DSL to an antiferromagnet and oriented our analysis towards its description. However, the results we obtained for the monopole scaling dimension would be the same with the introduction of a valley-dependent interaction $(\Psib \bm \mu \Psi)^2$. We could do a similar development to obtain the scaling dimension of monopole operators. By performing a large $N$ expansion with $N$ the number of spin components, the same scaling dimensions are found. Now, the monopole operator with the lowest scaling dimension is the monopole in the valley triplet with eigenvalue $-1$ under $\mu_z$. The results at leading order $1/\nf$ is also unchanged whether we pick the  $\SU(2)_{\rm spin}$ symmetric interaction $(\Psib \bm \sigma \Psi)^2$ or an $\SU(2)_{\rm spin}$ symmetry breaking interaction like $(\Psib \sigma_z \Psi)^2$. A similar adaptation can be made for a mixed spin-valley interaction  like  $(\Psib \mu_z \sigma_z \Psi)^2$. In this case,  the scaling dimension is also the same at leading order. 
 It would also be interesting to extend the computation and obtain $\mathcal{O}(\nf^0)$ corrections using the same methods as in Refs.~\cite{pufu_anomalous_2014, dyer_scaling_2015}.  

\begin{acknowledgements}
  We thank Yin-Chen He, Joseph Maciejko,  Subir Sachdev, Sergue\"i Tchoumakov and Chong Wang for useful discussions.
\'ED was funded by an Alexander Graham Bell CGS from NSERC. MP was funded by a Discovery Grant from NSERC. WWK was funded by a Discovery Grant from NSERC, a Canada Research Chair, and a ``\'Etablissement de nouveaux chercheurs et de nouvelles chercheuses universitaires" grant from FRQNT. 

\emph{Note added}--Shortly after this work was submitted, we became aware of Ref.~\onlinecite{zerf_critical_2019}, which partially overlaps with our work.
\end{acknowledgements}

\bibliographystyle{apsrev}
\bibliography{ref}

\appendix
\begin{widetext}
\section{Regularization of the integrated logarithm \label{app:reg}}
The  divergent integral $\int d\omega \log(\omega^2 + a^2)$ can be rewritten by using $\log A = - \dd A^{-s} / \dd s |_{s=0}$. The resulting expression integrates to an hyper-geometric function
\eqn{
\int_{-\infty}^{\infty} d\omega \log(\omega^2 + a^2) &= - \dv{}{s} \int d \omega \lb \omega^2 + a^2\rb^{-s} \bigg|_{s=0}
 =  2    \lb - \dv{}{s} a^{1-2s} \left [\omega \;\; _2F_1\left(\frac{1}{2},s;\frac{3}{2};-\omega^2\right) \right ]_{\omega=\infty} \bigg |_{s=0} \rb \,,
}
where it was assumed that $a>0$, which is the case in the main text where $a = \{\mg, \veps_{\ell}\}$. The analytic continuation then yields
\eqn{
\int_{-\infty}^{\infty} d\omega \log(\omega^2 + a^2) 
	&= 	2    \lb - \dv{}{s}a^{1-2s}  \lb \frac{\sqrt{\pi } \Gamma \left(-s-\frac{1}{2}\right)}{2 \Gamma (-s)} \rb \bigg |_{s=0} \rb  
	=	2  \lb  - a \dv{}{s} \lb -\pi  s+O\left(s^2\right) \rb \bigg |_{s=0} \rb 
 	= 	2 \pi     a \,. \label{eq:annexeA}
}
Alternatively, we can also regularize this integral by introducing a frequency UV cut-off $\Lambda$ and by keeping only the finite part as it is taken to infinity 
\eqn{
 \int_{-\Lambda}^{\Lambda} d\omega \log(\omega^2 + a^2) = 4 \Lambda \lb \log(\Lambda) - 1 \rb + 2 \pi a + \order{\Lambda^{-1}} \to 2 \pi a \,.
}

\section{Saddle point equations for $\mQED_3-\cHGN$ in a thermal setup \label{app:thermal}}
In Sec. \ref{sec:scaling}, we obtained the lowest scaling dimension $\Delta_q$ of monopole operators in $\mQED_3-\cHGN$. We did this by computing the leading order  free energy $F^{(0)}_{\g}$  \eqref{eq:F0_curv} on $S^2 \times \R$ with appropriate magnetic flux. We saw that in the case of a $\SU(2 \nf)$ symmetric interaction, the computation should be performed on a compactified ``time" direction $\R \to S^1_\beta$. Here, we repeat the computation in the former theory now using the ``thermal" setup,
\eqn{
F_q^{(0)} 
&=  - \frac{1}{\beta} \log \det \lc  \sl{D}_{-i\mu, A^q} + \mg  \sigma_z \rc  \\
&= - \frac{1}{\beta} \sum_{\sigma = \pm  1} \sum_{n \in \Z} \lc  d_{q}  \log \lc \omega_n - i  \mu + i \sigma \mg \rc   + \sum_{\ell = q+1}^\infty d_\ell \log  \lc \lb \omega_n - i \mu \rb^2 + \veps_\ell^2 \rc  
  \rc \,,
  }
  where as before $\varepsilon_\ell$ is given by Eq.\eqref{eq:veps_l} and $\omega_n$ are the Matsubara frequencies. First taking the sum over the magnetic spin degrees of freedom and then regularizing the sum on Matsubara frequencies, this expression becomes 
  \eqn{
F_q^{(0)} 
&= - \frac{1}{\beta}  \sum_{n \in \Z} \lc d_{q}   \log \lc  \lb \omega_n - i \mu\rb^2 + \mg^2 \rc + \sum_{\ell = q+1}^\infty  2 d_\ell \log  \lc \lb \omega_n - i \mu \rb^2 + \veps_\ell^2 \rc
 \rc \\
&= - \frac{1}{\beta}  \lc  d_{q} \log \lc 2 \lb \cosh(\beta \mg ) + \cosh(\beta \mu) \rb \rc  + \sum_{\ell = q+1}^\infty  2 d_\ell \log \lc 2 \lb \cosh(\beta \veps_\ell) + \cosh(\beta \mu) \rb\rc  \rc  \,.
\label{eq:Fq0_Heisenberg}
}
The saddle point equations are
\eqna{2}{
0 = \pdv{F_q^{(0)}}{\mu} 
&=   -d_{q} \lb \frac{\sinh(\beta \mu)}{\cosh(\beta \mg) + \cosh(\beta  \mu  )}\rb 
&&- \sum_{\ell = q+1}^\infty 2 d_\ell \lb \frac{\sinh(\beta  \mu )}{\cosh(\beta \veps_\ell) + \cosh(\beta \mu)} \rb  \,,\\
0 =  \pdv{F_q^{(0)}}{\mg} 
&=   - d_{q} \lb  \frac{\sinh(\beta \mg)}{\cosh(\beta \mg) + \cosh(\beta  \mu  )} \rb  
&&- \sum_{\ell = q+1}^\infty 2 d_\ell  \veps_\ell^{-1} \mg \lb  \frac{\sinh(\beta  \veps_\ell)}{\cosh(\beta \veps_\ell) + \cosh(\beta \mu)} \rb   \,.
}
The first saddle point equation is solved with $\mu = 0$. This is the reason why the formalism used in the present section is unnecessary, as it was mentioned in Sec.~\ref{sec:comparison}.  Setting $\mu=0$ in the second saddle point equation  yields
\eqn{
0 =  \pdv{F_q^{(0)}}{\mg} \bigg|_{\mu=0} &=   - d_{q} \lb \frac{\sinh(\beta \mg)}{1+\cosh(\beta \mg)} \rb - \sum_{\ell = q+1}^\infty 2 d_\ell  \veps_\ell^{-1} \mg \lb \frac{\sinh(\beta  \veps_\ell)}{1+\cosh(\beta \veps_\ell)}\rb  \,.
}
We get the interesting result that $\mg=0$ is a solution of this saddle point equation. This solution was not observed by working directly on $S^2 \times \R$.  Assuming $\mg>0$, the saddle point equation at leading order in $1/\beta$ yields the following condition
\eqn{
0 =  d_{q}  + \sum_{\ell = q+1}^\infty 2 d_\ell  \veps_\ell^{-1} \mg  \,,
\label{eq:gap_condition}
}
 A non trivial solution $\mg  \neq 0$ can be found numerically as in the main text. We distinguish the trivial and non-trivial solutions by studying the second derivatives of the free energy at these points.  The second derivative  in the $\mu$ direction designates imaginary fluctuations  of the gauge field and as such has no physical signification.  We only  compute  the second derivative in the $\mg$ direction 
\eqn{
\pdv[2]{F_q^{(0)}}{\mg} \bigg|_{\mu=0} 
&=    - d_{q} \lb \frac{\beta}{1+ \cosh(\beta \mg)} \rb - \sum_{\ell = q+1}^\infty 2 d_\ell \lb \frac{\lb \veps_\ell^2 - \mg^2 \rb  \sinh \lb \beta  \veps_\ell\rb+\beta  \mg^2 \veps_\ell}{\veps_\ell^3 \lb 1 + \cosh \lb \beta  \veps_\ell\rb\rb} \rb \,.
}
 We study the cases $\mg \neq 0$ and $\mg=0$ separately. We start with the former case. The second derivative at leading order in $1/\beta$ is 
\eqn{
\pdv[2]{F_q^{(0)}}{\mg} \bigg|_{\mu=0, \mg \neq 0} 
&= - 2   \sum_{\ell = q  + 1} \frac{d_\ell  \lb \veps_\ell^2 - \mg^2 \rb}{\veps_\ell^3}  \,.
}
Using the saddle point condition \eqref{eq:gap_condition} obtained for large $\beta$, this can be reformulated as 
\eqn{
\pdv[2]{F_q^{(0)}}{\mg} \bigg|_{\mu=0, \mg \neq 0}  =  d_{\g} \frac{1}{\mg} + 2  \sum_{\ell = q + 1}^\infty \frac{\mg^2}{\veps_\ell^3} > 0  \,.
}
The non trivial solution $\mg \neq 0$ thus corresponds to a minimum. The case of a vanishing mass  $\mg=0$ is now studied. In the large $\beta$ limit, the second derivative of the free energy at leading order in $1/\beta$ is given by
\eqn{
\pdv[2]{F_q^{(0)}}{\mg} \bigg|_{\mu=0, \mg = 0} = -\frac{\beta}{4} < 0  \,.
}
Thus,  when setting $\mu=0$, $\mg= 0$ is  a maximum.

\section{Feynman diagrams computation \label{app:dia}}
A RG study of the $\mQED_3-\cHGNY$ model was shown in Sec.\ref{sec:RG}. In the present section, we compute the one-loop corrections to the wavefunction normalization and coupling constants  of this QFT necessary to obtain the RG flow equations shown in the main text.  

\subsection{Feynman rules \label{sec:feyn_rules}}
We first write down the Feynman rules of the $\mQED_3 - \cHGNY$ field theory defined by the lagrangian \eqref{eq:lag_QED3-cHGNY}
\eqn{
\dia{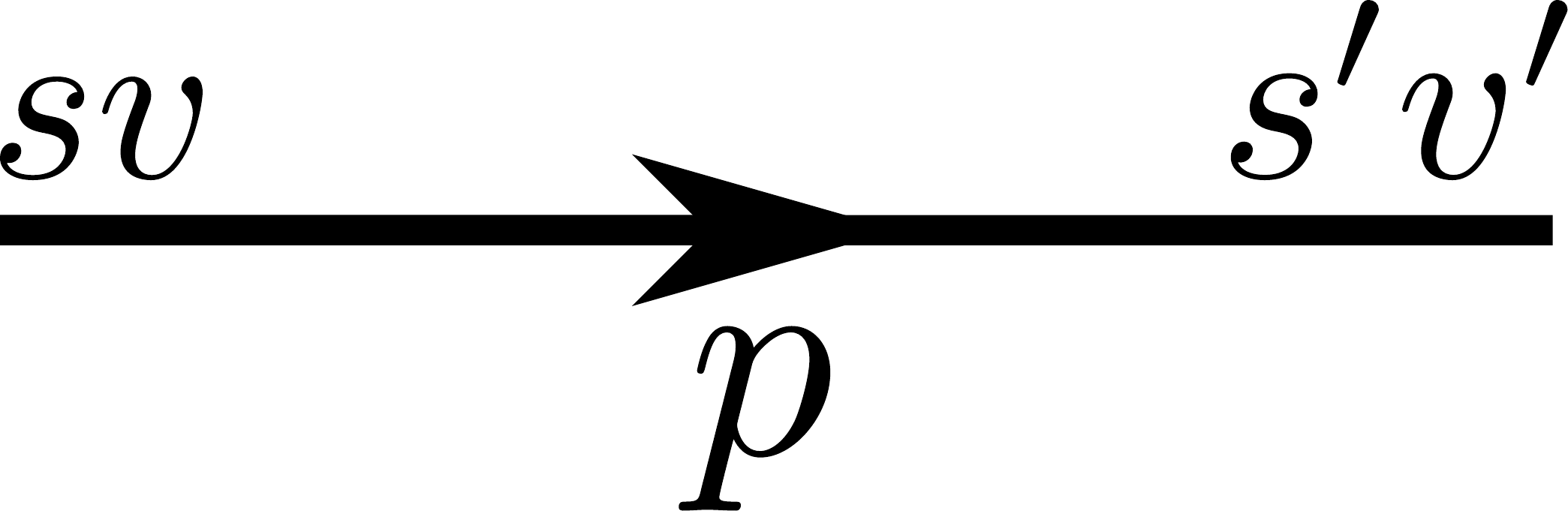}
&= G_{sv;s'v'}(p) \equiv \moy{\psi_{\s\v}(p) \psib_{\s' \v'}(p)} = \delta_{\s \s'} \delta_{\v \v'} \frac{\slashed{p}}{p^2} \,,\\
\dia{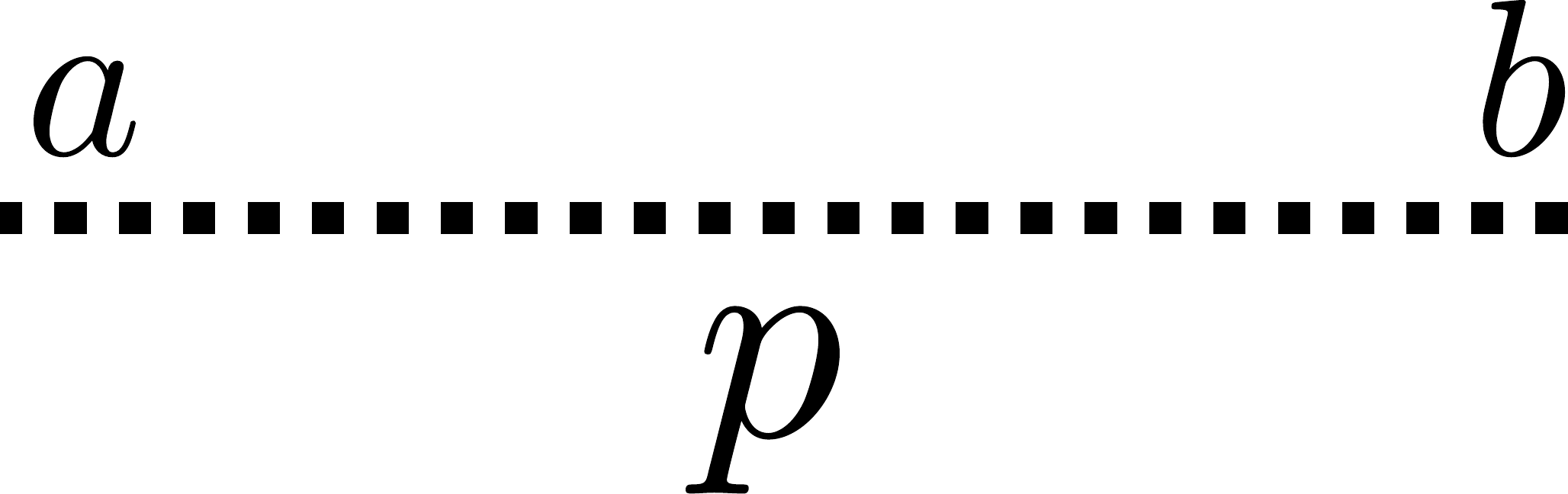}
&= D_{ab}(p) \equiv \moy{\phi_a(p) \phi_b(-p)} = \frac{\delta_{ab}}{p^2} \,,\\
\dia{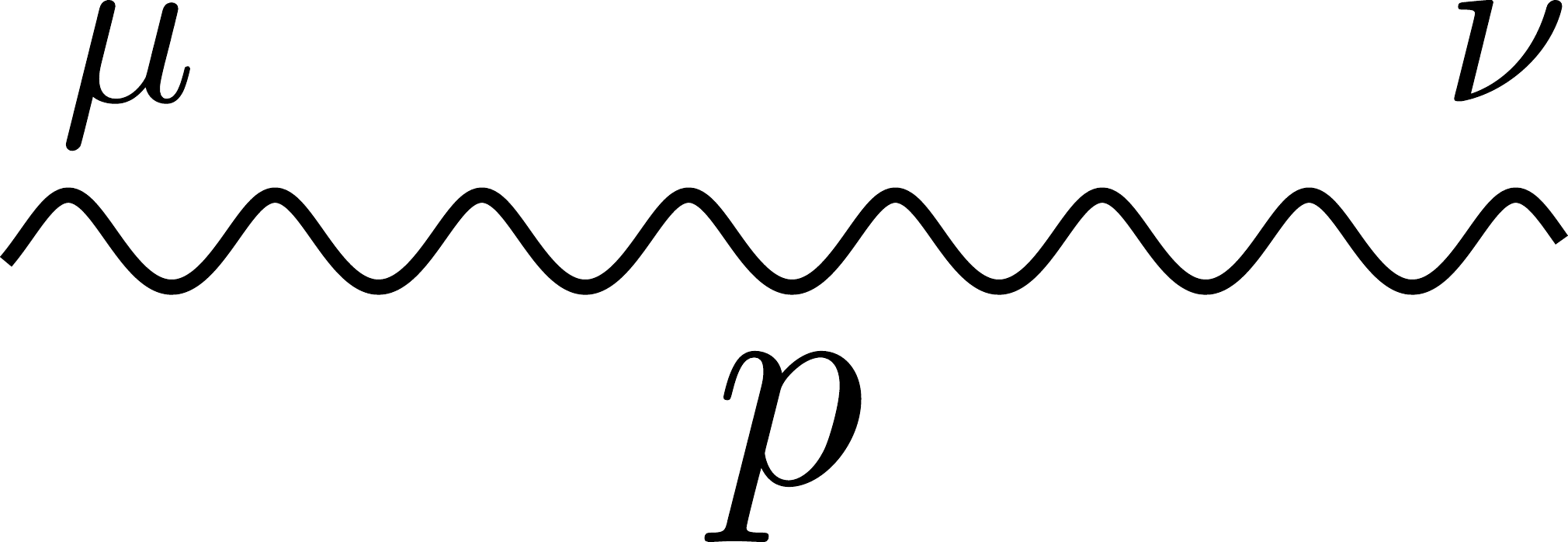}  
&= \Pi_{\mu \nu}(p) \equiv \moy{a_\mu(p) a_\nu(-p)} = \frac{1}{p^2} \lb \delta_{\mu \nu} + \lb \xi-1 \rb \frac{p_\mu p_\nu}{p^2} \rb \,, \\
\dia{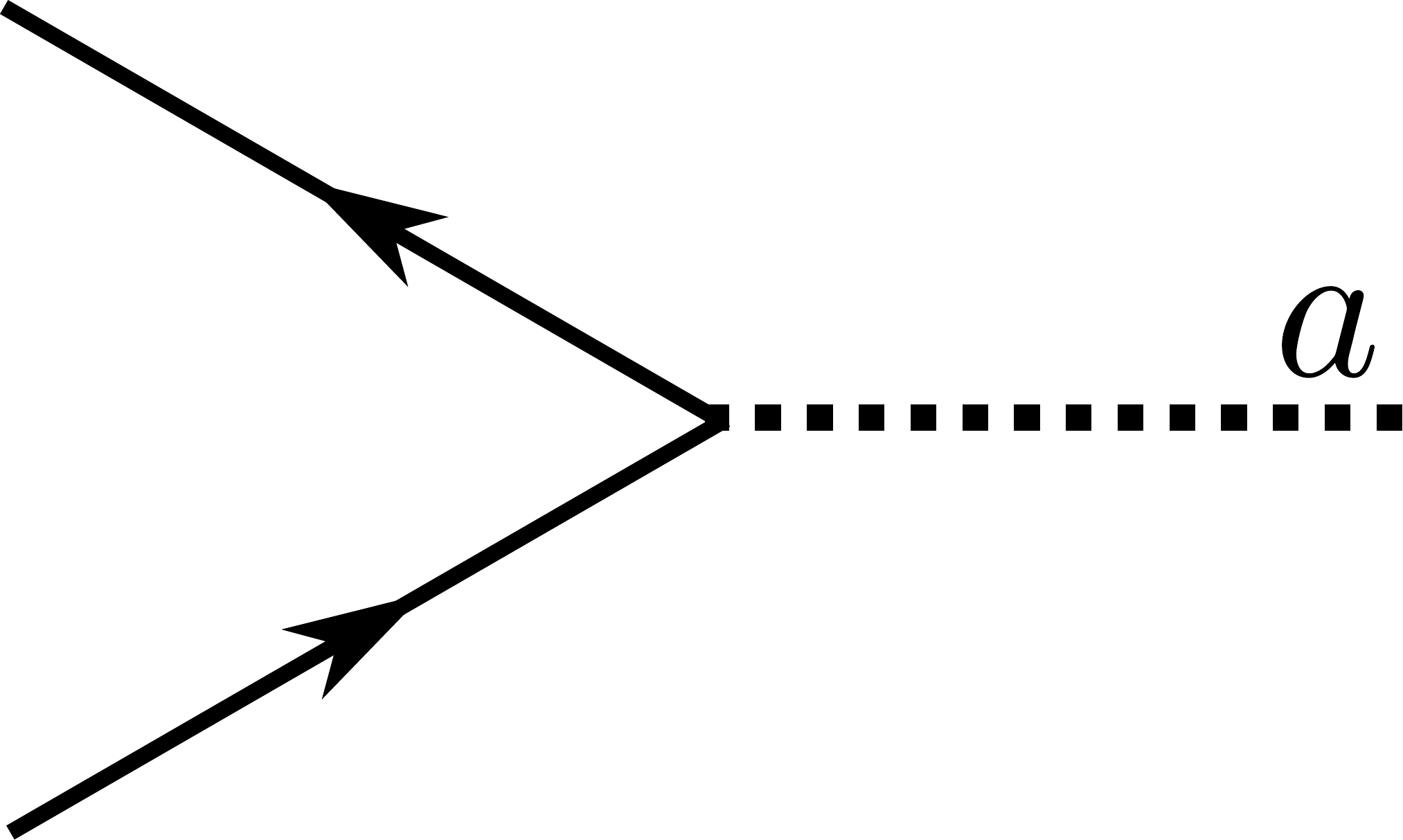} &= h \sigma_a \,,\\
\dia{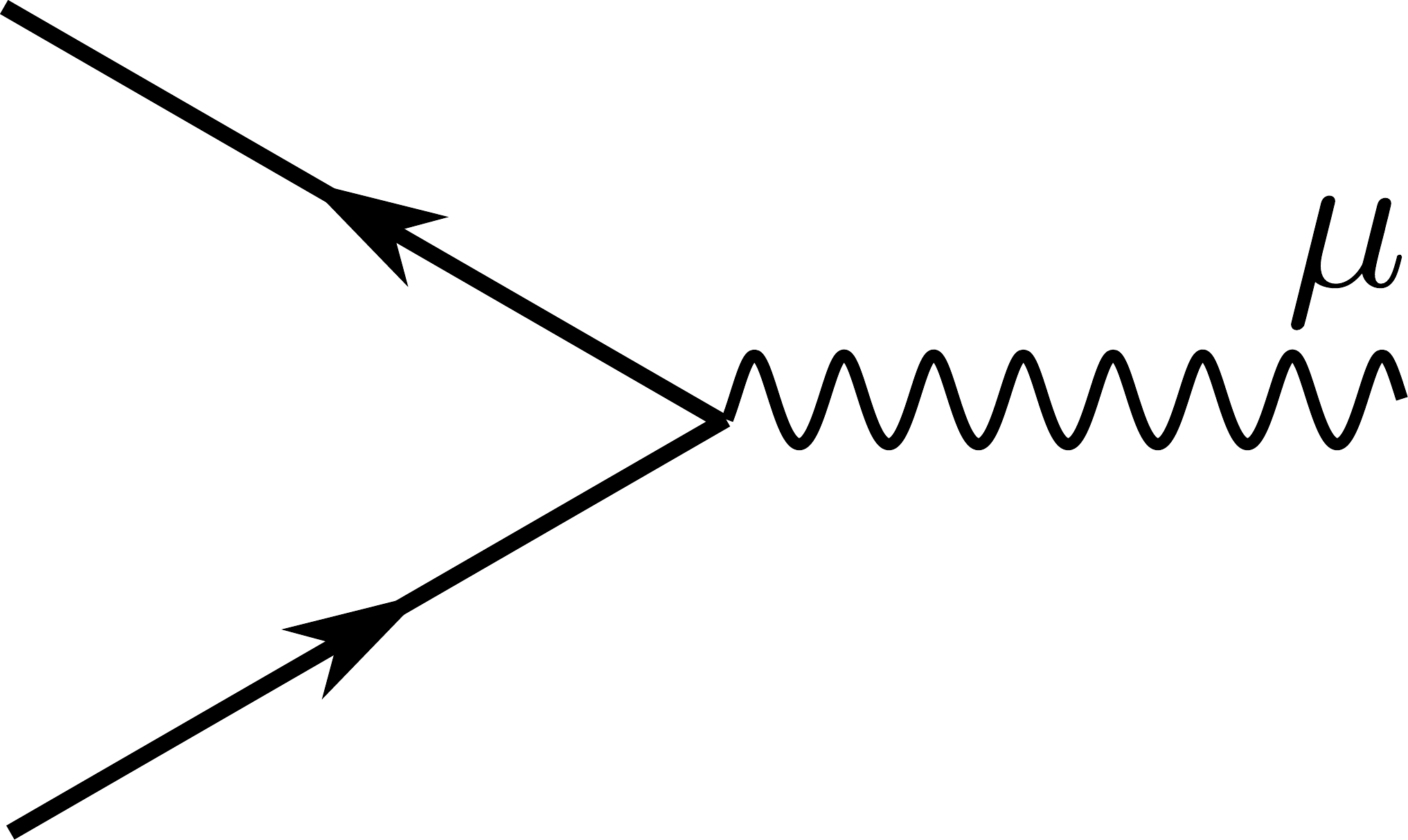}
&=  i e \gamma_\mu \,,\\
\dia{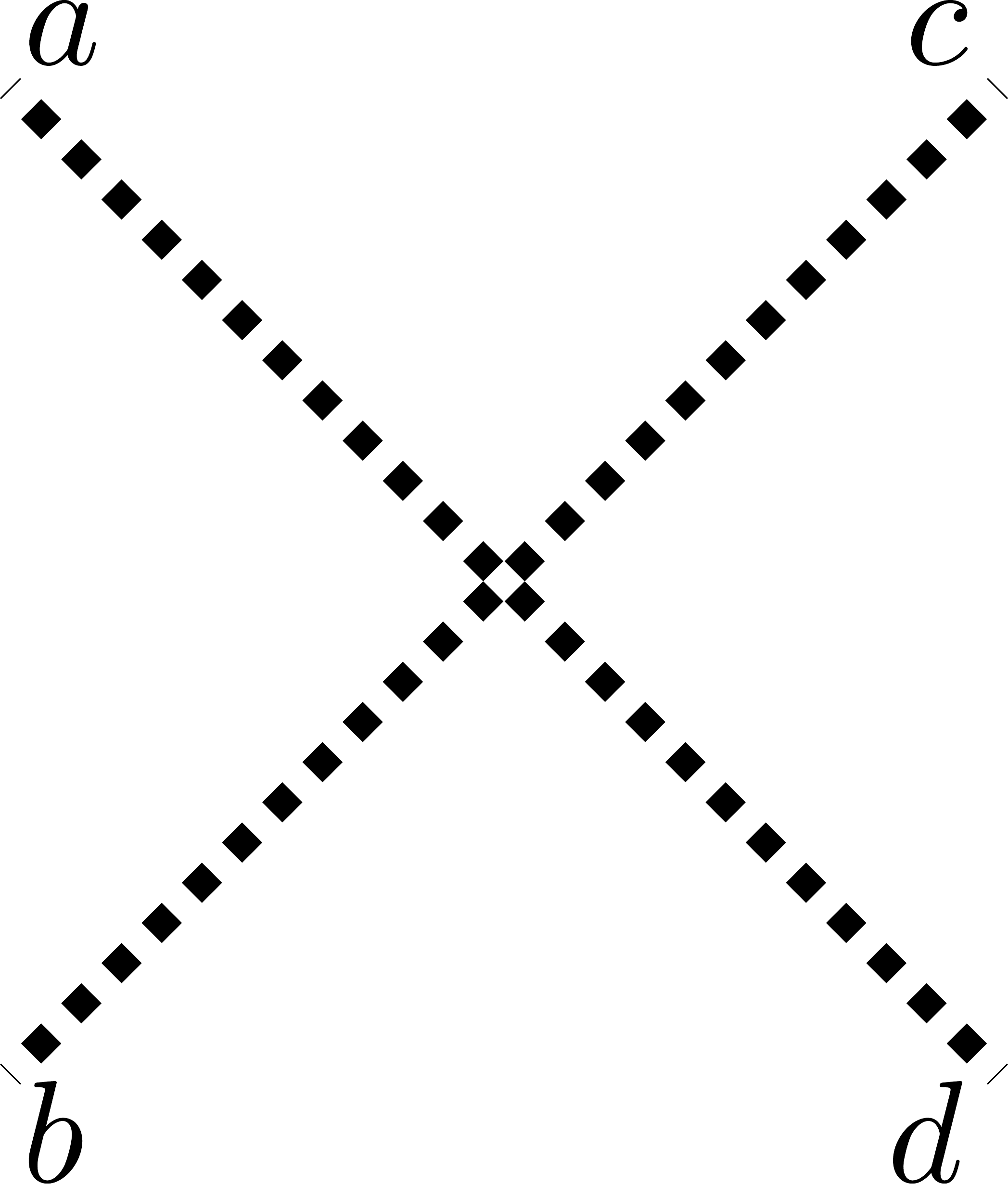}
&=  8 \lambda   \lb \delta_{ab}\delta_{cd} + \delta_{ac}\delta_{bd} + \delta_{ad}\delta_{bc}  \rb \,,
}
where $\xi$ is the gauge-fixing parameter for the gauge field $a_\mu$. Here, $\psi_{\s \v}$ is a two-component Dirac spinor with magnetic spin $\s$ and valley index $\v$. The mass perturbations can also be treated as interactions with the following Feynman rules
\eqn{
\diagen{0.075}{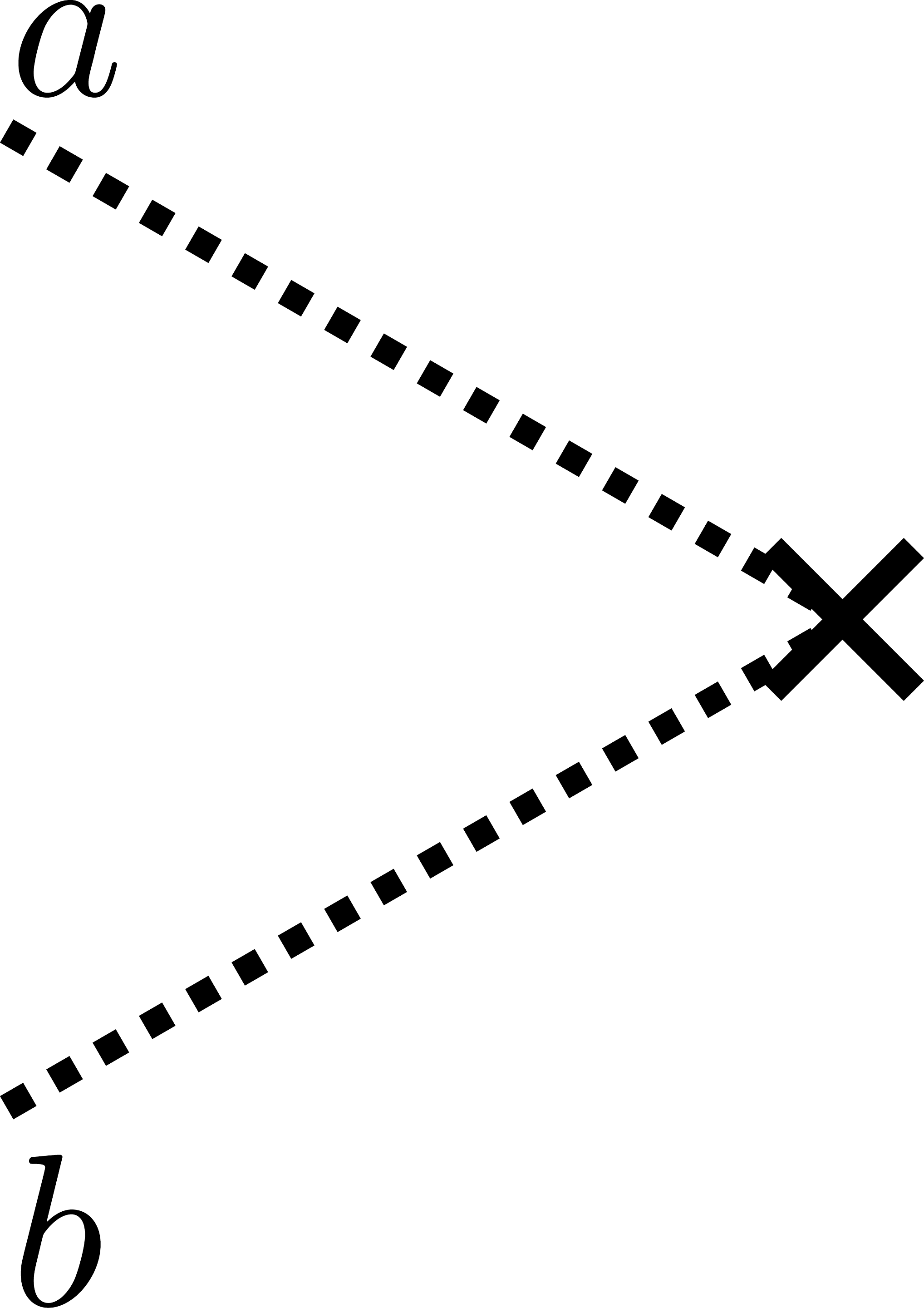}
&= m^2_\phi  \delta_{ab} \,, \\
\diagen{0.075}{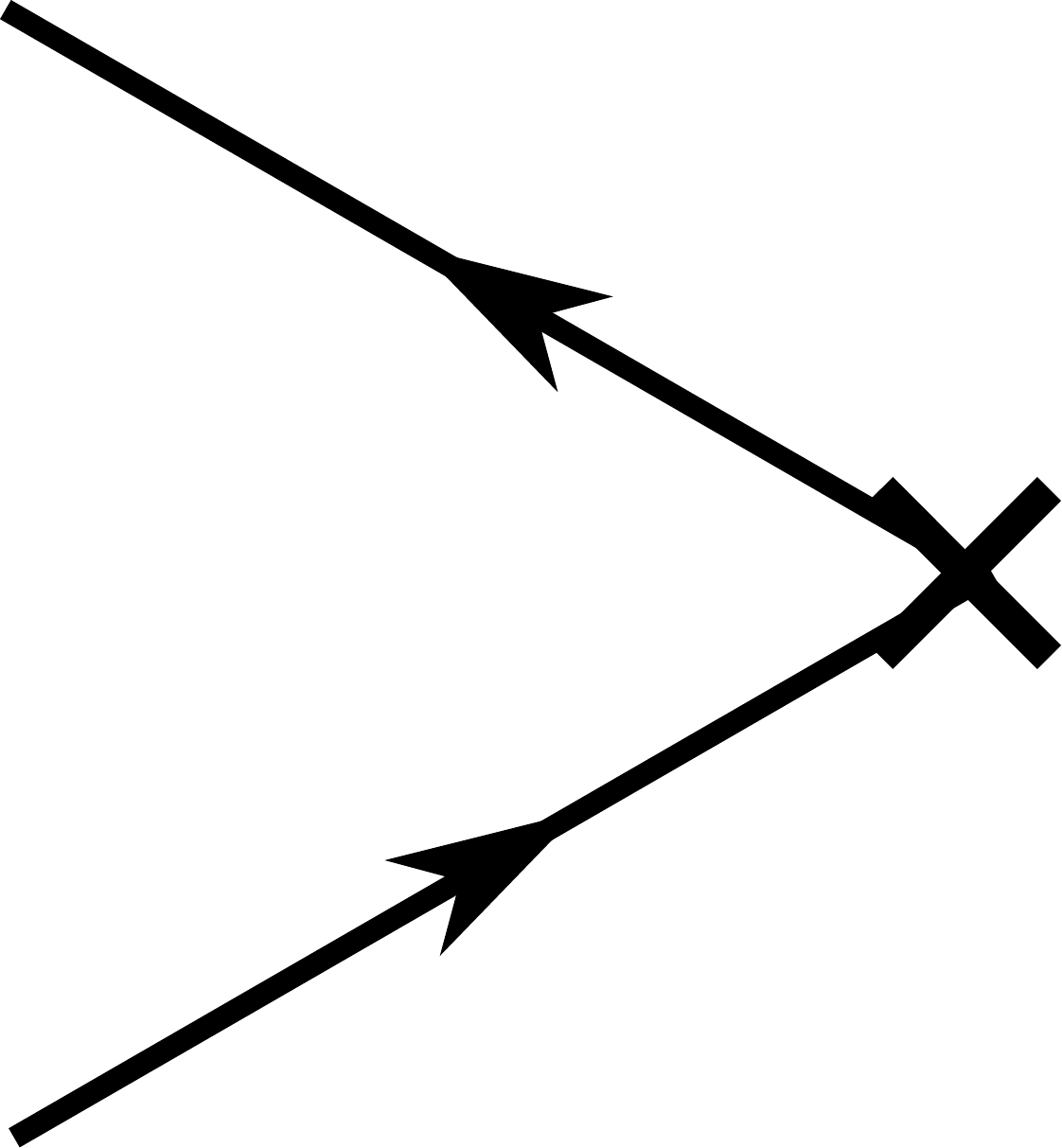}&= m_\psi \,,\\
\diagen{0.075}{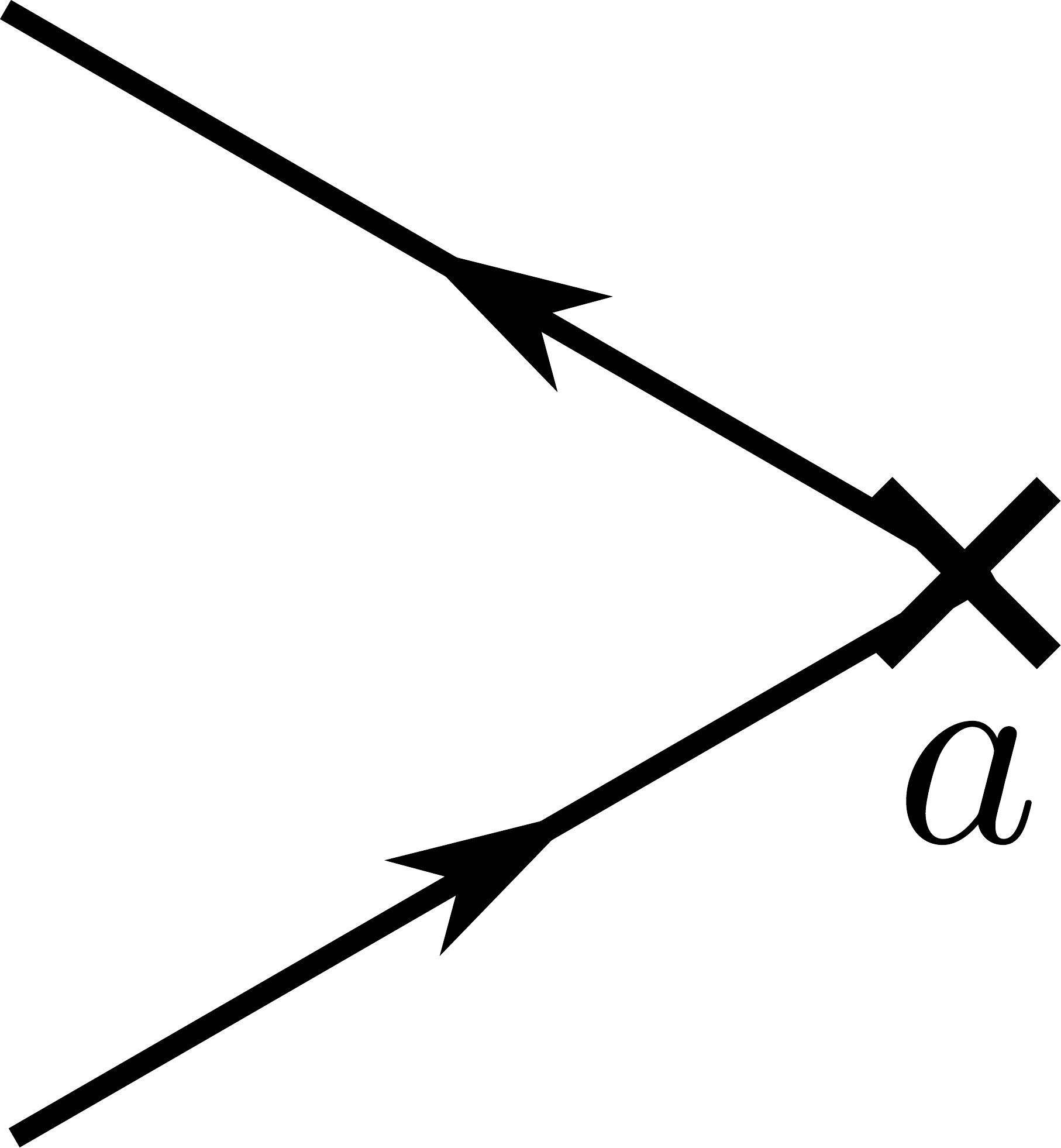} &= \tilde{m}_{\psi} \sigma_a  \,.
}

\subsection{Computation of one-loop Feynman diagrams}
We now compute the one-loop Feynman diagrams using the Feynman rules in Sec.\ref{sec:feyn_rules}. First, let us introduce shorthand notation.  The diagrams involve a momentum shell  loop integral factor $\tilde{I}$ 
\eqn{	
 \tilde{I}
& = 
 \int_{k=\Lambda e^{-l}}^{k=\Lambda} \frac{\dd^d k}{\lb 2 \pi \rb^d} \frac{1}{k^4} =  \lb \frac{\int \dd \Omega_{d-1}}{\lb 2\pi \rb^d} \Lambda^{d-4} \rb  \dd l  \,.
\label{eq:Ih}
}
The momentum space measure is also abbreviated as $\ddi{k} \equiv d^d k /\lb 2 \pi \rb^d$. As we write the Feynman diagrams, we identify  the corresponding correction to the normalization factors, $\delta Z_{x_i}$ with  $x_i \in \{\psi, \phi, a, e, h, \lambda\}$.
\subsubsection{Three-point functions}
We compute the vertex diagrams shown in Fig.~\ref{fig:vertex}
\begin{figure*}[ht!]
\centering
\subfigure[\label{fig:zhb}]
{\includegraphics[height=1.5cm]{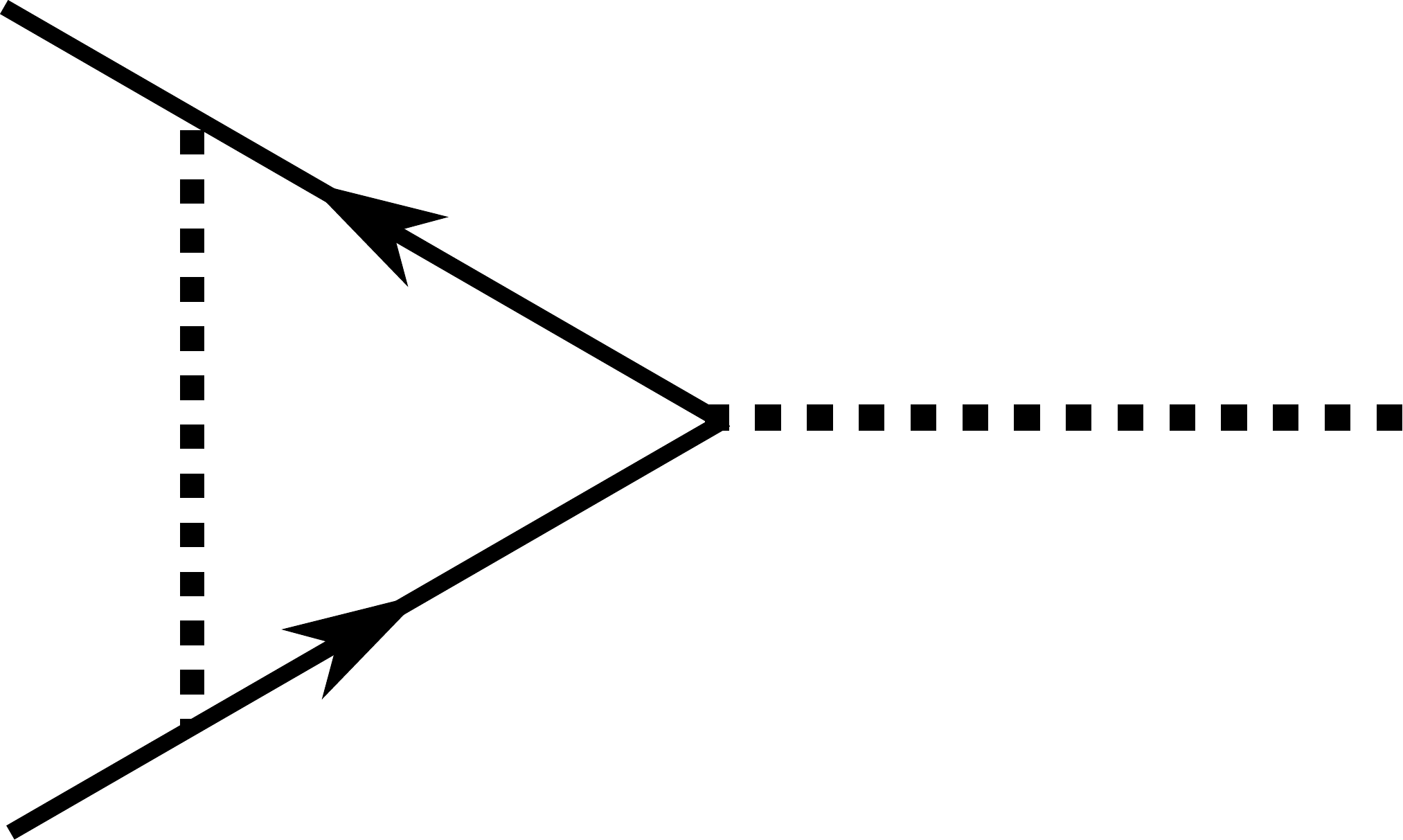}}
\qquad \qquad
\subfigure[\label{fig:zhp}]
{\includegraphics[height=1.5cm]{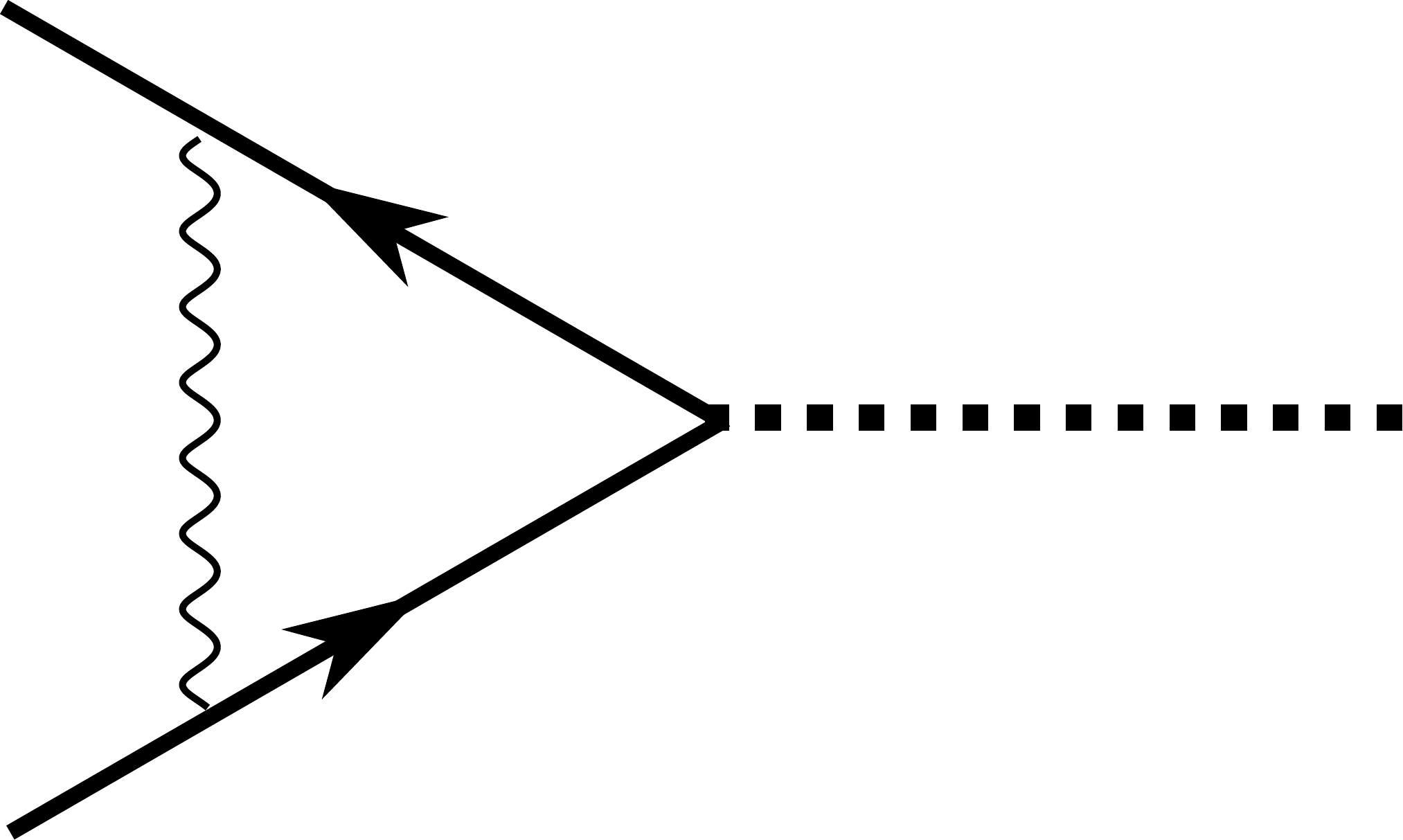}}
\qquad \qquad
\subfigure[\label{fig:zeb}]
{\includegraphics[height=1.5cm]{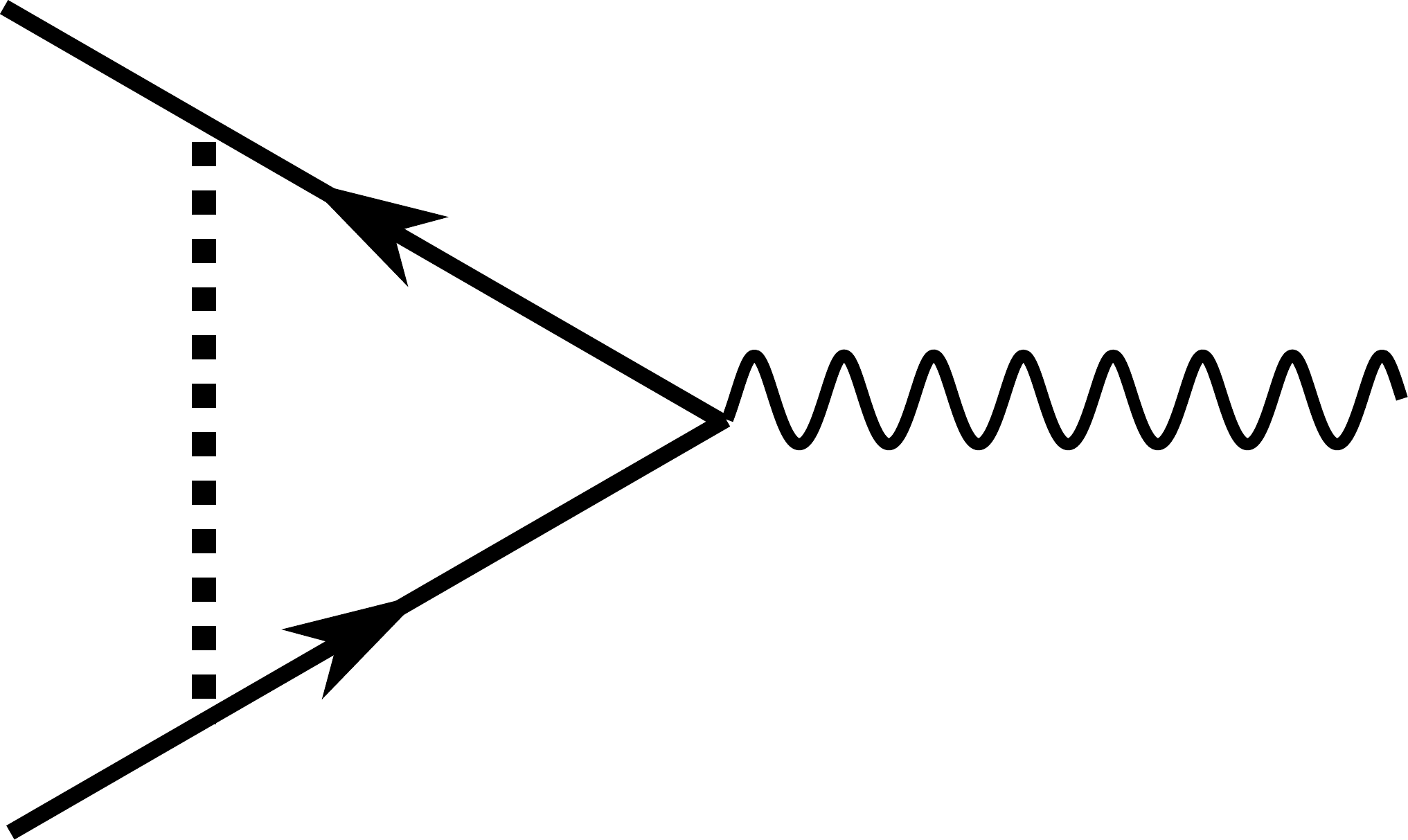}}
\qquad \qquad
\subfigure[\label{fig:zep}]
{\includegraphics[height=1.5cm]{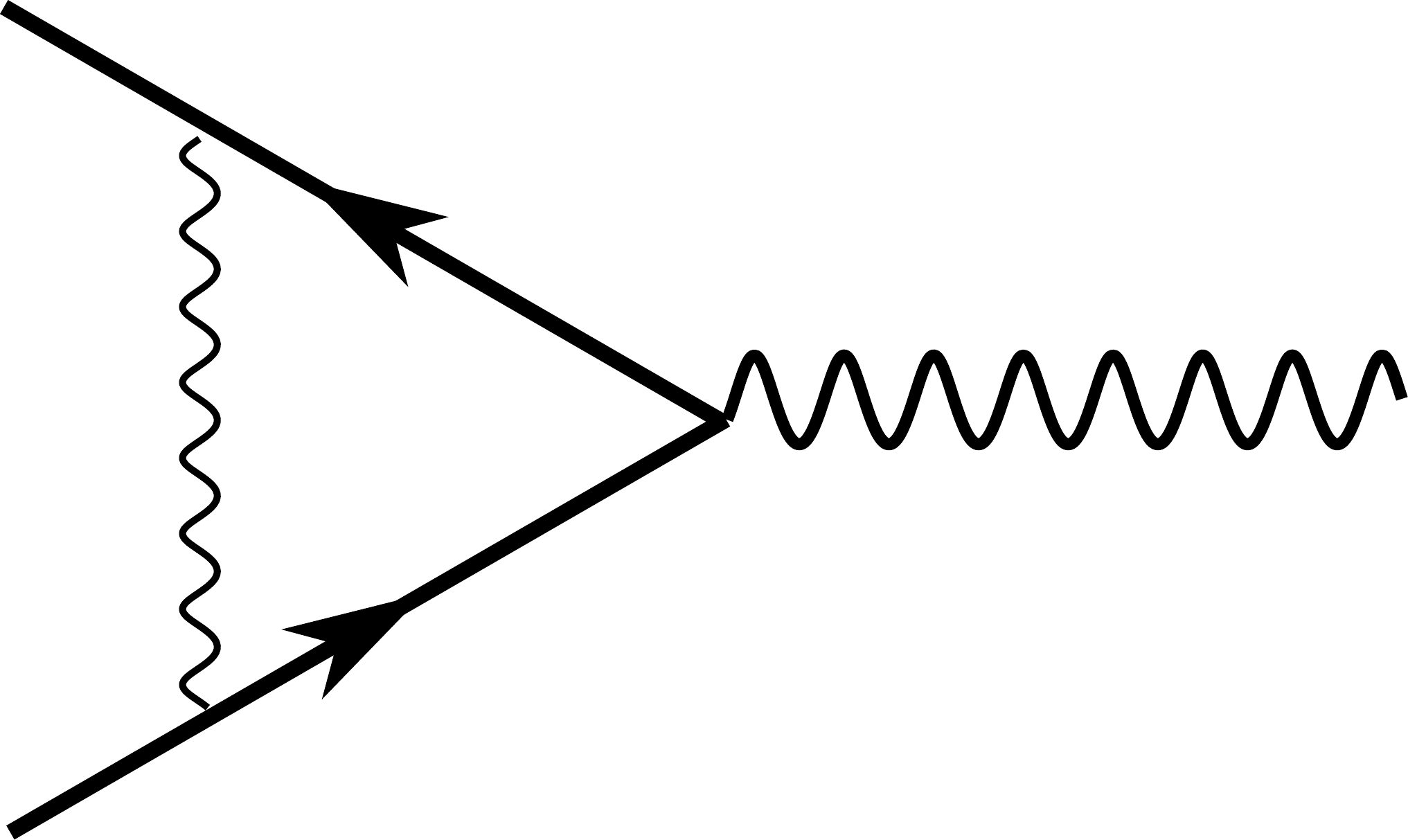}}
\caption{One-loop vertex diagrams. \label{fig:vertex}}
\end{figure*}
\eqna{3}{	
[\ref{fig:zhb}]
&= 
h \sigma_a \lb \delta Z_h^{(\phi)} \rb  
&&= 
h^3  \int \ddi{k} \sigma^b G(k) \sigma_a G(k) \sigma^c D_{bc}(k)
&&=
h \sigma_a \lb  - h^2 \lb \nb-2 \rb   \tilde{I} \rb \,, \\
[\ref{fig:zhp}]
&= 
h \sigma_a  \lb \delta Z_h^{(a)} \rb  
&&=   
h  \lb  i e \rb^2 \int \ddi{k} \gamma^\mu G(k) \sigma_a G(k) \gamma^\nu \Pi_{\mu \nu}(k) 
&&=
h \sigma_a \lb - e^2   \lb d + \xi-1 \rb    \tilde{I}  \rb \,,  \\
[\ref{fig:zeb}]
&=  i e \gamma^\alpha   \lb \delta Z_e^{(\phi)} \rb  
&&=  i e \lb - h \rb^2 \int \ddi{k} \sigma^b G(k) \gamma^\alpha G(k) \sigma^c D_{bc}(k) 
&&=
 i e  \gamma^\alpha  \lb -  h^2 \nb  \lb \frac{d-2}{d} \rb    \tilde{I} \rb \,.
}
The last diagram involves a longer computation
\eqna{2}{
[\ref{fig:zep}]
 &=   i e \gamma^\alpha  \lb  \delta Z_e^{(a)} \rb   
 &&=  i e \lb i e \rb^2 \int \ddi{k} \gamma^\mu G(k) \gamma^\alpha G(k) \gamma^\nu \Pi_{\mu \nu}(k) \nn 
 \\
&&&=
 i e \lb i e \rb^2 \int \ddi{k} \frac{1}{k^4}  \gamma^\mu \gamma^\lambda \gamma^\alpha \gamma^\rho \gamma^\nu \lc\frac{1}{d} \delta^{\mu \nu} \delta^{\lambda \rho} + \lb \frac{\xi-1}{d \lb d+2 \rb} \rb \lb \qgd{\mu}{\nu}{\lambda}{\rho}  \rb \rc 
\nn \\
&&&\uds{}{=}
 i e \lb i e \rb^2 \int \ddi{k} \frac{1}{k^4}  \lb \frac{\lb d-2 \rb^2}{d} \gamma^\alpha   +\lb \frac{\xi-1}{d \lb d+2 \rb} \rb \lb \lb d-2 \rb^2 \gamma^\alpha  + d^2 \gamma^\alpha + \gamma^\rho  \lb 4 \delta^{\alpha \rho} - \lb 4-d \rb \gamma^\alpha \gamma^\rho \rb   \rb \rb \nn \\
&&&=
i e \gamma^\alpha \lb - e^2 \lb d - 5  + \frac{4}{d} + \xi  \rb  \tilde{I} \rb \,,
}
where we used gamma identities 
 \eqn{	
\gamma^\mu \gamma^\alpha \gamma^\beta \gamma^\mu 
&=
4 \delta^{\alpha \beta} -\lb 4-d \rb \gamma^\alpha \gamma^\beta \,,\\
\gamma^\mu \gamma^\nu \gamma^\rho \gamma^\lambda \gamma^\mu 
&=
 - 2  \gamma^\lambda \gamma^\rho \gamma^\nu  + \lb 4-d \rb \gamma^\nu \gamma^\rho \gamma^\lambda \,,
}
and a symmetrization of momentum components appearing in loop integrals,
\eqn{	
k^\lambda k^\rho 
&\to 
\frac{k^2}{d} \delta^{\lambda \rho} \,, \\
k^\alpha k^\beta k^\gamma k^\delta 
&\to 
\frac{k^4}{d \lb d+2 \rb} \lb \delta^{\alpha \beta} \delta^{\gamma \delta} + \delta^{\alpha \gamma} \delta^{\beta \delta} + \delta^{\alpha \delta} \delta^{\beta \gamma} \rb \,.
}

\subsubsection{Two-point functions}
The two-point diagrams are shown in Fig.~\ref{fig:two}.
\begin{figure*}[ht!]
\centering
\subfigure[\label{fig:zfb}]
{\includegraphics[height=1cm]{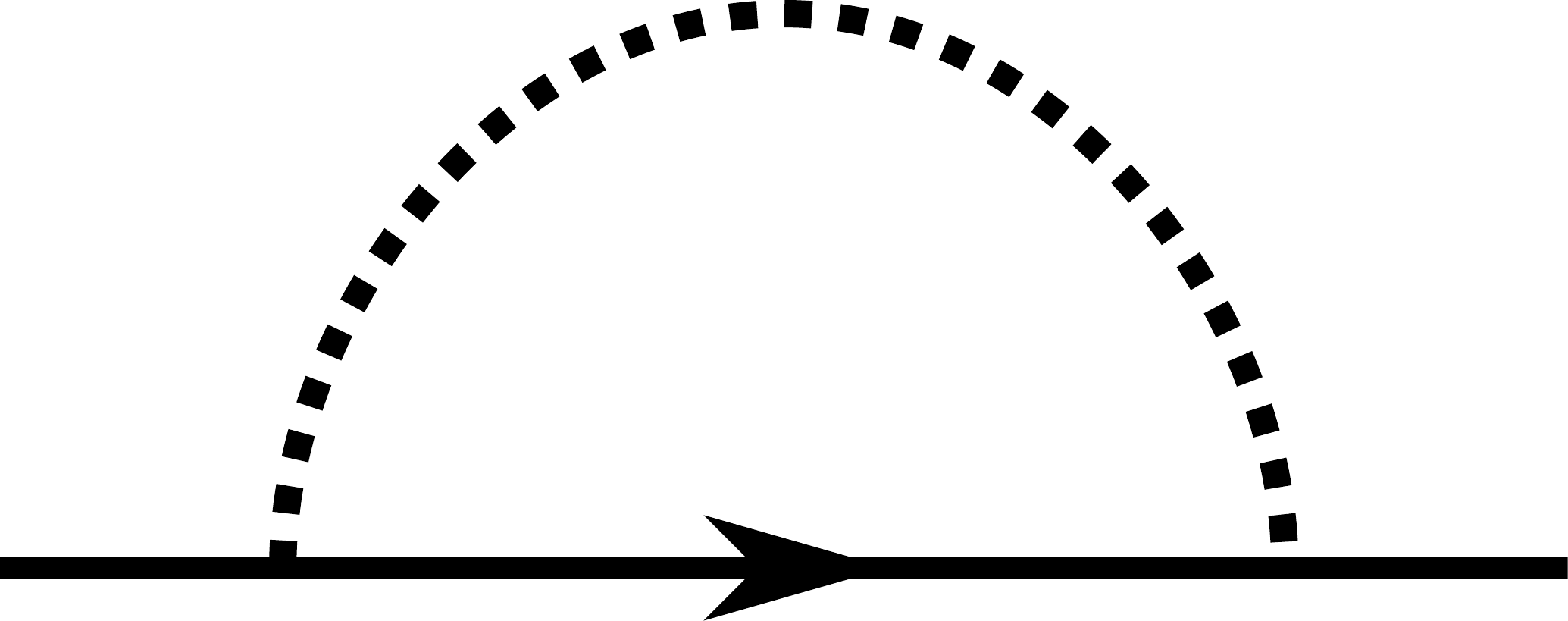}}
\qquad \qquad
\subfigure[\label{fig:zfp}]
{\includegraphics[height=1cm]{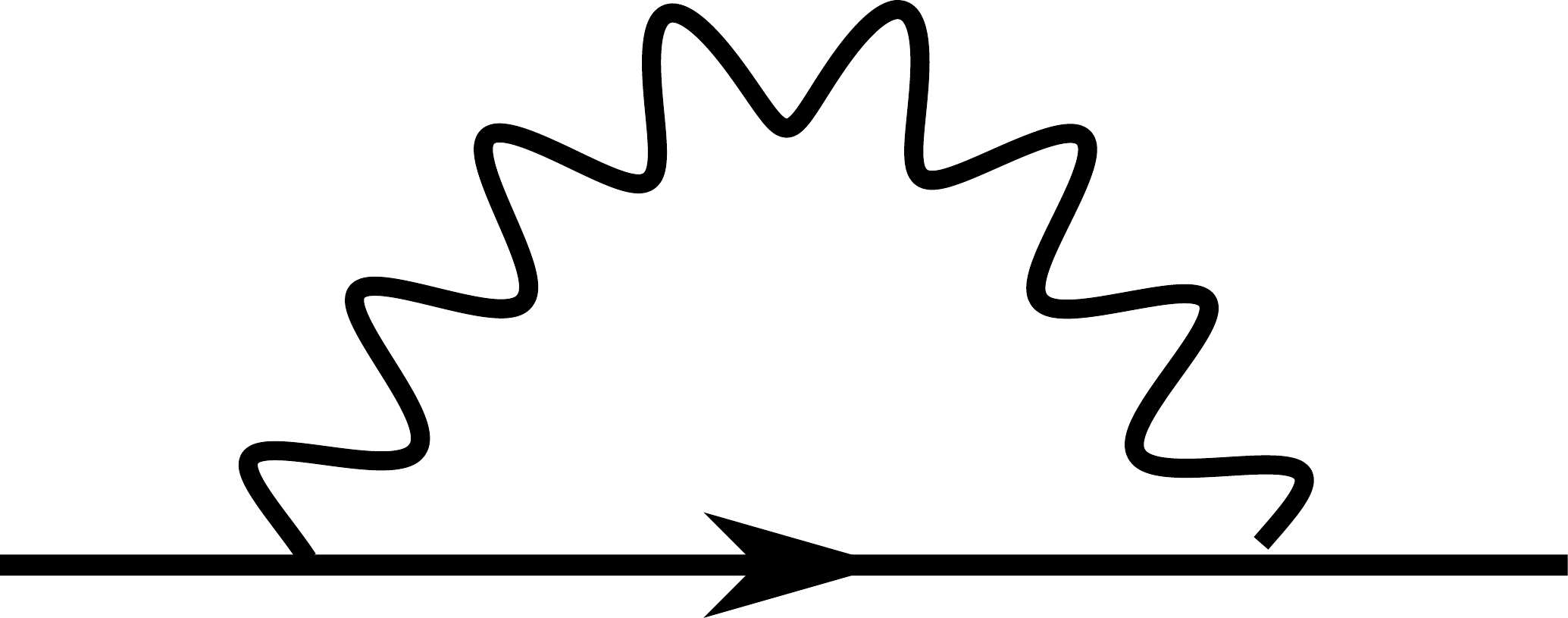}}
\qquad \qquad
\subfigure[\label{fig:zbf}]
{\includegraphics[height=1cm]{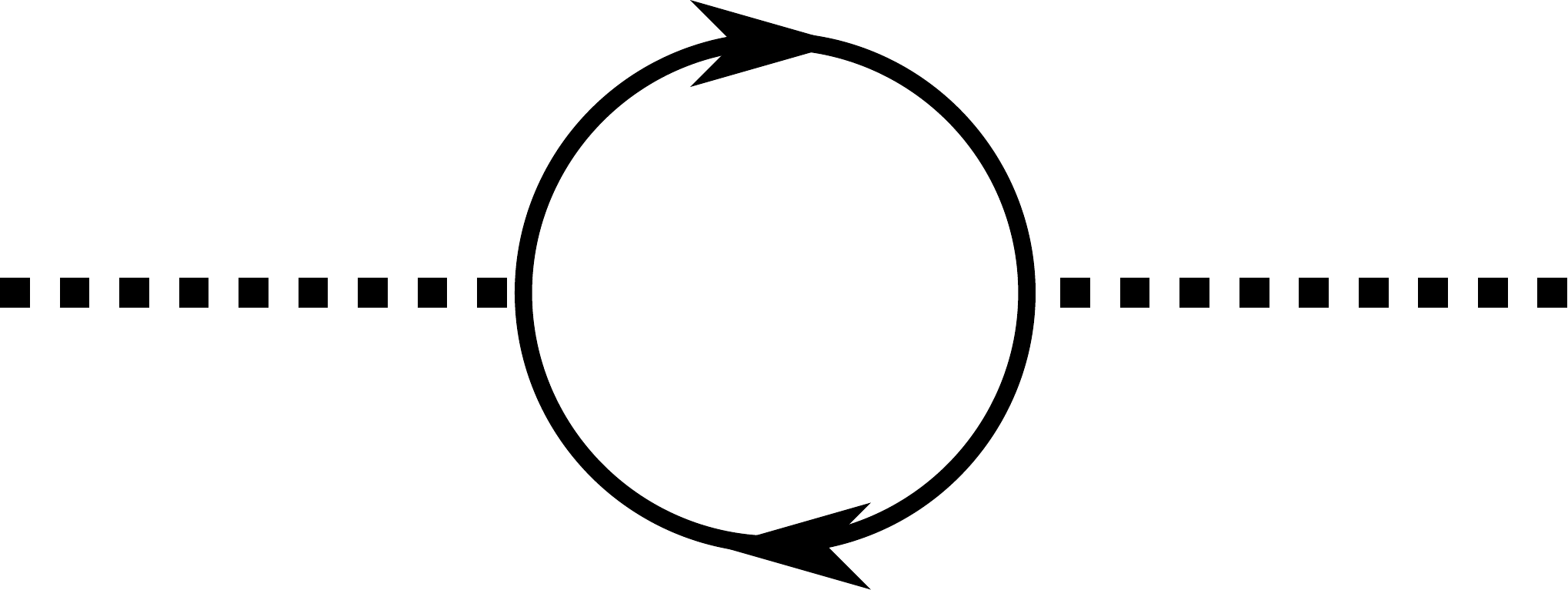}}
\qquad \qquad
\subfigure[\label{fig:zpf}]
{\includegraphics[height=1cm]{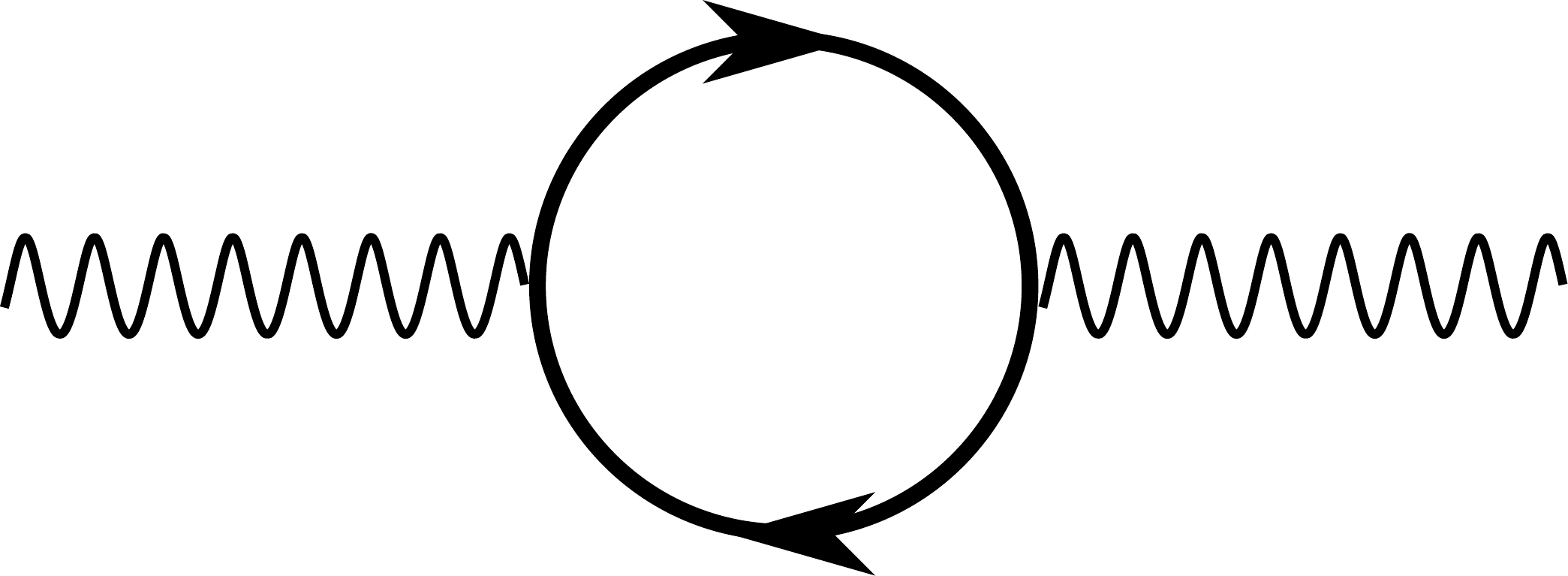}}
\caption{One-loop two-point diagrams. \label{fig:two}}
\end{figure*}
The boson contribution to the fermion self-energy is 
\eqn{
\delta Z_\psi^{(\phi)} = -\pdv{}{\sl{p}} [\ref{fig:zfb}]
&= -\lb \gamma^\mu \rb^{-1}\pdv{}{p^\mu} \lb \lb h \rb^2 \int \ddi{k}  \sigma^b G(p-k) \sigma^c D_{bc}(k)  \rb
\nn \\
&=  \lb \gamma^\mu \rb^{-1}  \lb h \rb^2 \int \ddi{k}  \sigma^b G(p-k) \pdv{G^{-1}(p-k)}{p^\mu} G(p-k) \sigma^c D_{bc}(k) 
\nn \\
&=  \lb \gamma^\mu \rb^{-1}  \lb h \rb^2 \int \ddi{k}  \sigma^b G(p-k) \gamma^\mu  G(p-k) \sigma^c D_{bc}(k) 
\nn \\
&\uds{p \to 0}{=} \delta Z_e^{(\phi)} \,.
}
Similarly, the gauge field  contribution is related to another  vertex correction
\eqn{
\delta Z_\psi^{(A)} = - \pdv{}{\sl{p}} [\ref{fig:zfp}]  = \delta Z_e^{(A)} \,.
}
These results are the  one-loop version of the Ward identity. As for the boson $\bm \phi$ self-energy, it only gets a one-loop contribution by the fermion
\eqn{
[\ref{fig:zbf}]
= 
p^2 \delta_{ab}  \lb  - \delta Z_\phi^{(\psi)}  \rb 
&= -\lb h \rb^2 \int \ddi{k} \tr{\sigma_a G(k) \sigma_b G(k+p)}\nn \\
&=
 -\lb h \rb^2 \lb d_\sigma d_\mu d_\gamma \rb \delta_{ab}  \int \ddi{k} \frac{k \cdot \lb k+p \rb}{k^2 \lb k+p \rb^2} \nn \\
&=
 -\lb h \rb^2 \lb d_\sigma d_\mu d_\gamma \rb \delta_{ab}  \int_0^1 \dd x \int \ddi{k} \frac{ \lb k - px \rb \cdot \lb k+p \lb 1-x \rb \rb}{\lc k^2 + p^2 x \lb 1-x \rb \rc^2} \nn \\
&= p^2 \delta_{ab} \lb 2 \nf  h^2 \tilde{I} \rb \,,
}
where $d_\sigma$, $d_\mu$ and $d_\gamma$ are the size of the respective $\SU(2)$ representations, respectively $2, \nf$ and $2$, and where we introduced a Feynman parameter $x$. The gauge field also only gets a correction from the fermion
\eqn{
[\ref{fig:zpf}] = \lb p^2 \delta_{\mu \nu} - p_\mu p_\nu /p^2 \rb \lb - \delta Z_a^{(\psi)} \rb  &=  - \lb i e \rb^2 \int \ddi{k} \tr{\gamma^\mu G(k) \gamma^\nu G(k+p)} \nn \\
&=   e^2 \int \dd x \int \ddi{k} \frac{\tr{\gamma^\mu \lb \sl{k} - \sl{p} x \rb  \gamma^\nu \lc  \lb \sl{k} +\sl{p} \lb 1-x \rb  \rb \rc }}{\lc k^2 + p^2 x \lb 1-x \rb \rc^2} \nn\\
&=
  e^2 \int \dd x \int \ddi{k} \frac{\tr{\frac{\lb 2-d \rb}{d} k^2 \gamma^\mu  \gamma^\nu  - x \lb 1-x \rb  p^\alpha p^\beta \gamma^\mu \gamma^\alpha \gamma^\nu \gamma^\beta }}{\lc k^2 + p^2 x \lb 1-x \rb \rc^2} \nn \\
&=
 4 \nf  e^2 \int \dd x \int \ddi{k} \frac{\frac{\lb 2-d \rb}{d} k^2 \delta_{\mu \nu}  - x \lb 1-x \rb  p^\alpha p^\beta \lb \qg{\mu}{\alpha}{\nu}{\beta} \rb }{\lc k^2 + p^2 x \lb 1-x \rb \rc^2} \nn \\
&=
  4 \nf e^2 \int \dd x \int \ddi{k}  \frac{ \delta_{\mu \nu} \lc  \frac{\lb 2-d \rb}{d} k^2 - p^2 \rc     +2 x \lb 1-x \rb  \lb \delta_{\mu \nu} p^2 - p^\mu p^\nu  \rb }{\lc k^2 + p^2 x \lb 1-x \rb \rc^2}  \,.
}
The polarization  tensor should be proportional to the transverse tensor $\delta_{\mu \nu} - p_\mu p_\nu / p^2$. This form cannot be achieved with a cut-off regularization which breaks Lorentz invariance.  Terms that don't satisfy this form are removed. Since a $p_\mu p_\nu$ term cannot be generated by expansion of the denominator, the $p_\mu p_\nu$ term in the numerator serves as a reference to single out the polarization tensor
\eqn{
[\ref{fig:zpf}]
= \lb p^2 \delta_{\mu \nu} - p_\mu p_\nu /p^2 \rb \lb - \delta Z_a^{(\psi)} \rb 
&=
\lb p^2 \delta_{\mu \nu} - p_\mu p_\nu /p^2 \rb \lb  4 \nf e^2  \lb \int \dd x  \,\,  2x \lb 1-x \rb  \rb \tilde{I} \rb \nn \\
&=\lb p^2 \delta_{\mu \nu} - p_\mu p_\nu /p^2 \rb  \lb  \frac{4 \nf}{3} e^2  \tilde{I} \rb \,.
}

\subsubsection{Four-point function}
The four-point diagrams are shown in Fig.~\ref{fig:four}.
\begin{figure*}[ht!]
\centering
\subfigure[\label{fig:zlb}]
{\includegraphics[width=0.25\linewidth]{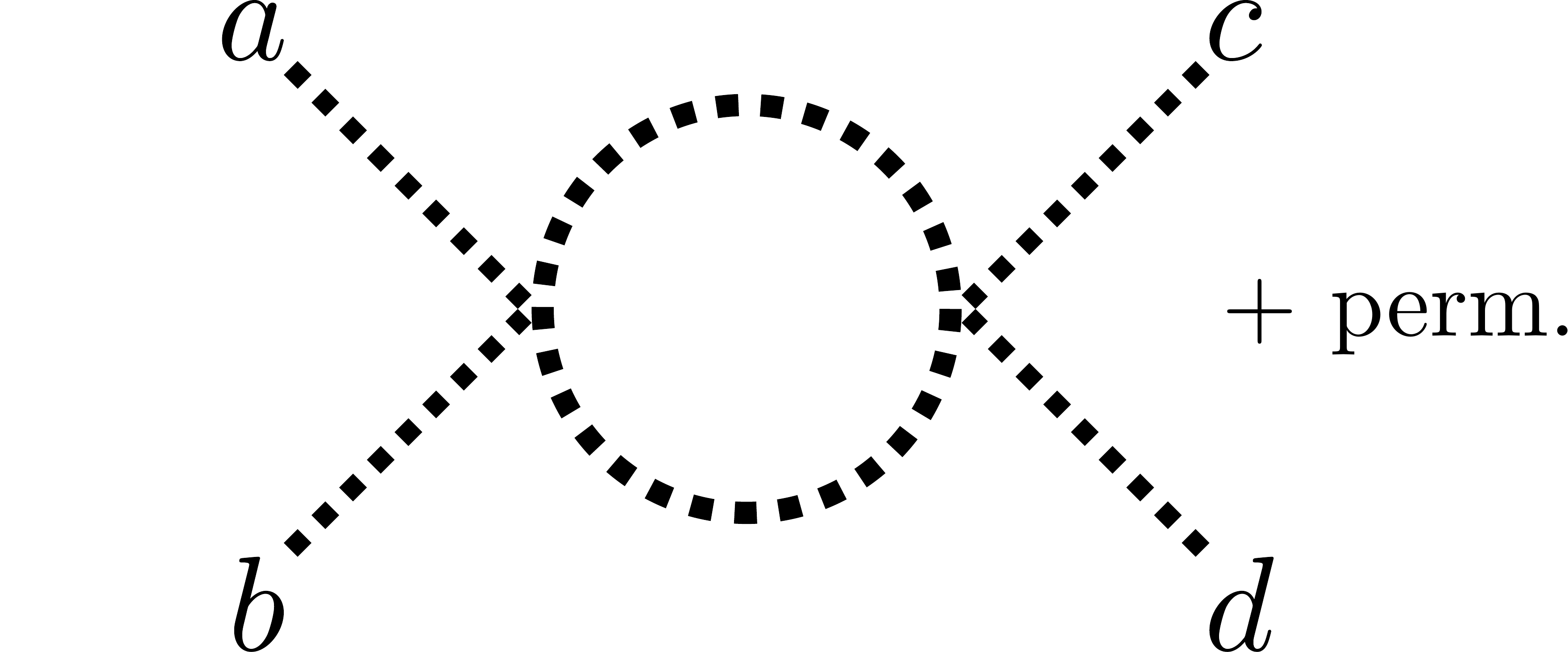}} 
\qquad \qquad
\subfigure[\label{fig:zlf}]
{\includegraphics[width=0.22\linewidth]{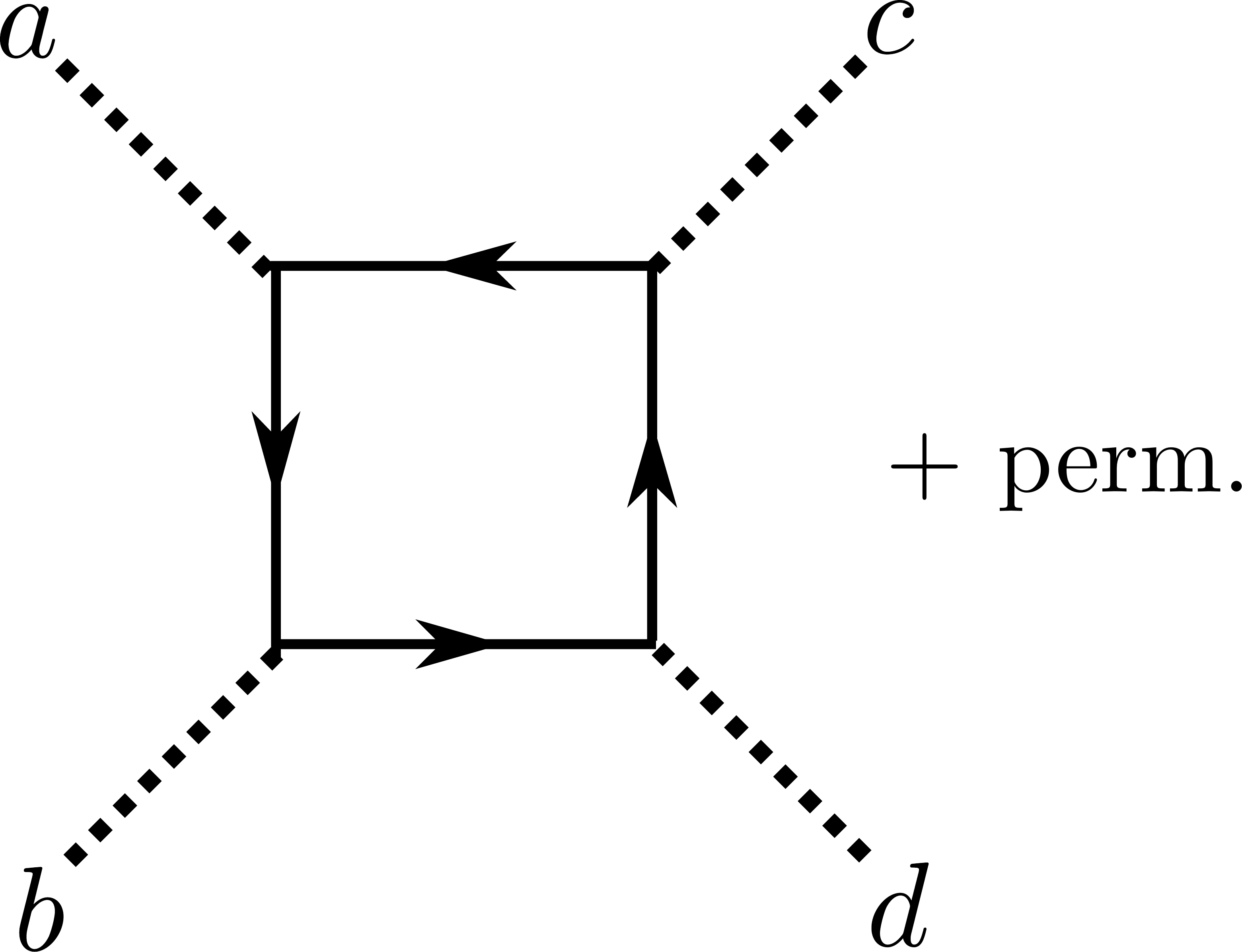}}
\caption{One-loop four-point diagrams. \label{fig:four}}
\end{figure*}

  The boson autointeraction's correction has a contribution from the boson
\eqn{
[\ref{fig:zlb}] =  8 \lambda  \lb \delta_{ab}\delta_{cd} + \delta_{ac}\delta_{bd} + \delta_{ad}\delta_{bc}  \rb \lb  \delta Z_\lambda^{\phi} \rb
&= 
\half 
\lc  8 \lambda  \lb \delta_{ab}\delta_{ef} + \delta_{ae}\delta_{bf} + \delta_{af}\delta_{be}  \rb \rc  
\lc 8 \lambda \lb \delta_{cd}\delta_{ef} + \delta_{ce}\delta_{df} + \delta_{cf}\delta_{de}  \rb   \rc 
 \tilde{I} + \text{perm.} \nn \\
&=
  8 \lambda \lb \delta_{ab}\delta_{cd} + \delta_{ac}\delta_{bd} + \delta_{ad}\delta_{bc}  \rb  \lb 4 \lb \nb + 8 \rb \lambda \tilde{I} \rb \,,
}
where the factor $1/2$ is  a symmetry factor. The fermion also contributes to the correction
\eqn{
[\ref{fig:zlf}]
=
  8 \lambda   \lb \delta_{ab}\delta_{cd} + \delta_{ac}\delta_{bd} + \delta_{ad}\delta_{bc}  \rb \lb \delta Z_{\lambda}^{\psi}  \rb  &=
-h^4  \lb \tr{\sigma_a \sigma_b \sigma_c \sigma_d} + \tr{\sigma_d \sigma_c \sigma_b \sigma_a} \rb \tilde{I}\nn \\
&=
 8 \lambda \lb \delta_{ab}\delta_{cd} + \delta_{ac}\delta_{bd} + \delta_{ad}\delta_{bc}  \rb \lb - h^4 \lambda^{-1} \nf \tilde{I} \rb \,.
}

\subsubsection{Mass perturbations}
Diagrams corresponding to mass perturbation are shown in Fig.~\ref{fig:mass}.
\begin{figure*}[ht!]
\centering
\subfigure[\label{fig:zmb}]
{\includegraphics[height=1.5cm]{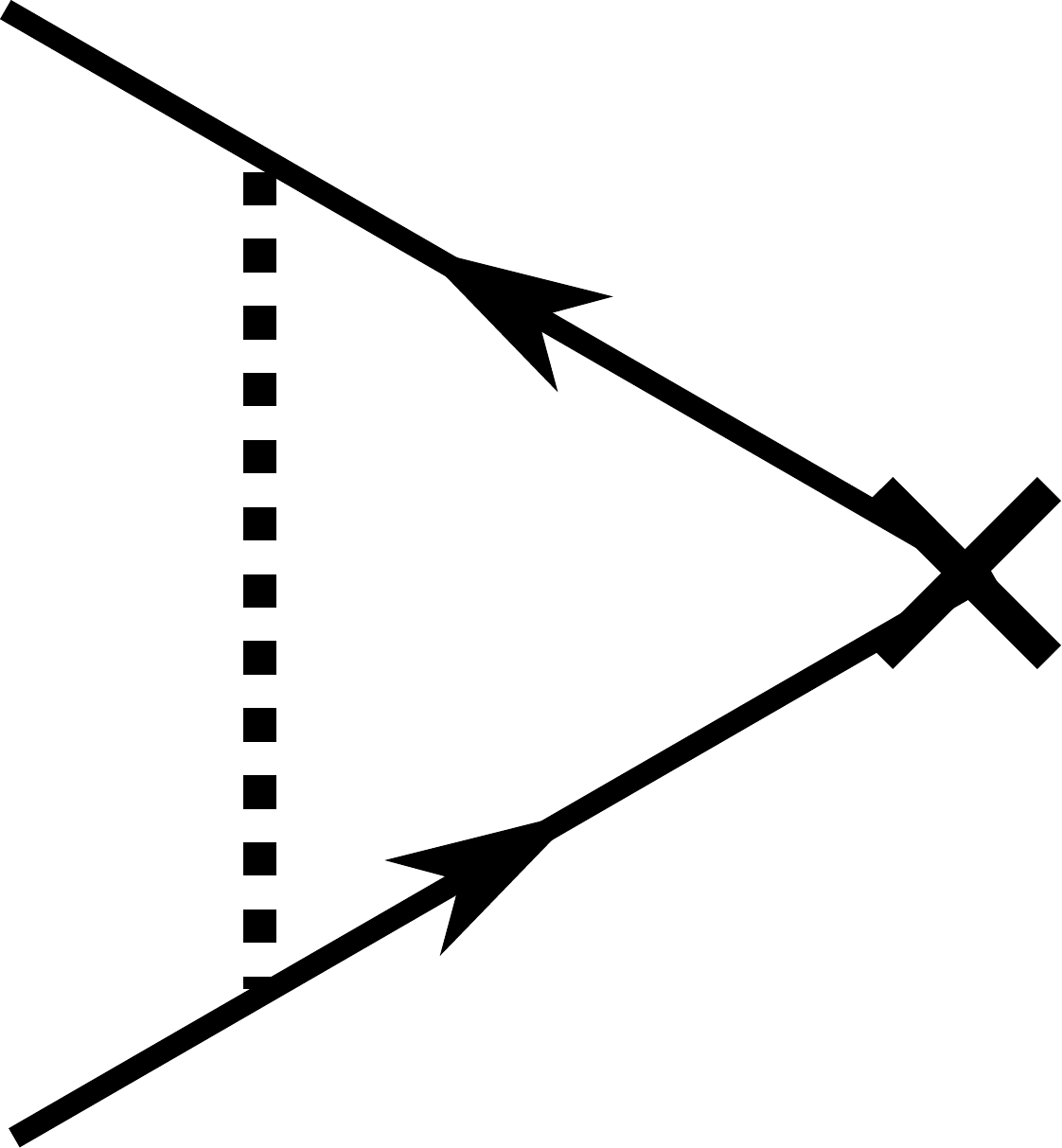}}
\qquad \qquad
\subfigure[\label{fig:zmp}]
{\includegraphics[height=1.5cm]{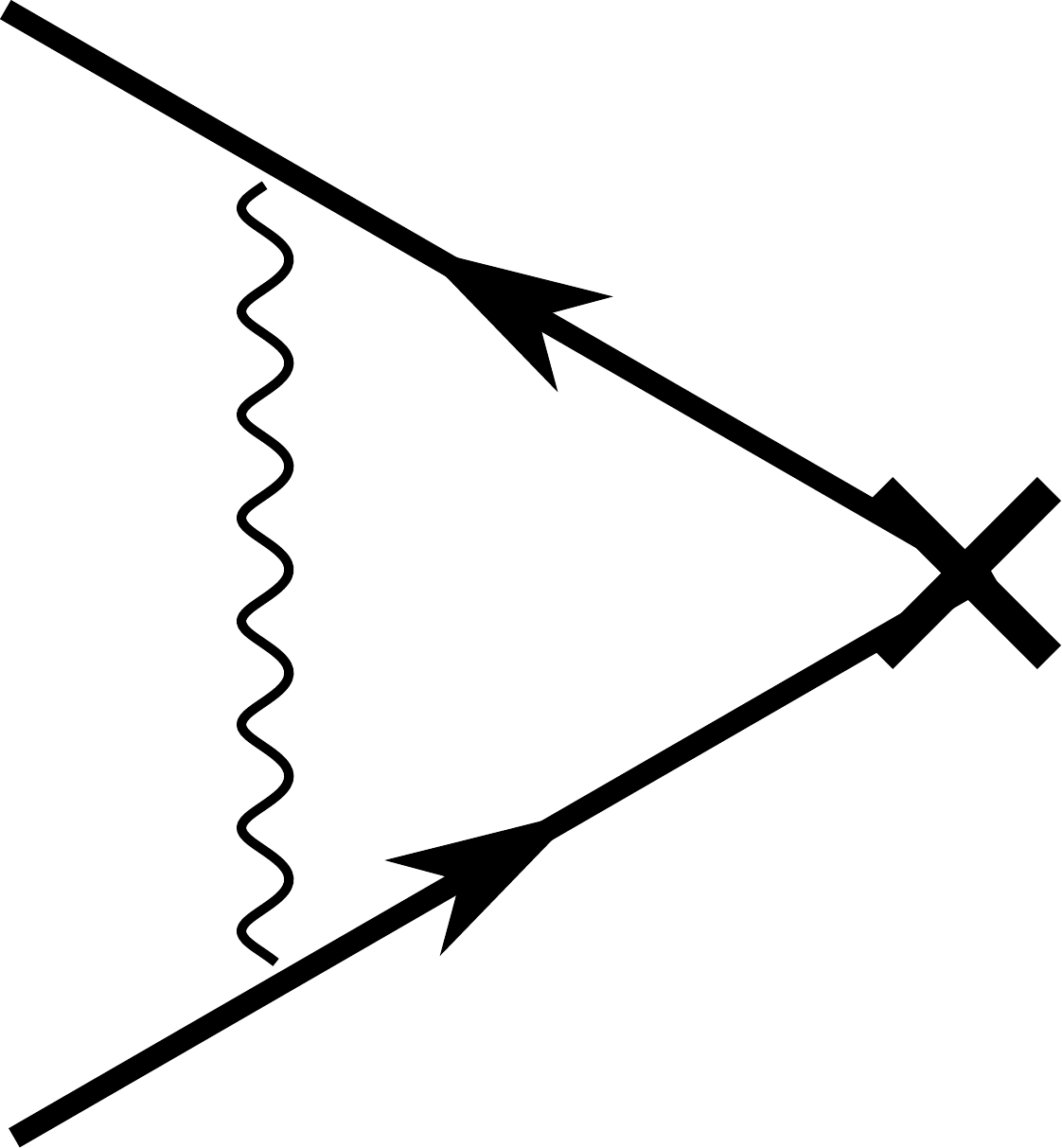}}
\qquad \qquad
\subfigure[\label{fig:zmsb}]
{\includegraphics[height=1.5cm]{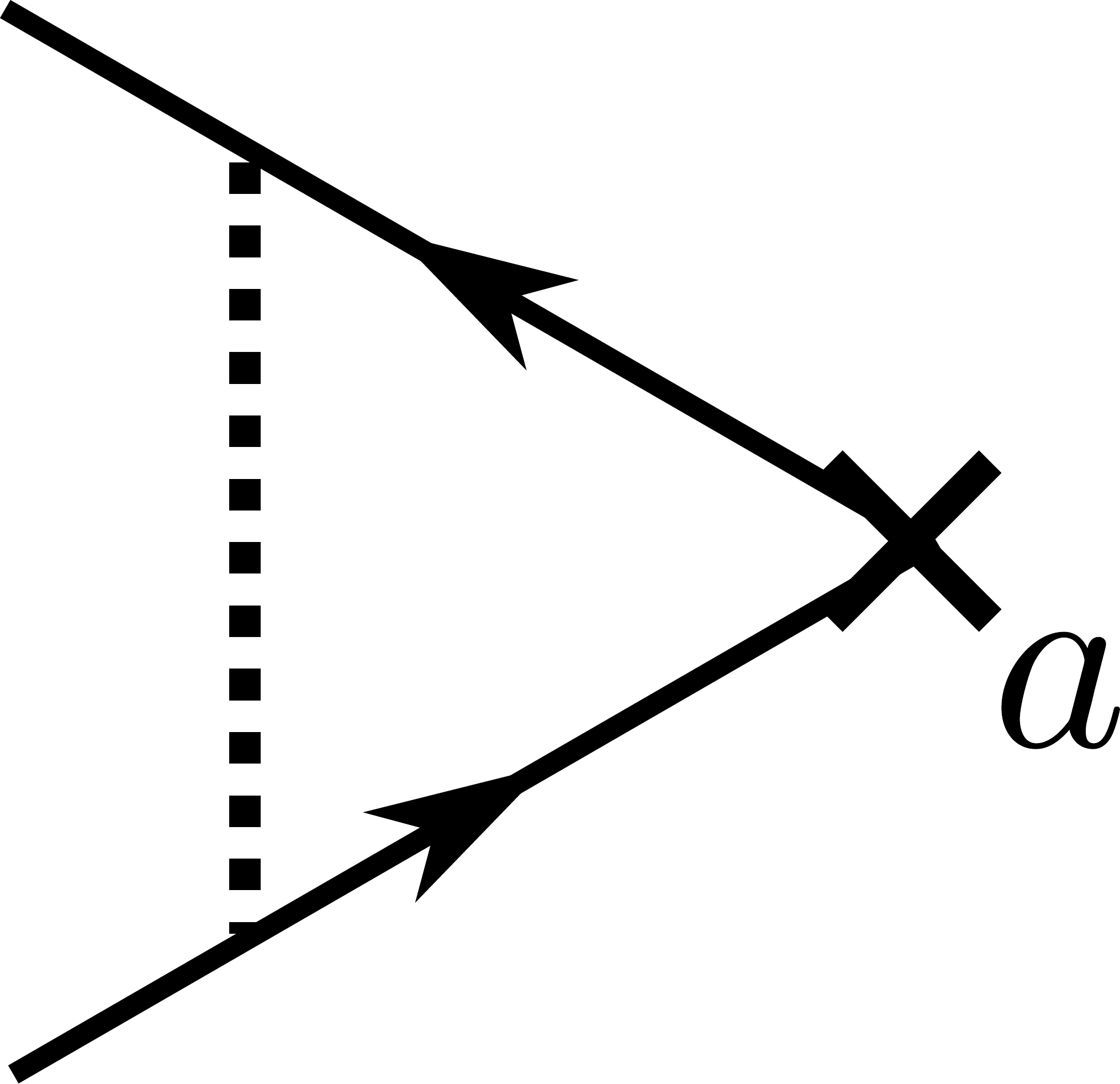}}
\qquad \qquad
\subfigure[\label{fig:zmsp}]
{\includegraphics[height=1.5cm]{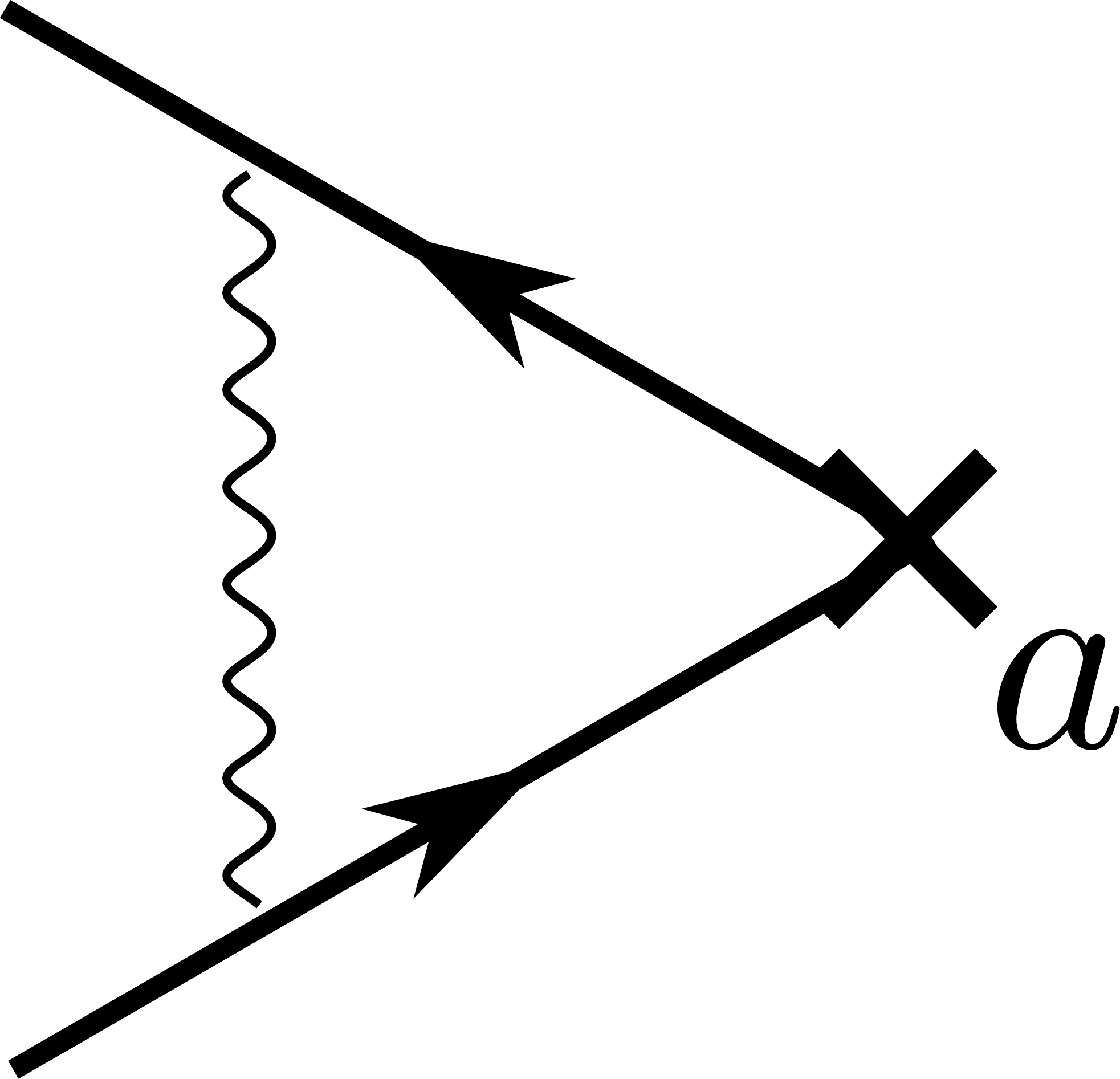}}
\qquad \qquad
\subfigure[\label{fig:zmbb}]
{\includegraphics[height=1.5cm]{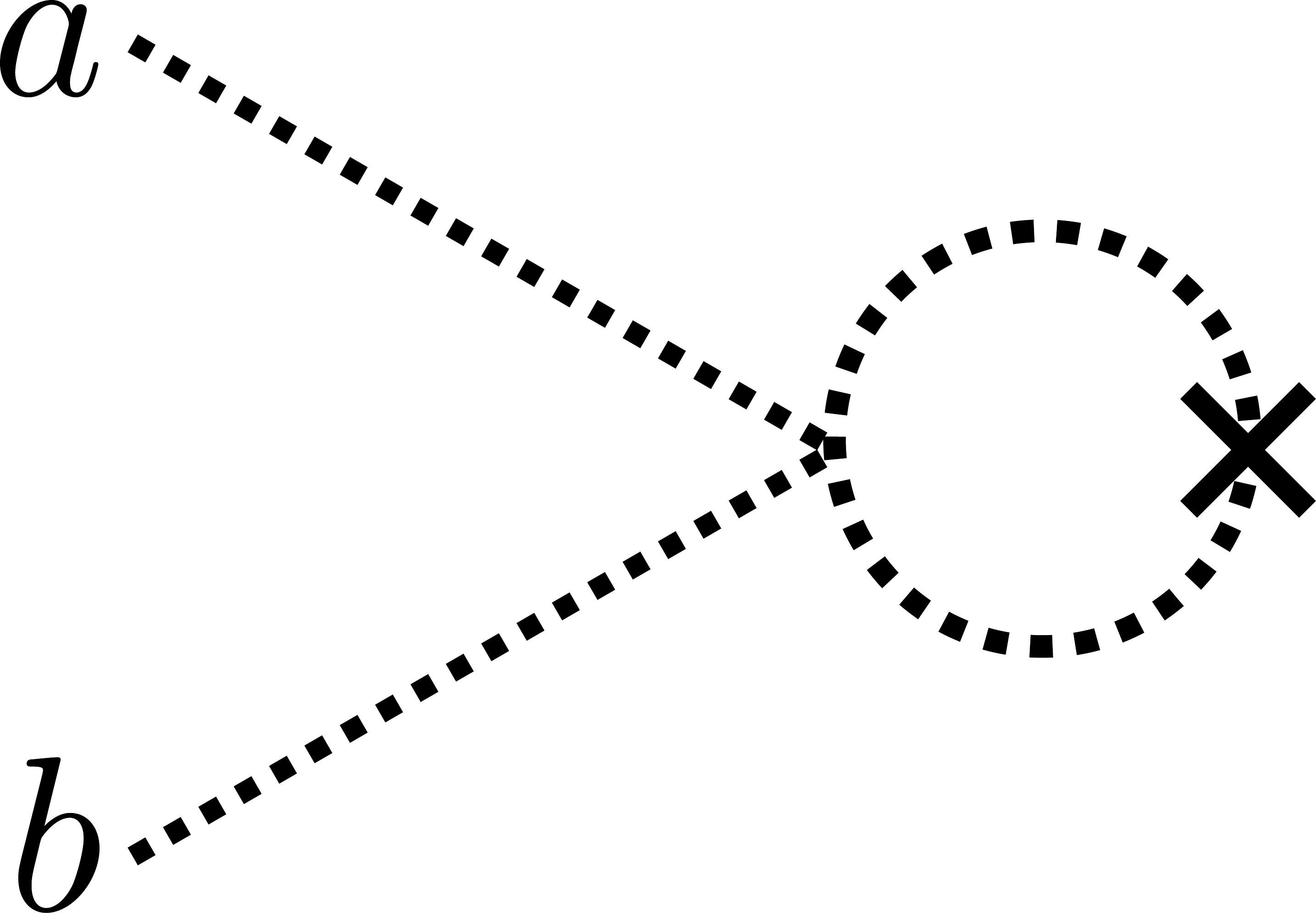}}
\caption{One-loop diagrams corresponding to mass perturbations. \label{fig:mass}}
\end{figure*}
Some of the one-loop diagrams for the bilinear perturbations can be related to  previously computed diagrams 
\eqna{3}{	
[\ref{fig:zmb}]
&= m_{\psi} \lb  \delta Z_{m_\psi}^{(\phi)} \rb
&&= m_\psi \nb \frac{1}{\lb \lb 2-\nb \rb h\tr{\Id} \rb} \tr{ \sigma_a [\ref{fig:zhb}]_a} 
&&= m_\psi \lb \nb  h^2 \tilde{I} \rb \,,   \\
[\ref{fig:zmp}]
&= m_{\psi} \lb  \delta Z_{m_\psi}^{(a)} \rb
&&= m_\psi \frac{1}{\lb h\tr{\Id} \rb} \tr{\sigma_a [\ref{fig:zhp}]_a} 
&&= m_\psi \lb  -\lb d+\xi-1 \rb e^2 \tilde{I} \rb \,,
\\
[\ref{fig:zmsb}]
&= \tilde{m}_{\psi} \sigma_a\lb  \delta Z_{\tilde{m}_{\psi}}^{(\phi)} \rb
&&= \tilde{m}_{\psi}  h^{-1} [\ref{fig:zhb}]_a
&&= \tilde{m}_{\psi} \sigma_a  \lb -  \lb \nb - 2 \rb h^2 \tilde{I} \rb \,,  \\
[\ref{fig:zmsp}] 	
&= \tilde{m}_{\psi} \sigma_a\lb \delta Z_{\tilde{m}_{\psi}}^{(a)} \rb 
&&= \tilde{m}_{\psi} h^{-1} [\ref{fig:zhp}]_a
&&=  \tilde{m}_{\psi}  \sigma_a \lb -  \lb d+\xi-1 \rb e^2 \tilde{I} \rb \,.  
} 
Only the $\bm \phi$ field mass correction requires a new computation
\eqn{
[\ref{fig:zmbb}]
= m_{\phi}^2 \delta_{ab} \lb  \delta Z_{m_{\phi}^2}^{(\phi)} \rb
&= \half \, 8 \lambda \lb \delta_{ab} \delta_{cd} + \delta_{ac} \delta_{bd} + \delta_{ad} \delta_{bc} \rb \lb m_\phi^2 \delta_{cd} \rb \int \ddi{k} \frac{1}{k^4} \nn \\
&=    m_\phi^2 \delta_{ab} \lb 4  \lb \nb + 2 \rb \lambda \rb  \tilde{I}  \,.
}

\subsection{Results}
The renormalization constants are obtained by summing the  corrections found above, $ Z_{x_i} =  1 + \sum_{\Phi = \phi, \psi,  a} \delta Z_{x_i}^{(\Phi)}$ where $x_i \in \{\psi, \phi, a, e, h, \lambda\}$. In turn, this allows us to obtain the coefficients $\gamma_{x_i} = - \dd \ln Z_{x_i} / \dd l \approx - \dd Z_{x_i} / \dd l$  defined in the main text in Eq.~\eqref{eq:gamma} which are given by
\eqn{
\gamma_\phi			&=  2 \nf h^2   \,,			 \\
\gamma_\psi 		&=  \lb d - 5 + \frac{4}{d} + \xi \rb e^2   +    \nb \lb \frac{d-2}{d} \rb   h^2   \,, \\
\gamma_a			&=   \frac{4\nf}{3}e^2    \,,  \\
\gamma_e			&= \gamma_\psi \,,\\
\gamma_h			&=   \lb d + \xi - 1 \rb e^2 +  \lb \nb - 2 \rb h^2       \,, \\
\gamma_\lambda 		&=  - 4 \lb \nb + 8 \rb \lambda   +  \nf h^4 \lambda^{-1}     \,,\\
\gamma_{m_\phi^2} 	&=   - 4 (\nb +2) \lambda \,    \,,\\
\gamma_{m_\psi} 	&=    (d + \xi -1 )e^2 - \nb h^2     \,,\\
\gamma_{\tilde{m}_\psi}	&=  (d+\xi -1) e^2  +  \left(N_b-2\right) h^2     \,,
}
where we have eliminated the loop integral factor by rescaling the coupling constants 
\eqn{
(e^2, h^2, \lambda) \to  \lb \frac{\int \dd \Omega_{d-1}}{\lb 2\pi \rb^d} \Lambda^{d-4} \rb^{-1} (e^2, h^2, \lambda) \,.
}

\end{widetext}
\end{document}